\documentclass[preprint,superscriptaddress,nofootinbib]{revtex4-1}
\usepackage{hyperref}
\usepackage{bbm}
\usepackage{amsfonts}
\usepackage{mathrsfs}
\usepackage{mathtools}
\usepackage{bbold}

\usepackage{amsmath,amssymb}

\usepackage{epsfig}
\usepackage{graphicx}               
\usepackage{url}
\usepackage{hyperref}
\usepackage{float}
\usepackage{pstricks}
\usepackage{color}
\usepackage{multirow}
\usepackage{slashed}

\newcommand{\lsim}{\!\mathrel{\hbox{\rlap{\lower.55ex \hbox{$\sim$}} \kern-.34em \raise.4ex \hbox{$<$}}}}
\newcommand{\gsim}{\!\mathrel{\hbox{\rlap{\lower.55ex \hbox{$\sim$}} \kern-.34em \raise.4ex \hbox{$>$}}}}

\def\be{\begin{equation}}
\def\ee{\end{equation}}
\def\bea{\begin{eqnarray}}
\def\eea{\end{eqnarray}}
\def\bit{\begin{itemize}}
\def\eit{\end{itemize}}

\newcommand{\x}{\chi}
\newcommand{\xbar }{\overline{\chi}}
\newcommand{\sfe}{\tilde{f}}
\newcommand{\sq}{\tilde{q}}
\newcommand{\nl}{\nonumber \\}

\newcommand{\Dslash}{D\!\!\!\!/\,\,}
\newcommand{\vslash}{v\!\!\!/\,}
\newcommand{\order}{ {\cal O} }

\def\to{\rightarrow}

\definecolor{ms}{rgb}{0.8, 0.3,0.3}

\begin{document}
\title{The Bino Variations: Effective Field Theory Methods for Dark Matter Direct Detection}
\author{Asher Berlin}
\affiliation{
Department of Physics, Enrico Fermi Institute \\
\vskip -8 pt
University of Chicago, Chicago, IL 60637 USA}
\author{Denis S. Robertson}
\affiliation{Theoretical Physics Group \\ \vskip -8 pt Lawrence Berkeley National Laboratory, Berkeley, CA 94709 USA}
\affiliation{Berkeley Center for Theoretical Physics \\ \vskip -8 pt University of California, Berkeley, CA 94709 USA}
\affiliation{Instituto de F\'isica, Universidade de S\~ao Paulo \\ \vskip -8 pt
R. do Mat\~ao, 187, S\~ao Paulo, SP 05508-900, Brazil}
\author{Mikhail P. Solon}
\affiliation{Theoretical Physics Group \\ \vskip -8 pt Lawrence Berkeley National Laboratory, Berkeley, CA 94709 USA}
\affiliation{Berkeley Center for Theoretical Physics \\ \vskip -8 pt University of California, Berkeley, CA 94709 USA}
\author{Kathryn M. Zurek}
\affiliation{Theoretical Physics Group \\ \vskip -8 pt Lawrence Berkeley National Laboratory, Berkeley, CA 94709 USA}
\affiliation{Berkeley Center for Theoretical Physics \\ \vskip -8 pt University of California, Berkeley, CA 94709 USA}

\begin{abstract}
We apply effective field theory methods to compute bino-nucleon scattering, in the case where tree-level interactions are suppressed and the leading contribution is at loop order via heavy flavor squarks or sleptons. We find that leading log corrections to fixed-order calculations can increase the bino mass reach of direct detection experiments by a factor of two in some models. These effects are particularly large for the bino-sbottom coannihilation region, where bino dark matter as heavy as 5-10 TeV may be detected by near future experiments.  For the case of stop- and selectron-loop mediated scattering, an experiment reaching the neutrino background will probe thermal binos as heavy as 500 and 300 GeV, respectively.  We present three key examples that illustrate in detail the framework for determining weak scale coefficients, and for mapping onto a low energy theory at hadronic scales, through a sequence of effective theories and renormalization group evolution. For the case of a squark degenerate with the bino, we extend the framework to include a squark degree of freedom at low energies using heavy particle effective theory, thus accounting for large logarithms through a ``heavy-light current." Benchmark predictions for scattering cross sections are evaluated, including complete leading order matching onto quark and gluon operators, and a systematic treatment of perturbative and hadronic uncertainties. 
\end{abstract}

\preprint{}

\maketitle

\tableofcontents

\section{\textbf{Introduction}}
\label{sec:introduction}

Decades of technological advances and increased detector sizes have led to impressive projected sensitivities of on-going and future dark matter (DM) direct detection experiments \cite{Cushman:2013zza,Akerib:2012ys,Akerib:2015cja,Aprile:2011dd}.  For DM with mass $10^2 - 10^4$ GeV, the LUX-ZEPLIN (LZ) experiment is projected to reach cross sections as small as $ \sigma_{\rm SI} \sim 10^{-47} - 10^{-48} \mbox{ cm}^2$, tantalizingly close to the neutrino background, residing at cross sections an order of magnitude smaller. As these experiments extend their reach, they will push through a number of important benchmarks in the hunt for Weakly Interacting Massive Particles (WIMPs). 

Current experiments are in fact already probing rates several orders of magnitude below ``weak-scale'' cross sections: constraints from LUX and Xenon100 reach as low as $\sigma_{\rm SI} \sim 10^{-45} \mbox{ cm}^2$, while a simple estimate suggests that the spin-independent (SI) scattering cross section through the $Z$-boson is $\sigma_{\rm SI}  \sim 10^{-39} \mbox{ cm}^2$. The scattering of a WIMP on nucleon targets, however, depends strongly on its identity.  While a scalar electroweak doublet has a large cross section through the $Z$-boson, Majorana fermions have no vector coupling, and the axial-vector interactions are either $v^2$-suppressed or lead to spin-dependent (SD) scattering.

At tree-level, this leaves scattering through the Higgs boson as the process for leading SI interactions. For neutralinos, the size of the scattering through the Higgs boson depends on its electroweak composition. Triplet (``wino''), doublet (``higgsino''), and singlet (``bino'') states mix with each other, allowing the lightest stable neutral WIMP, $\chi$, to couple to the Higgs at tree-level: $\lambda_\chi ~ h ~ {\bar \chi} \chi$~. This gives rise to a typical scattering cross section $\sigma_{\rm SI} \sim \left(\frac{\lambda_\chi}{0.1}\right)^2 10^{-45} \mbox{ cm}^2 $. Thus, the currently running and next generation ton-scale experiments are probing tree-level ``Higgs-interacting'' massive particles.

Pure electroweak states (wino, Higgsino, or bino), however, do not couple to the Higgs at tree-level.  For these cases, the evaluation of direct scattering of the lightest electrically-neutral state on nucleon targets requires the analysis of loop amplitudes at leading order. Assuming weak-scale mediators, a simple estimate of the scattering cross section is given by $\sigma_{\rm SI}  \sim \alpha_w^4 {m_N^4 / m_{\rm weak}^6}  \sim 10^{-46} \, {\rm cm}^2$~, where $m_N$ is the nucleon mass and $m_{\rm weak} \sim 100$ GeV. The prospects for wino and Higgsino dark matter, however, are challenged by an accidental cancellation between amplitudes, leading to cross sections smaller by a few orders of magnitude \cite{Hisano:2011cs, Hill:2011be, Hill:2013hoa,Hisano:2015rsa}.  For the wino, the cross section was found to be $\sigma_{\rm SI} \sim  10^{-47} \mbox{ cm}^2$, while for the Higgsino, the cancellation gives rise to an unreachably small scattering cross section. Nonetheless, it is remarkable that in some cases, while the tree-level cross section may be absent, ton-scale direct detection experiments are becoming sensitive to one-loop interactions. 

\begin{figure}[t]
\begin{center}
\includegraphics[width=0.6\textwidth]{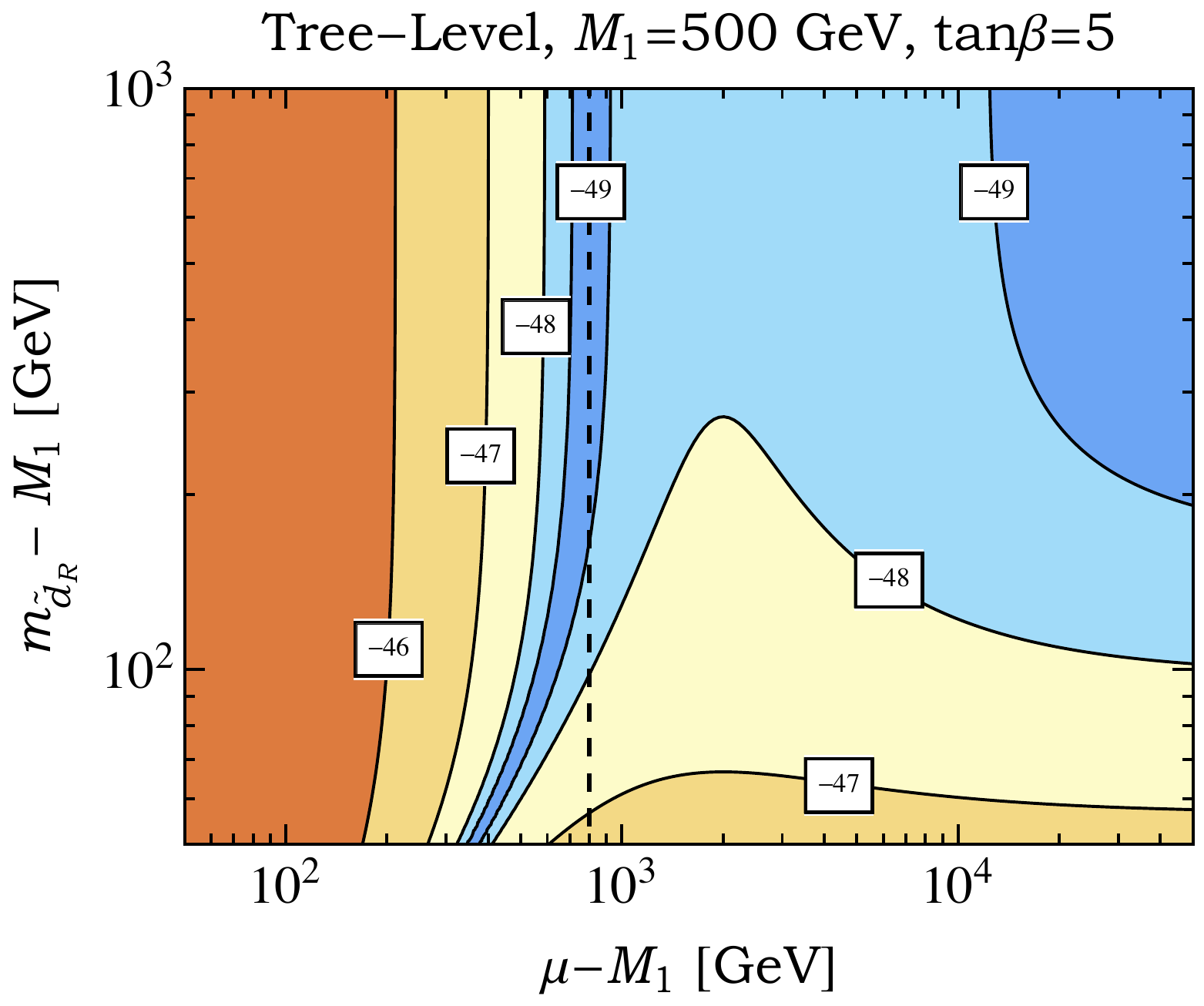} 
\caption{\label{fig:intro} SI nucleon cross sections from tree-level Higgs and squark exchange in the Higgsino and sdown mass plane for a bino mass of $M_1= 500$ GeV and $\tan{\beta}=5$. The labeled contours correspond to values of $\log_{10}{(\sigma_\text{SI}/\text{cm}^2)}$, while the vertical black-dashed line denotes the precise value of $\mu$ at which the lightest neutralino's coupling to the Higgs vanishes at tree-level.}
\end{center}
\end{figure}

Similar to the wino and Higgsino, bino scattering through the Higgs boson vanishes at tree-level.  If heavy flavor squarks or sleptons are nearby in the spectrum, however, loop processes are induced.  In this case, prospects for detection are improved through direct coupling to colored scalars. The interplay of a number of effects, such as power suppression if the new states are heavy compared to the electroweak scale, enhancement from on-shell-poles, and sizable mixing between colored scalars, could impact this. We assume that light flavor squarks and the Higgsino are decoupled from the low-energy spectrum since tree-level amplitudes would otherwise dominate over loops. To quantify the degree to which these must be decoupled, we show in Fig.~\ref{fig:intro} the SI cross section as a function of the Higgsino mass $\mu$ and the sdown mass $m_{\tilde{d}_R}$, when the lightest supersymmetric particle (LSP) is a bino-like neutralino that interacts with the Standard Model (SM) Higgs and a right-handed down-squark ($\tilde{d}_R$). Sufficient decoupling occurs when the leading order scattering rate in Fig.~\ref{fig:intro} drops below $\sigma_{\rm SI} \sim 10^{-49}$ cm$^2$~.

Processes relevant for one-loop bino scattering cross sections and related simplified models have already been considered in the literature~\cite{Djouadi:2001kba,Drees:1993bu,Djouadi:2000ck,Hisano:2010ct,Hisano:2011um,Gondolo:2013wwa,Hisano:2015bma,Berlin:2015ymu,Ibarra:2015fqa,Chang:2013oia,Chang:2014tea,Garny:2015wea}. At the same time, a great deal of effective field theory (EFT) machinery has recently been developed for systematically integrating out heavy particle thresholds and running Wilson coefficients to the low scales characteristic of the processes in direct detection experiments \cite{Hill:2014yka,Hill:2014yxa,D'Eramo:2014aba}. Our aim is to apply these techniques, focusing on QCD effects, to the case of bino DM where the SM is extended with a Majorana gauge singlet, and a few sfermions with the same quantum numbers as either left- or right-handed quarks or leptons.  

We capture a number of effects that have been previously neglected.  First, we are able to systematically incorporate the multiple scales involved in direct scattering, accounting for potentially large contributions, $ \sim \alpha_s \log { m_t \over  1 \, {\rm GeV}}$~. Second, we are able to include additional states at low energies, beyond those of $n_f$-flavor QCD. For example, when the mass difference between the bino and sbottom is much less than the weak scale, both are active degrees of freedom at low energies, and we use heavy particle techniques to describe their interactions with soft bottom quarks. Third, we are able to assess the uncertainties from both higher-order perturbative corrections and hadronic inputs.

In addition to incorporating renormalization group evolution (RGE), we also go beyond previous fixed-order computations that have focused on the parameter space for either purely left- or right-handed sfermions. We explore a larger part of the minimal supersymmetric standard model (MSSM) parameter space by considering the impact of mixing between left- and right-handed third generation squarks.  We also perform a complete leading order matching at the weak scale, considering  contributions such as the spin-2 gluon operator (significant when a sbottom is close in mass to the bino), and the anapole operator from photon exchange.

While we adopt the nomenclature and explicit couplings of the MSSM for definiteness, key components of our analysis, such as the results for loop amplitudes and RGE solutions, are generic, and can be readily applied to investigate the phenomenology of other models that incorporate interactions of DM with scalars charged under the SM.  For example, many of the effects considered here may also be applied to the case of suppressed tree-level scattering (``blind-spots''), where loop corrections are necessary to meaningfully compare theory and experiment  \cite{Cheung:2012qy,Cheung:2013dua,Huang:2014xua,Crivellin:2015bva}.

The remainder of the paper is structured as follows.  In Sec.~\ref{sec:fixedorder}, we review the standard fixed-order approach in the literature for determining amplitudes for WIMP-nucleon scattering. This lays the groundwork for the effective theory framework described in Sec.~\ref{sec:cases}. There we discuss the factorization of the scattering amplitude into contributions from the relevant physical scales, and illustrate the techniques for matching, renormalization, and coefficient evolution by presenting three detailed examples of increasing intricacy: a bino coupled to {\em (i)} a right-handed stop, {\em (ii)} a heavier right-handed sbottom, and {\em (iii)} a nearly mass degenerate right-handed sbottom. The reader interested in the phenomenological results may go straight to Sec.~\ref{sec:pheno}, where we evaluate cross sections for models with stop, sbottom, and slepton mediators. The most promising case for detection is a bino interacting with a nearly degenerate right-handed sbottom: a bino as heavy as 10 TeV may be detected at LZ if the mass splitting is a few GeV. On the other hand, a bino nearly degenerate with a right-handed stop is only detectable above the neutrino background for masses below about $500$ GeV. 

We collect the technical results in the appendices. In Appendix~\ref{sec:model}, we set up our conventions for the sfermion mass matrices, as well as the DM-fermion-sfermion interactions. Appendices~\ref{app:hadinputs} and~\ref{app:R&M} contain the hadronic form factors and the running and matching matrices employed in our numerical analysis. In Appendix~\ref{sec:app1}, we present details of the Wilson coefficients for all relevant amplitudes, such as tree-level sbottom exchange, one-loop Higgs, $Z$, and $\gamma$ exchange, one-loop diagrams involving charged electroweak gauge bosons, and one-loop contributions to the gluon coefficients. We compute these keeping all fermion and sfermion masses explicit, and allowing for left-right sfermion mixing. We note for each diagram where our results differ from previous literature.

\section{\textbf{Fixed Order Approach to WIMP-Nucleon Scattering}}
\label{sec:fixedorder}

Amplitudes for WIMP-nucleon scattering involve energy scales that span several orders of magnitude, ranging from the masses of the new particles and the mediating SM particles ($\gtrsim 100 \, {\rm GeV}$), to the scales of heavy quark thresholds and of hadronic physics ($\gtrsim 1 \, {\rm GeV}$), and the typical momentum transfers relevant for direct detection ($ \sim \, {\rm MeV}$). A standard approach in the DM literature is to determine these amplitudes at ``fixed order," treating this broad range of physical scales at a single scale. In this section, we review this matching procedure between the full theory of the SM and its extension, specified at high energies $E \gtrsim 100 \, {\rm GeV}$, and an EFT for WIMP-nucleon scattering, specified at low energies $E \gtrsim 1 \, {\rm GeV}$.

At high energies, $E \gtrsim 100 \, {\rm GeV}$, the basic interaction that we consider is of a single sfermion ($\sfe$) with a bino LSP ($\x$) and a SM fermion ($f$), adopting the following notation:
\be\label{eq:fulltheory}
\mathcal{L} \supset  \tilde{f} ~ \bar{f} \left( \alpha_f + \beta_f \gamma^5 \right) \x + \text{ h.c.} ~~ .
\ee
The couplings $\alpha_f,~\beta_f$ are parametrized in terms of the SM hypercharge coupling $g^\prime$ and the sfermion mixing angles of Eqs.~(\ref{eq:SfermionMass}) and (\ref{eq:couplings}). To simplify the discussion in this section and the next, we illustrate  general methods for the case where $\sfe$ constitutes a single right-handed stop or sbottom and $f$ the corresponding top or bottom quark, and assuming the theory in Eq.~(\ref{eq:fulltheory}) is defined at the weak scale $\sim 100 \, {\rm GeV}$. The impact (from RGE) of considering couplings defined at an even higher scale is illustrated in Sec.~\ref{sec:RHsbottom}. Examples pertaining to mixed stops and sbottoms, and sleptons, are treated in a similar way, and we discuss them in Secs.~\ref{sec:mixed} and~\ref{sec:slepton}. 

The hadronic matrix elements necessary for describing WIMP-nucleon scattering are determined, e.g., from lattice measurements, at low energies $E \sim {1 \, {\rm GeV}}$, in a theory with three quark flavors. At these energies, an effective theory captures the interactions of the WIMP with the degrees of freedom of 3-flavor QCD. For the bino, a gauge-singlet Majorana fermion, a set of operators for low-velocity scattering is
\begin{align}\label{eq:lowbasis}
{\cal L} &= 
\sum_{q=u,d,s} \Bigg\{
 c_{q}^{(0)} ~ {\bar \chi } \chi ~ O_q^{(0)}
+ c_{q}^{(1)} ~{\bar \chi} \gamma_\mu \gamma^5 \chi ~ O_q^{(1)\mu}
 +  \frac{c_{q}^{(2)}}{m_\x^2}~{\bar \chi} i \partial_\mu i \partial_\nu \chi  ~ O_q^{(2)\mu \nu} \Bigg\} 
\nl &\quad
 + c_{g}^{(0)}~ {\bar \chi} \chi ~ O_g^{(0)} 
+  \frac{c_{g}^{(2)}}{m_\x^2}~{\bar \chi} i \partial_\mu i \partial_\nu \chi ~ O_g^{(2)\mu \nu} \,,
\end{align}
where the relevant QCD currents are
\begin{align}\label{eq:ops}
O^{(0)}_{q} &= m_q~\bar{q} q \,, \quad  O^{(1)\mu}_q = {\bar q} \gamma^\mu \gamma^5 q \,, \quad  O^{(2)\mu\nu}_{q} = \frac12 ~ \bar{q}\bigg[ \gamma^{\{\mu} iD_-^{\nu\}} 
- \frac{1}{d} g^{\mu\nu} i\Dslash_- \bigg] q \,,  \quad  \nl
O^{(0)}_g &= (G^A_{\mu\nu})^2 \,, \quad O^{(2)\mu\nu}_g = -G^{A \mu \lambda} G^{A \nu}_{\phantom{A \nu} \lambda} + \frac{1}{d}~ g^{\mu\nu} ~(G^A_{\alpha\beta})^2 \,,
\end{align} 
with $G^A_{\mu\nu}$ the gluon field strength and $d=4-2\epsilon$ the spacetime dimensions. We adopt the notation $D_- \equiv \overrightarrow{D} -\overleftarrow{D}$ and $A^{\{\mu}B^{\nu\}} \equiv (A^\mu B^\nu + A^\nu B^\mu)/2$~, and have neglected operators that lead to kinematically suppressed contributions. Leading order SI scattering is given by the scalar ($O_{q,g}^{(0)}$) and spin-2 ($O_{q,g}^{(2) \mu \nu}$) quark and gluon currents, while leading order SD scattering is given by the quark axial current ($O_q^{(1)\mu}$). We neglect the operator ${\bar \chi} \gamma^\mu \gamma^5 \chi ~ {\bar q} \gamma_\mu q$ involving the quark vector current, which leads to SI scattering that is power-enhanced relative to the scalar and spin-2 contributions, but is velocity suppressed. We have reduced the operators to a linearly independent set; e.g., the operators ${\bar \chi} i \partial_\mu  \gamma_\nu \chi ~ O_{q,g}^{(2)\mu \nu}$ and ${\bar \chi} i \partial_\mu i \partial_\nu \chi  ~ O_{q,g}^{(2)\mu \nu}$ are redundant in the forward scattering limit. We ignore flavor non-diagonal operators, whose nucleon matrix elements have an additional weak-scale suppression relative to those considered. We will not be concerned here with operators involving leptons. 

\begin{figure}[t]
\begin{center}
\includegraphics[width=1\textwidth]{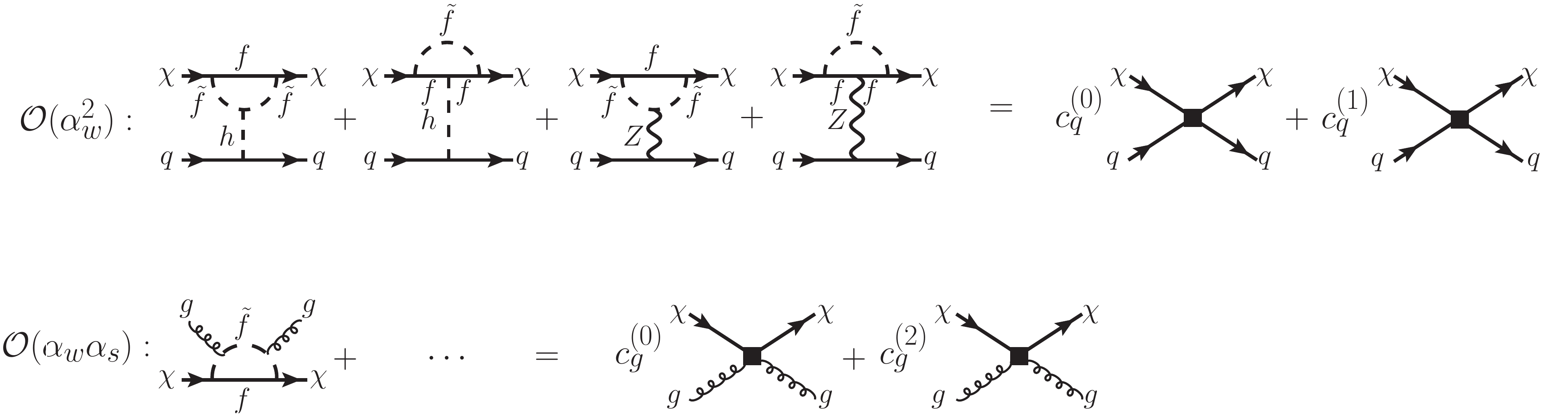} 
\caption{\label{fig:matchingfixedorder} Matching conditions for a fixed-order calculation. Charge-reversed diagrams are not shown. Here, $\tilde{f}$ denotes a right-handed stop or sbottom, and $q$ refers to the quarks of 3-flavor QCD. In the bottom line, the ellipsis denotes similar diagrams where the insertion of the gluon legs vary (see Appendix~\ref{app:glue}).}
\end{center}
\end{figure}

In the standard fixed-order approach, the full theory in Eq.~(\ref{eq:fulltheory}) is matched onto the effective theory in Eq.~(\ref{eq:lowbasis}), by integrating out the sfermion ${\tilde f}$, the gauge bosons $Z,\, W^\pm$, the Higgs $h$, the Goldstones $G, G^\pm$, and the heavy quarks $t, b, c$, altogether at a \emph{single} scale. The matching condition for the case of a right-handed stop or sbottom (denoted as $\tilde{f}$) is shown in Fig.~\ref{fig:matchingfixedorder}. The leading contributions to the quark and gluon coefficients are at $\order(\alpha_w^2)$ and $\order(\alpha_w \alpha_s)$, respectively.

Once the Wilson coefficients are determined, the hadronic matrix elements are evaluated. We adopt the definitions and values from Sec.~4 of Ref.~\cite{Hill:2014yxa} for the hadronic matrix elements of the QCD currents in Eq.~(\ref{eq:ops}). For completeness, we collect their definitions here:
\begin{align}\label{eq:matrixelements}
\langle N | O_q^{(0)} | N \rangle &\equiv m_N ~ f_{q,N}^{(0)}\,, \quad  \frac{-9 \alpha_s(\mu)}{8 \pi}  \langle N | O_g^{(0)}(\mu) | N \rangle \equiv ~ m_N ~ f_{g,N}^{(0)}(\mu) \,, \nl
\langle N(k) | O_{q }^{(1)\mu} (\mu) | N(k) \rangle &\equiv s^\mu f_{q,N}^{(1)}(\mu) \,, \nl
\langle N(k) | O_{i}^{(2)\mu \nu}(\mu) | N(k) \rangle &\equiv {1 \over m_N}\left( k^\mu k^\nu - \frac{1}{4} m_N^2 g^{\mu \nu} \right) f_{i,N}^{(2)}(\mu)\,,
\end{align}
where $N=p,n$ for proton or neutron, $i=q,g$ for quark or gluon, and the spin vector $s^\mu = \bar{u}(k) \gamma^\mu \gamma^5 u(k) $ satisfies $k \cdot s =0$ and $s^2 = -1$, assuming non-relativistic normalization for the spinor $u(k)$.

The axial form factors, $f_{q,N}^{(1)}$, are extracted from hyperon semileptonic decay, from $\nu p$ scattering, or from observables of polarized deep inelastic scattering. The scalar quark form factors, $f_{q,N}^{(0)}$, are extracted from lattice measurements, while the scalar gluon form factor is obtained through the leading order relation~\cite{Shifman:1978zn}
\begin{align}
\label{eq:gluemassfrac}
f_{g,N}^{(0)} = 1 - \sum\limits_{q=u,d,s} f_{q,N}^{(0)} + \order(\alpha_s) ~ .
\end{align}
The quark and gluon spin-2 form factors, $f_{q,N}^{(2)} \,, f_{g,N}^{(2)}$, are extracted from the second moment of parton distribution functions (PDFs). In Appendix~\ref{app:hadinputs}, we collect the values employed in our numerical analysis.

These nucleon matrix elements, together with the Wilson coefficients, define the SI and SD amplitudes
\begin{align}
\label{eq:effnucleon}
{\cal M}_{\text{SI}, N} &= m_N \left\{ \sum_{q=u,d,s} \left[  f_{q,N}^{(0)} ~ c_q^{(0)} + \frac{3}{4}  f_{q,N}^{(2)} ~ c_q^{(2)}  \right] - {8 \pi \over 9 \alpha_s } f^{(0)}_{g,N} ~ c_g^{(0)} + \frac{3}{4}f_{g,N}^{(2)} ~ c_g^{(2)} \right\} \,, \nl
{\cal M}_{\text{SD},N} &= \sum_{q=u,d,s} f_{q,N}^{(1)} ~ c_q^{(1)} \,,
\end{align}
and, finally, the cross sections for SI and SD scattering on a nucleon target are obtained,
\be\label{eq:cross}
\sigma_{\text{SI}} = \frac{4}{\pi} 
\left( \frac{m_\x m_N}{m_\x + m_N} \right)^2 
\left| {\cal M}_{\text{SI}, N} \right|^2~ \,, 
\quad \sigma_\text{SD} = \frac{12}{\pi}  \left( \frac{m_\x m_N}{m_\x + m_N} \right)^2  \left| {\cal M}_{\text{SD}, N} \right|^2 \,.
\ee

This is a straightforward strategy for determining WIMP-nucleon scattering cross sections, with, however, limitations that motivate a more thorough analysis. First, there are potentially large perturbative corrections, $\sim \alpha_s \log { m_t \over  1 \, {\rm GeV}}$~, inherent in treating a multiscale process at a single scale. For example, while the  Wilson coefficients are determined at the weak scale employing $\alpha_s(\sim 100 \, {\rm GeV})$, the leading order scalar gluon form factor in Eq.~(\ref{eq:gluemassfrac}) is subject to sizable corrections due to the large size of $\alpha_s(\sim 1 \, {\rm GeV})$. Second, determining higher order corrections in a fixed-order framework is difficult; e.g., at NLO two- or three-loop amplitudes are required. Theoretical control of perturbative corrections would allow us to estimate their numerical impact, and, in the event of a detection, to systematically improve predictions for WIMP-nucleon scattering. In the next section, we lay out the effective theory framework to deal with these issues head on.

\section{\textbf{Effective Theory Approach to WIMP-Nucleon Scattering}}
\label{sec:cases}

As mentioned in the previous section, WIMP-nucleon scattering involves a multitude of physical scales, and the separation between the weak scale, $\sim 100 \, {\rm GeV}$, and the hadronic scale, $\sim 1 \, {\rm GeV}$, may lead to large uncertainties when employing the fixed-order framework. In this section, we discuss the ``effective theory" approach, which factorizes the scattering amplitudes into contributions from different physical scales by constructing a sequence of EFTs from the weak scale down to the hadronic scale, and connecting them through RGE and matching. This allows for the separate analysis of perturbative corrections at each energy threshold and for the resummation of large logarithms, e.g., $\sim \alpha_s \log {m_t \over 1 \, {\rm GeV} }$~. 

\begin{figure}[top]
\centering
\includegraphics[scale=0.4]{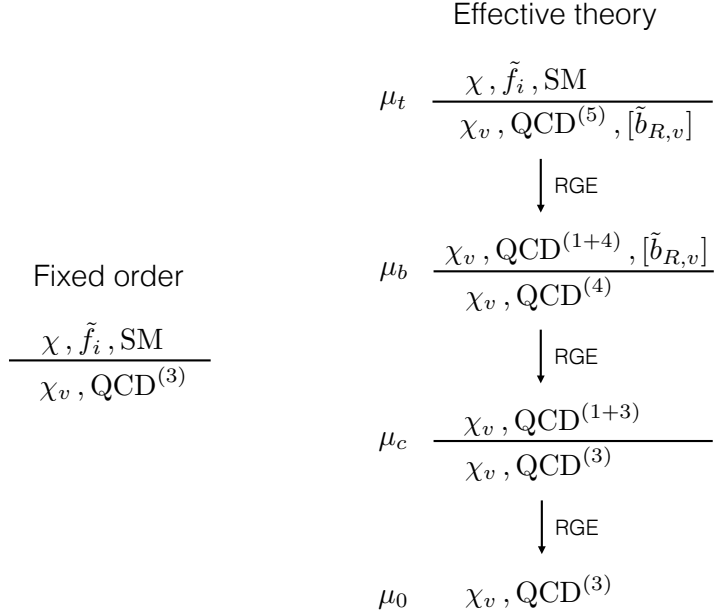}
\caption{\label{fig:formalism}
In the fixed-order approach (left), the full theory is directly matched onto the low energy theory with 3-flavor QCD. 
In the effective theory approach (right), the full theory is matched onto the low energy theory with 3-flavor QCD by systematically passing through a sequence of effective theories defined at the weak scale ($\mu_t \sim m_t$), the bottom mass scale ($\mu_b \sim m_b$), the charm scale ($\mu_c \sim m_c$), and the hadronic scale ($\mu_0 \sim 1 \, {\rm GeV}$). The matching and running between these effective theories are discussed in the main text. If the mass splitting between a sbottom (${\tilde b}_R$) and the bino is much smaller than the weak scale, then the effective theory setup is modified to include a heavy sbottom field ${\tilde b}_{R,v}$, accounting for sbottom-bino interactions at low energies. The subscript $v$ denotes a heavy particle field as defined through the field redefinitions in Eqs.~(\ref{eq:HPTredefinition}) and~(\ref{eq:HPTredefinition3}).}
\end{figure}

This framework is depicted in Fig.~\ref{fig:formalism}. To further elaborate on its general features, let us present the corresponding factorized amplitude, and briefly discuss its components in turn; a more detailed discussion is given in the subsections below. In the EFT approach, the scattering amplitude is determined as 
\begin{align}\label{eq:factorize}
{\cal M} =  {\boldsymbol f}^T(\mu_0) {\boldsymbol R}(\mu_0,\mu_c)\, {\boldsymbol M}(\mu_c)\, {\boldsymbol R}(\mu_c,\mu_b)\, {\boldsymbol M}(\mu_b)\, {\boldsymbol R}(\mu_b, \mu_t)\, {\boldsymbol c}(\mu_t) \,,
\end{align}
where the renormalization scales $\mu_t$, $\mu_b$, $\mu_c$, and $\mu_0$ correspond respectively to the weak scale $\sim m_t$, the bottom quark threshold $\sim m_b$, the charm quark threshold $ \sim m_c$, and the hadronic scale $\sim 1 \, {\rm GeV}$, where nucleon matrix elements are defined. The vector ${\boldsymbol c}(\mu_t)$ collects the Wilson coefficients determined at the scale $\mu_t$ by integrating out weak scale degrees of freedom, and matching onto a theory with five quark flavors. The matrix ${\boldsymbol R}(\mu_b, \mu_t)$ implements coefficient running from $\mu_t$ down to $\mu_b$, while the matrix ${\boldsymbol M}(\mu_b)$ implements coefficient matching across the bottom quark threshold, between the theory with five and four quark flavors. The matrices $ {\boldsymbol R}(\mu_c,\mu_b)$ and ${\boldsymbol M}(\mu_c)$ are analogously defined, implementing running in 4-flavor QCD and matching across the charm quark threshold. Finally, the coefficients are run down to the hadronic scale in 3-flavor QCD, using ${\boldsymbol R}(\mu_0,\mu_c)$, and the matrix elements are evaluated through multiplication of the (transposed) vector ${\boldsymbol f}^T(\mu_0)$, which collects the form factors $f_{q,g}$ defined in Eq.~(\ref{eq:matrixelements}). 

Clearly, Eq.~(\ref{eq:factorize}) has separation of scales, with components ${\boldsymbol c}(\mu_t)$, ${\boldsymbol M}(\mu_b)$, ${\boldsymbol M}(\mu_c)$, and ${\boldsymbol f}(\mu_0)$ depending only on scales of a similar order. The logarithms in the amplitude are resummed through the RGE factors ${\boldsymbol R}$, and additional perturbative corrections to each component can be separately and systematically analyzed without having to evaluate the whole amplitude at higher loop order. Note that $\alpha_s \log {m_b \over m_c }$ does not constitute a large logarithm, and hence integrating out the bottom and charm quarks at a single scale would suffice. Nonetheless, since $\alpha_s(1\, {\rm GeV})$ is sizable, higher-order corrections may have significant impact, and we may conveniently employ known results for the matrices ${\boldsymbol M}(\mu_b)$, ${\boldsymbol M}(\mu_c)$, and ${\boldsymbol R}$ to include them. 

Note also that the PDFs relevant for the spin-2 matrix elements defined in Eq.~(\ref{eq:matrixelements}) are available at a high-scale, e.g., $\order(100)$ GeV, and thus allows us to evaluate the amplitude without running down these Wilson coefficients to a low-scale. The running, however, would be relevant for relating the spin-2 current to low-energy effective DM-nucleon contact operators (see e.g., Refs.~\cite{Fan:2010gt,Fitzpatrick:2012ix}), and for including the impact of multi-nucleon effects (see e.g., Refs.~\cite{Prezeau:2003sv,Cirigliano:2012pq}). In the present analysis, we RG evolve all Wilson coefficients as a default, but have checked that our results are consistent, up to uncertainties, with an evaluation at the high scale. We find that the additional perturbative uncertainty from running the spin-2 coefficients increases the overall uncertainty by less than 10$\%$.

The factorization in Eq.~(\ref{eq:factorize}) is a general result of our effective theory analysis, and in the following subsections we provide further details on each of its components. Section~\ref{sec:HPT} considers formalism for representing the relevant degrees of freedom in the low energy theory, and for matching at the weak scale $\mu_t \sim m_t$ . In Secs.~\ref{sec:case1},~\ref{sec:case2}, and~\ref{sec:lightfsmalldelta}, we go into explicit detail by applying the effective theory framework to three examples, classified according to the mass, $m_f$, of the fermion partnered to the sfermion, and the mass splitting, $\delta_{\sfe} = m_{\tilde f} - m_\x$, between the sfermion and bino. Case I considers $m_f \gtrsim \mu_t$ and arbitrary $\delta_{\sfe}$~, case II considers $m_f \ll \mu_t \lesssim \delta_{\sfe}$~, and case III considers $\delta_{\sfe}~,~ m_f \ll \mu_t$~. These examples illustrate, in increasing complexity, the key ingredients of the effective theory framework. Case I goes through the basic computational pipeline involving the components ${\boldsymbol c}$, ${\boldsymbol R}$, ${\boldsymbol M}$, and ${\boldsymbol f}$ of Eq.~(\ref{eq:factorize}). Case II presents an example where nontrivial renormalization of the bare coefficients arises. Finally, for case III, a heavy sfermion field ${\tilde f}_v$ (denoted as ${\tilde b}_{R,v}$ in Fig.~\ref{fig:formalism}) is included in the low-energy theory to account for sfermion-bino interactions.

\subsection{Integrating out the Mass but Not the Particle}
\label{sec:HPT}

A key step in the effective theory approach involves integrating out weak scale degrees of freedom by matching onto a low energy theory of the bino $\chi$ and the quarks and gluons of 5-flavor QCD. In this procedure, the gauge, Higgs, and Goldstone bosons, as well as the stop and top, are integrated out. However, the bino, despite having a weak scale mass, $m_\chi \gtrsim 100 \, {\rm GeV}$, is not integrated out -- the goal of calculating a WIMP-nucleon scattering cross section requires that it is kept in the low energy theory. Moreover, the same applies to a sbottom whose mass is close to that of the bino: despite $m_{\tilde b} \approx m_\chi \gtrsim 100 \, {\rm GeV}$, the sbottom should not be integrated out since the bottom quark is an active degree of freedom in the low energy theory and bino-sbottom interactions are thus allowed.

How do we integrate out the \emph{mass} of a field without integrating out the \emph{field} itself? The idea is simple and can be pictured by considering the following parametrization of the bino momentum at low energies: $p^\mu = m_\chi v^\mu + k^\mu$, where $v^\mu$ is a reference time-like unit vector and $k^\mu \ll m_\chi v^\mu$. The interactions of the heavy bino with the much lighter quarks and gluons of 5-flavor QCD involve only soft momenta of $\order(k^\mu)$, while the large momentum component $m_\chi v^\mu$, corresponding to its mass, plays no role and can be integrated out. This procedure is formally done by going from a relativistic description of the field to a ``heavy particle" description, order-by-order in the small parameter $|k|/m_\chi$~. The technique is called ``heavy particle effective theory," and is known from applications for heavy quark physics (for a review see, e.g., Ref.~\cite{Manohar:2000dt}). 

We may pass from a relativistic to a heavy particle description for the bino (Majorana fermion) by making the field redefinition
\be
\label{eq:HPTredefinition}
\x =  \sqrt{2} e^{-i m_\x v \cdot x}  (\x_v + X_v)  \,,
\ee
where the spinors obey $\vslash \chi_v =\chi_v$ and $\vslash X_v = - X_v$~. In terms of the momentum decomposition discussed above, the phase $e^{-i m_\x v \cdot x}$ extracts the large momentum component $m_\x v^\mu$.
Upon introducing this field redefinition into the kinetic term $\frac12 {\bar \chi} \left( i \slashed{\partial} - m_\x \right) \x$ , we find that the component $X_v$ has mass $2m_\x$ , and is thus integrated out, e.g., at tree-level by solving its equation of motion. The remaining component $\x_v$ describes the heavy bino degree of freedom with the (canonically normalized) kinetic term $ {\bar \chi}_v  i v \cdot \partial \x_v$ , depending only on the soft momentum $k^\mu$. The Majorana condition $\chi = \chi^c$ allows us to write the field redefinition~(\ref{eq:HPTredefinition}) alternatively as
\be
\label{eq:HPTredefinition2}
\x =\sqrt{2} e^{i m_\x v \cdot x}  (\x_v^c + X_v^c) \,,
\ee
where charge conjugation is denoted by $\psi^c = {\cal C} \psi^*$ with the unitary and symmetric matrix ${\cal C}$ obeying ${\cal C}^\dagger \gamma^\mu {\cal C} = - \gamma^{\mu*}$. This implies an invariance of the heavy particle Lagrangian for $\x_v$ under the simultaneous transformations~\cite{Kopp:2011gg,Heinonen:2012km} 
\be\label{eq:invariance}
v \to -v \,, \quad \chi_v \to \chi_v^c ~ .
\ee
This invariance and the form of the field redefinition in Eq.~(\ref{eq:HPTredefinition2}) will be useful in Sec.~\ref{sec:lightfsmalldelta} for considering the interactions of a heavy bino with a heavy sbottom.

Instead of introducing the field redefinition~(\ref{eq:HPTredefinition}) into a basis of relativistic operators, we may also proceed in the spirit of effective theory, employing building blocks to directly write down low energy operators consistent with symmetries. For our low-energy theory, the building blocks are the usual relativistic degrees of freedom (quarks and gluons), the reference vector $v^\mu$, and the heavy bino field $\chi_v~$. Thus, for a Majorana dark matter particle whose mass satisfies $m_\chi \gg m_b~$, the basis of operators describing its interactions with 5-flavor QCD is
\begin{align}
\label{eq:LWIMPSM}
 {\cal L}_{\chi_v} / 2 &= 
\sum_{q=u,d,s,c,b}  \Bigg\{ 
 c_{q}^{(0)} ~ \bar{\chi}_v\chi_v ~ O_{q}^{(0)}
 +
 c_q^{(1)} ~ \bar{\chi}_v \gamma^\perp_\mu \gamma^5 \chi_v ~ O_{q}^{(1) \mu}
 +
 c_{q}^{(2)} ~ \bar{\chi}_v\chi_v ~ v_\mu v_\nu ~ O_{q}^{(2)\mu\nu} \Bigg\} \nl
&\qquad + c_{g}^{(0)} ~ \bar{\chi}_v\chi_v ~ O_{g}^{(0)} + c_{g}^{(2)} ~ \bar{\chi}_v\chi_v ~ v_\mu v_\nu ~ O_{g}^{(2)\mu\nu} 
+ \dots \,,
\end{align} 
where the ellipsis denotes higher dimension operators, and the relevant QCD currents are given in Eq.~(\ref{eq:ops}). Here, we have subtracted off the component of $\gamma_\mu \gamma^5$ which vanishes between the heavy particle bilinear, defining $\gamma_\mu^\perp = \gamma^\mu - v^\mu \vslash$. Alternatively, Eq.~(\ref{eq:LWIMPSM}) is obtained by making the substitution~(\ref{eq:HPTredefinition}) into the basis of operators in Eq.~(\ref{eq:lowbasis}). We have introduced a conventional factor of $1/2$ on the left-hand side of Eq.~(\ref{eq:LWIMPSM}) since the field redefinition (\ref{eq:HPTredefinition}) would otherwise lead to a factor of 2 discrepancy between the coefficients in Eqs.~(\ref{eq:lowbasis}) and (\ref{eq:LWIMPSM}). 


In the relativistic basis of Eq. (2), $c_q^{(0)}$ and $c_q^{(2)}$ are treated on equal footing, despite corresponding to operators whose mass dimensions differ by two, i.e., seven and nine, respectively. As a result, power counting is possible but not \emph{manifest} (leading order SI scattering involves operators of dimension seven and nine), and it is less straightforward how the basis extends beyond leading order. In contrast, power counting is manifest in Eq.~(\ref{eq:LWIMPSM}), and thus the operators relevant at each order are known without having first to evaluate the full theory amplitudes. In particular, leading order low-velocity SI (SD) scattering is obtained from dimension seven (six) operators, and subleading corrections can be systematically computed. In the remainder of the paper, when referring to Wilson coefficients, we assume the form given in Eq.~(\ref{eq:LWIMPSM}).

Having discussed the formalism for incorporating both relativistic and heavy particle degrees of freedom at low energies, let us now turn to the computation of weak scale coefficients ${\boldsymbol c}(\mu_t)$ of Eq.~(\ref{eq:factorize}). At the scale $\mu_t \sim m_t~$, we match the full relativistic theory of Eq.~(\ref{eq:fulltheory}), with six quark flavors and a relativistic bino $\chi$~, onto the low energy theory of Eq.~(\ref{eq:LWIMPSM}), with five quark flavors and a heavy particle bino $\chi_v$~. The full theory diagrams are computed using standard relativistic Feynman rules, while the effective theory diagrams are computed using the Feynman rules of Eq.~(\ref{eq:LWIMPSM}). 

This matching procedure determines the \emph{bare} Wilson coefficients, and may involve loop contributions from the low energy effective theory. It is simplest to compute the full theory amplitudes setting all mass scales much lighter than the weak scale to zero, and regulating infrared divergences in $4-2\epsilon$ dimensions. The weak scale coefficients ${\boldsymbol c}(\mu_t)$ then depend only on the weak scale masses $m_W$, $m_Z$, $m_h$, $m_t$, $m_\chi$, and $m_{{\tilde f}}~$, and are determined up to corrections of $\order(m_b/m_t)$. Of course, for matching a full theory amplitude onto the scalar quark current $O_q^{(0)}$ of Eq.~(\ref{eq:lowbasis}), the leading $m_q$ factor should be retained. In dimensional regularization, the loop integration measure has scaling dimension [mass]$^{4-2\epsilon}$~, and therefore any loop integral is dimensionful. A loop integral that has no mass scale to soak up this dimensionality must vanish by consistency. This is the well-known statement that scaleless integrals vanish in dimensional regularization. With light quark masses set to zero, the effective theory loop contributions are scaleless, and hence vanish. Alternatively, keeping light quark masses nonzero would regulate infrared divergences, but would require the computation of non-vanishing effective theory loop amplitudes. An explicit example involving such effective theory loop contributions will be presented in Sec.~\ref{sec:case2}.

The remaining $1/\epsilon$ poles in the bare coefficients are UV divergences of the low energy theory, and are renormalized accordingly. For a detailed discussion on the renormalization of the QCD currents in Eq.~(\ref{eq:ops}), we refer the reader to Sec.~3 of Ref.~\cite{Hill:2014yxa}. Here, we will simply quote the results. At leading order in $\alpha_s$~, the scalar and axial-vector coefficients are trivially renormalized, i.e., $c(\mu) = c^\text{bare}$, while the spin-2 coefficients are renormalized as
\begin{align} \label{eq:renormalize}
c_q^{(2)}(\mu) & = c_q^{(2){\rm bare}} + \order(\alpha_s) \,,
\quad
c_g^{(2)}(\mu) = \sum_{q} {1\over \epsilon}{\alpha_s \over 6\pi}  c_q^{(2){\rm bare}} + c_g^{(2){\rm bare}} 
+ \order(\alpha_s^2) \,,
\end{align}
where the sum runs over the active quark flavors, i.e., $q=u,d,s,c,b$ in 5-flavor QCD.
The $\order(\epsilon^0)$ terms of the coefficients $c_q^{(2)\text{bare}}$ introduce a $1/\epsilon$ pole in $c_g^{(2)}(\mu)$ that is cancelled by the $1/\epsilon$ pole in $c_g^{(2){\rm bare}}$. Note that the nontrivial renormalization also requires the $\order(\epsilon^1)$ terms of the coefficients $c_{q}^{(2) {\rm bare}}$.  We will see an explicit example of this renormalization in Sec.~\ref{sec:case2} when $c_g^{(2)\text{bare}}$ is divergent due to gluons emitted from massless quarks. 

As mentioned above, a sfermion that is nearly degenerate in mass with the bino should be a degree of freedom in the low energy theory if sfermion-bino interactions with light fermions are present. Hence, only the sfermion mass is integrated out (encoded in Wilson coefficients through the full theory amplitudes), and a heavy sfermion field is included at low energies. In particular, a so-called ``heavy-light current" describes the interactions of the heavy bino with the heavy sfermion and light fermion. This is described in Sec.~\ref{sec:lightfsmalldelta}.

Let us now move on to three cases that illustrate in explicit detail the general aspects of the EFT approach discussed above. Previous works have focused on fixed-order calculations~\cite{Gondolo:2013wwa,Ibarra:2015nca,Ibarra:2015fqa,Hisano:2010ct,Djouadi:2001kba} or on the EFT treatment of the scalar gluon coupling~\cite{Hisano:2015bma}. In the present analysis, we perform leading order matching onto the complete set of operators in Eq.~(\ref{eq:LWIMPSM}), including contributions to quark operators from exchanges of electroweak bosons. For example, we find that the Higgs-exchange diagrams are numerically relevant, significantly improving the projected reach of LZ (e.g., compared to those found in Ref.~\cite{Ibarra:2015nca}). Moreover, the following subsections present a pedagogical discussion of the EFT framework, illustrating aspects such as matching and the infrared pole structure, and the application of the heavy-light current. The case of a sfermion nearly degenerate in mass with the bino discussed in Sec.~\ref{sec:lightfsmalldelta} is new and physically relevant.

\subsection{Case I: Right-Handed Stop}\label{sec:case1}

\begin{figure}[t]
\begin{center}
\includegraphics[width=1\textwidth]{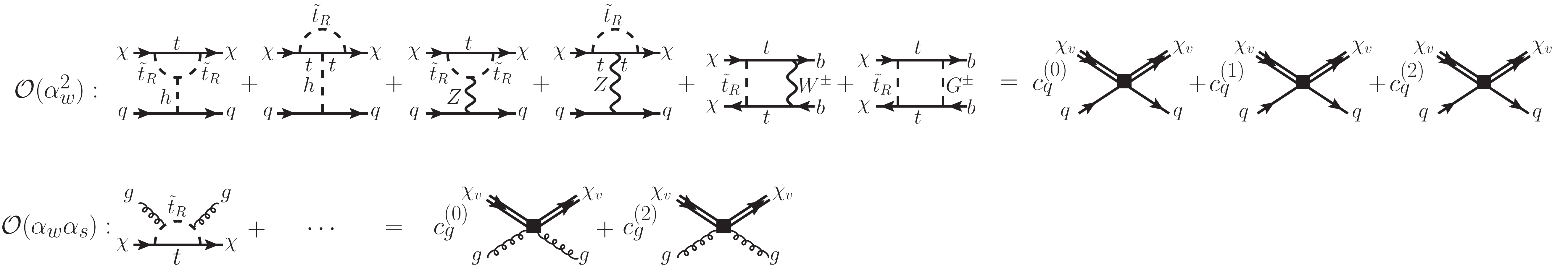} 
\caption{\label{fig:MatchStop} Weak scale matching conditions for the case of a right-handed stop. Crossed and charge-reversed diagrams are not shown. Here, $q$ refers to the quarks of 5-flavor QCD. In the bottom line, the ellipsis denotes similar diagrams where the insertion of the gluon legs vary (see Appendix~\ref{app:glue}). Single (double) lines correspond to relativistic (heavy particle theory) fields. We have omitted the label ``bare" on the coefficients on the right-hand side.}
\end{center}
\end{figure}

The simplest example arises when the mass of the fermion partnered to the sfermion is of order or greater than the weak scale, $m_f \gtrsim \mu_t$~. Although this case broadly applies to many models, for concreteness, we will restrict to the case of a single right-handed stop ($\tilde{t}_R$) interacting with the bino ($\x$) and a top quark ($t$). Let us discuss in turn the ingredients ${\boldsymbol c}$, ${\boldsymbol R}$, ${\boldsymbol M}$, and ${\boldsymbol f}$ of the factorization presented in Eq.~(\ref{eq:factorize}).

\underline{Weak scale coefficients ${\boldsymbol c}(\mu_t)$} : The matching condition at the weak scale $\mu_t \sim m_t$ is shown in Fig.~\ref{fig:MatchStop}. The full theory amplitudes are computed using the Lagrangian in Eq.~(\ref{eq:fulltheory}), while the effective theory amplitudes are computed using the Lagrangian in Eq.~(\ref{eq:LWIMPSM}). The weak scale particles $W^\pm\,, Z \,, h \,, G^\pm \,, t \,, {\tilde t}_R$ are highly virtual at low energies and are thus integrated out. Their effects are encoded into the Wilson coefficients of an effective theory describing a heavy bino $\chi_v$ interacting with the quarks and gluons of 5-flavor QCD.

The contributions to the quark and gluon coefficients begin at $\order(\alpha_w^2)$ and $\order(\alpha_w \alpha_s)$, respectively. The $h$-exchange diagrams contribute to the scalar coefficient $c_q^{(0)}$, while the $Z$-exchange diagrams contribute to the axial-vector coefficient $c_q^{(1)}$. The box diagrams exchanging $W^\pm$ or $G^\pm$ contribute to $c_b^{(0)}$, $c_b^{(1)}$, and $c_b^{(2)}$. The explicit results for the relevant diagrams are collected in Eqs.~(\ref{eq:DegStopHiggs}),~(\ref{eq:DegStopZ}),~(\ref{eq:DegStopBox}), and~(\ref{eq:DegStopGlue}). Working consistently at leading order, the gluon matching condition does not include contributions from effective theory diagrams involving loops of quarks since these are $\order(\alpha_w^2 \alpha_s)$. Accordingly, we also drop the $\order(\alpha_w^2 \alpha_s)$ terms in the renormalization condition in Eq.~(\ref{eq:renormalize}), and thus all bare Wilson coefficients are trivially renormalized for this example, i.e., $c_{q,g}(\mu_t) = c_{q,g}^{\rm bare}$. We collect the renormalized Wilson coefficients in the vectors
\begin{align}\label{eq:weakscaleC}
 {\boldsymbol c}_{\text{SI}}^T(\mu_t) = \left\{ c_q^{(0)}(\mu_t) ~,~ c_g^{(0)}(\mu_t) ~,~ c_q^{(2)}(\mu_t) ~,~ c_g^{(2)}(\mu_t) \right\}\,, \quad
{\boldsymbol c}_{\text{SD}}^T(\mu_t) &= \left\{ c_q^{(1)}(\mu_t) \right\}
~,
\end{align}
where $c_q^{(0,1,2)}$ is representative of the five quark flavors, i.e., $q=u,d,s,c,b$ , and hence the vectors $ {\boldsymbol c}_{\text{SI}}$ and $ {\boldsymbol c}_{\text{SD}}$ have twelve and five components, respectively. The coefficients are collected into two vectors in anticipation of evaluating the SI and SD amplitudes separately. 

\underline{Running and matching matrices ${\boldsymbol R}$ and ${\boldsymbol M}$} : For cases where the degrees of freedom below the weak scale are a gauge singlet (under $SU(3)_c \times U(1)_\text{EM}$) DM particle and the quarks and gluons of $n_f$-flavor QCD, the relevant matrices for running and matching are specified by loop-level matrix elements of the QCD currents in Eq.~(\ref{eq:ops}). We adopt the results from Tables~5 and~6 of Ref.~\cite{Hill:2014yxa}, and collect their leading order forms in Appendix~\ref{app:R&M} for completeness. In practice, we work at leading log (LL) order. For the axial current, the corrections to coefficient evolution and threshold matching begin at $\order(\alpha_s^2$), and are therefore subleading~\cite{Larin:1993tq,Grozin:1998kf,Grozin:2006xm}. In particular, this implies that the weak scale coefficients $c_{u,d,s}^{(1)}$ contribute to the amplitude, while $c_{c,b}^{(1)}$ may be neglected. Nonetheless, we will keep the discussion of weak scale coefficients ${\boldsymbol c}(\mu_t)$ general, including the determination of $c_{c,b}^{(1)}$.

\underline{Nucleon matrix elements ${\boldsymbol f}(\mu_0)$} : Let us collect the nucleon matrix elements defined in Eq.~(\ref{eq:matrixelements}) in the following vectors:
\begin{align}\label{eq:matrixelementsF}
{\boldsymbol f}_{{\rm SI},N}^T(\mu_0) &= m_N \left\{ f_{q,N}^{(0)} ~,~ {-8\pi \over 9 \alpha_s (\mu_0)} f_{g,N}^{(0)}(\mu_0) ~,~  \frac{3}{4} f_{q,N}^{(2)}(\mu_0) ~,~  \frac{3}{4} f_{g,N}^{(2)}(\mu_0) \right\}  \,, \nl  {\boldsymbol f}_{{\rm SD},N}^T(\mu_0) &= \left\{ f_{q,N}^{(1) } (\mu_0)\right\}  \,, 
\end{align}
where $f_{q,N}^{(0,1,2)}$ is representative of the three light quark flavors, i.e., the vectors ${\boldsymbol f}_{{\rm SI},N}$ and ${\boldsymbol f}_{{\rm SD},N}$ have eight and three components, respectively. To be consistent with the higher order effects included in the running and matching matrices ${\boldsymbol R}$ and ${\boldsymbol M}$, we must also include higher order corrections to the leading order gluon scalar matrix element of Eq.~(\ref{eq:gluemassfrac}). From the nucleon mass sum rule that links the gluon and quark scalar form factors (see, e.g., Ref.~\cite{Hill:2014yxa}), we have
\be
\label{eq:gluemassfrac2}
f_{g,N}^{(0)} (\mu)= \frac{-\alpha_s(\mu)}{4 \pi}~\frac{9}{\tilde{\beta}(\mu)}~\Big[ 1 - \Big( 1 - \gamma_m(\mu) \Big) \sum\limits_{q=u,d,s} f_{q,N}^{(0)} \Big] 
~,
\ee
where $\tilde{\beta} = \beta/g_s$ with $\beta$ the QCD beta function, and $\gamma_m$ is the quark mass anomalous dimension. In our numerical analysis, we include terms in $\tilde{\beta}$ and $\gamma_m$ through $\order(\alpha_s)$ (see Eq.~(\ref{eq:glueNLO})).

With all ingredients specified, we may now evaluate the amplitudes as in Eq.~(\ref{eq:factorize}). The result can be expressed as
\be
{\cal M}_{\text{SI},N} =  {\boldsymbol f}_{{\text{SI}},N}^T(\mu_0) ~ {\boldsymbol c}_\text{SI}(\mu_0) ~,~
 {\cal M}_{\text{SD},N} =  {\boldsymbol f}_{{\text{SD}},N}^T(\mu_0) ~ {\boldsymbol c}_\text{SD}(\mu_0) ~, 
\ee
which when expanded takes the form in Eq.~(\ref{eq:effnucleon}). The vectors ${\boldsymbol c}_{\text{SI}, \text{SD}}(\mu_0)$ contain the low energy coefficients properly mapped from the weak scale through the running and matching factors:
\be
\label{eq:lowscaleC}
{\boldsymbol c}(\mu_0) =  {\boldsymbol R}(\mu_0,\mu_c)\, {\boldsymbol M}(\mu_c)\, {\boldsymbol R}(\mu_c,\mu_b)\, {\boldsymbol M}(\mu_b)\, {\boldsymbol R}(\mu_b, \mu_t)\, {\boldsymbol c}( \mu_t) \,. 
\ee
These vectors are defined as in Eq.~(\ref{eq:weakscaleC}) but with the light quarks ($u,d,s$) and gluon of 3-flavor QCD. In practice, we will not evolve the coefficients after integrating out the charm quark at $\mu_c$, and hence we take $\mu_0 = \mu_c$~. Finally, the cross section is determined as in Eq.~(\ref{eq:cross}). Note that Eq.~(\ref{eq:cross}) applies for a relativistic Majorana field $\chi$, but is also valid for our heavy particle field $\chi_v$, given the conventional factor of $1/2$ on the left-hand side of Eq.~(\ref{eq:LWIMPSM}).

\subsection{Case II: Right-Handed Sbottom, Large Mass Splitting}\label{sec:case2}

An example similar to the previous one, but slightly more involved due to the interplay between quark and gluon coefficients, is when the mass of the fermion partnered to the sfermion is much lighter than the weak scale, $m_f \ll m_t$~, and the mass splitting between the sfermion and the bino is comparable to or greater than the weak scale, $\delta_{\tilde{f}} = m_{\tilde f} - m_\chi \gtrsim m_t$~. Although the procedure described here applies to a wide variety of models, for definiteness, we focus on the case of a right-handed sbottom ($\tilde{b}_R$) interacting with the bino ($\chi$) and bottom quark ($b$). Let us discuss in turn the ingredients ${\boldsymbol c}$, ${\boldsymbol R}$, ${\boldsymbol M}$, and ${\boldsymbol f}$ of the factorization presented in Eq.~(\ref{eq:factorize}).

\begin{figure}[t]
\begin{center}
\includegraphics[width=0.9\textwidth]{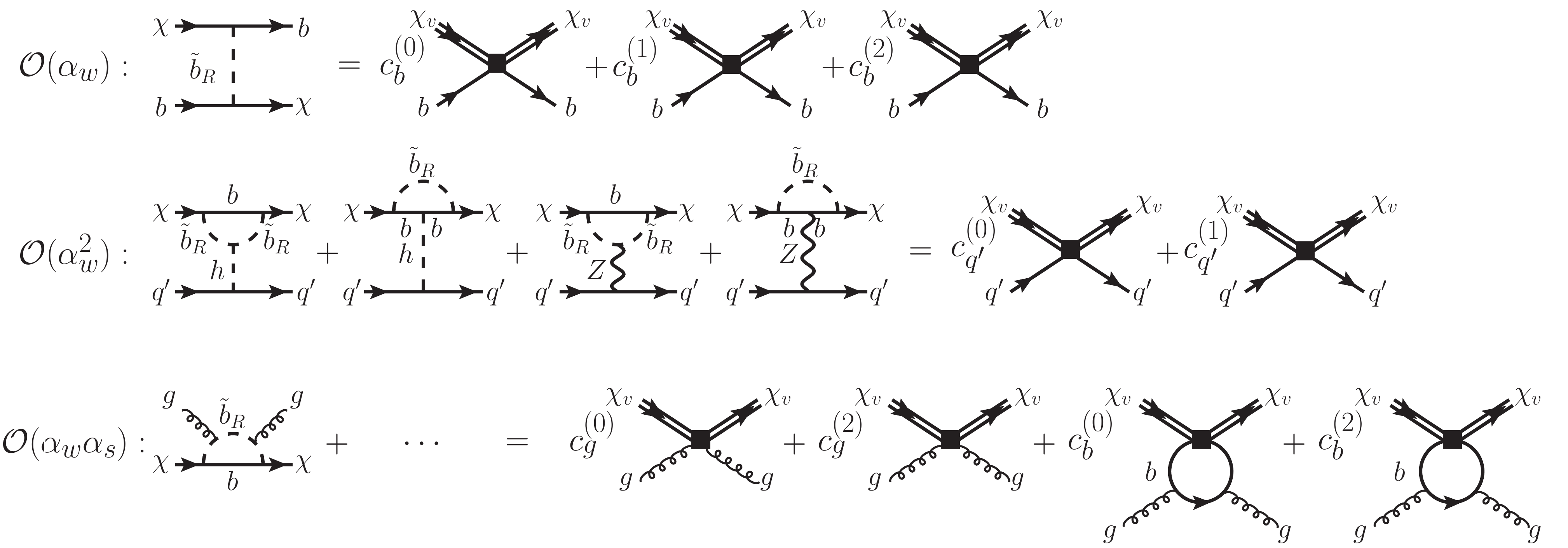} 
\caption{\label{fig:MatchHeavySbottom} Weak scale matching conditions for the case of a right-handed sbottom that is much heavier than the bino. Crossed and charge-reversed diagrams are not shown. In the full theory diagrams, $q^\prime$  refers to $u,d,s,c$~. 
The ellipsis denotes similar diagrams where the insertion of the gluon legs vary (see Appendix~\ref{app:glue}). Single (double) lines correspond to relativistic (heavy particle theory) fields. We have omitted the label ``bare" on the coefficients on the right-hand side.}
\end{center}
\end{figure}

\underline{Weak scale coefficients ${\boldsymbol c}(\mu_t)$} : The matching condition at the weak scale $\mu_t \sim m_t$ is shown in Fig.~\ref{fig:MatchHeavySbottom}. As in the previous example, the full theory amplitudes are computed using the Lagrangian in Eq.~(\ref{eq:fulltheory}), while the effective theory amplitudes are computed using the Lagrangian in Eq.~(\ref{eq:LWIMPSM}). The weak scale particles $W^\pm\,, Z \,, h \,, G^\pm \,, t \,, {\tilde b}_R$ are integrated out, and their effects are encoded in Wilson coefficients of the effective theory describing a heavy bino $\chi_v$ interacting with the quarks and gluons of 5-flavor QCD.

As in the previous example, the leading contributions to the coefficients $c_{u,d,s,c}^{(0,1,2)}$ are $\order(\alpha_w^2)$ loop diagrams.
What distinguishes this case is the presence of a tree-level, $\order(\alpha_w)$, contribution to the bottom quark coefficients $c_b^{(0,1,2)}$ and the associated loop-level, $\order(\alpha_w \alpha_s)$, effective theory contributions to the gluon coefficients $c_g^{(0,2)}$. As discussed in Sec.~\ref{sec:HPT}, we adopt the scheme where all mass scales much lighter than the weak scale (such as $m_b$) are set to zero, and employ dimensional regularization. The full theory contribution to $c_g^{(2) {\rm bare} }$ is IR divergent due to gluons emitted off of a massless bottom quark. The effective theory contributions from a bottom quark loop, shown on the right-hand side of Fig.~\ref{fig:MatchHeavySbottom}, are scaleless, and thus vanish. In the low energy theory, the remaining $1/\epsilon$ pole of the bare coefficient is regarded as an UV divergence that is renormalized according to Eq.~(\ref{eq:renormalize}). For illustration, we present the explicit pole structure of the contributions to the renormalized spin-2 gluon coefficient:
\begin{align}\label{eq:poles}
c_g^{(2)} (\mu) &= c_g^{(2)\rm{FT}}  - c_g^{(2)\rm{EFT}}  +  c_b^{(2)} ~ \frac{\alpha_s}{6 \pi} ~ \frac{1}{\epsilon_\text{UV}}  + \order(\alpha_s^2)  \nl
&= \Bigg[ \frac{- \alpha_s \alpha^\prime m_\x}{27 \left(m_{\tilde{b}_R}^2-m_\x^2\right)^2} ~ \frac{1}{\epsilon_\text{IR}}   + \mbox{finite} \Bigg] - \Bigg[ c_b^{(2)} ~ \frac{\alpha_s}{6\pi} \left( \frac{1}{\epsilon_\text{UV}} - \frac{1}{\epsilon_\text{IR}} \right) \Bigg] \nl
&\quad +  c_b^{(2)} ~ \frac{\alpha_s}{6 \pi} ~ \frac{1}{\epsilon_\text{UV}}   + \order(\alpha_s^2)  \,,  
\end{align}
where $c_g^{(2)\rm{FT}}$ ($c_g^{(2)\rm{EFT}}$) is the the full (effective) theory loop contribution appearing on the left (right) side of the gluon matching condition in Fig.~\ref{fig:MatchHeavySbottom}, and the last term comes from the renormalization prescription of Eq.~(\ref{eq:renormalize}). We have omitted the label ``bare" on the coefficients on the right-hand side, and expressed the vanishing effective theory contribution, $c_g^{(2)\rm{EFT}}$, in terms of canceling UV and IR poles. Note the required consistency between $c_b^{(2)}$ (given in Eq.~(\ref{eq:HeavySbottomTree})) and the infrared pole of the full theory contribution $c_g^{(2)\rm{FT}}$ (given in Eq.~(\ref{eq:SbottomGlue})) to yield a finite renormalized coefficient $c_g^{(2)} (\mu)$. The other coefficients $c_q^{(0,1)}$ and $c_g^{(0)}$ are simply renormalized as $c (\mu) = c^\text{bare}$. 

As before, we collect the renormalized Wilson coefficients in the vectors
\begin{align}\label{eq:weakscaleC2}
 {\boldsymbol c}_{\text{SI}}^T(\mu_t) = \left\{ c_q^{(0)}(\mu_t) ~,~ c_g^{(0)}(\mu_t) ~,~ c_q^{(2)}(\mu_t) ~,~ c_g^{(2)}(\mu_t) \right\}\,, \quad
{\boldsymbol c}_{\text{SD}}^T(\mu_t) &= \left\{ c_q^{(1)} \right\}
~,
\end{align}
where $c_q^{(0,1,2)}$ is representative of the five quark flavors, i.e., $q=u,d,s,c,b$, so that these two vectors are 12 and 5 dimensional, respectively.  Note that $c_q^{(2)}(\mu_t)$ is non-zero only for $q = b$. In general, $Z$-exchange contributes to the SD interaction $c_q^{(1)}$, but when $m_b =0$ and the sbottom is purely right-handed, this amplitude vanishes at leading order in momentum transfer by gauge invariance (Eq.~(\ref{eq:SbottomZ})). The loop diagram where the Higgs is radiated off the bottom quark also vanishes, while the one where the Higgs is radiated off the sbottom contributes to $c_q^{(0)}$ (Eq.~(\ref{eq:SbottomHiggs})).

\underline{Running and matching matrices ${\boldsymbol R}$ and ${\boldsymbol M}$, and nucleon matrix elements ${\boldsymbol f}(\mu_0)$} : 
Since the theory below the weak scale is again given by Eq.~(\ref{eq:LWIMPSM}), the mapping of the weak scale coefficients to the hadronic scale is identical to the previous example of Sec.~\ref{sec:case1}. In particular, the components ${\boldsymbol R}$ and ${\boldsymbol M}$ implement RGE and matching across heavy quark thresholds, respectively, while  ${\boldsymbol f}$ applies nucleon matrix element form factors.

\subsection{Case III: Right-Handed Sbottom, Small Mass Splitting}
\label{sec:lightfsmalldelta}

Finally, we consider the case where both the mass of the fermion partnered to the sfermion and the mass splitting between the sfermion and the bino are much lighter than the weak scale, $\delta_{\tilde{f}}\,, m_f \ll m_t$~. For definiteness, we focus on the case of a right-handed sbottom ($\tilde{b}_R$) interacting with the bino ($\chi$) and bottom quark ($b$). 

\begin{figure}[t]
\begin{center}
\includegraphics[width=1\textwidth]{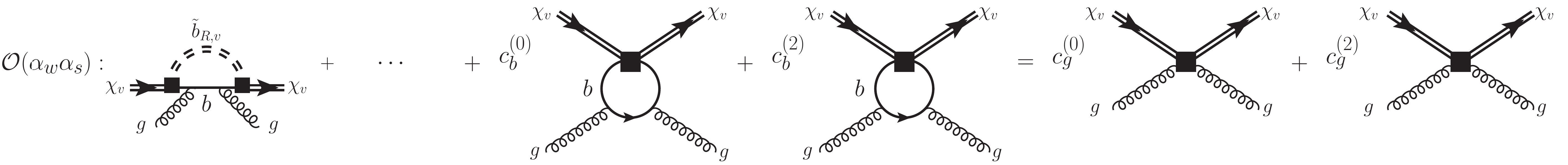} 
\caption{\label{fig:MatchLightSbottom2} Matching condition at the bottom quark threshold for a heavy particle effective theory of a right-handed sbottom that is nearly degenerate with the bino. Single (double) lines correspond to relativistic (heavy particle theory) fields. The ellipsis denotes similar diagrams where the insertion of the gluon legs vary (see Appendix~\ref{app:HPTglue}).}
\end{center}
\end{figure}

In this example, the sbottom is not highly virtual at low energies since the small sbottom-bino mass splitting kinematically allows for sbottom-bino interactions through a soft bottom. Weak-scale physics is still integrated out by matching onto 5-flavor QCD, but \emph{both} the bino and sbottom are kept as heavy fields in the effective theory (valid for $m_{\tilde{b}_R}\,, m_\x  \gg m_b$). The relevant interactions may be obtained from the full theory by introducing the field redefinition of Eq.~(\ref{eq:HPTredefinition2}) for the relativistic bino field $\chi$, and 
\begin{align}\label{eq:HPTredefinition3}
{\tilde b}_R = {1 \over \sqrt{2 m_\chi}} ~e^{-i m_\x v \cdot x}~ {\tilde b}_{R,v}
\end{align}
for the relativistic sbottom field ${\tilde b}_R$~. The field $X_v$ from Eq.~(\ref{eq:HPTredefinition2}) is again integrated out, and upon employing the invariance described in Eq.~(\ref{eq:invariance}) for heavy self-conjugate fields, we obtain
\begin{align}\label{eq:heavylight}
{\cal L} \supset  {\tilde b}^*_{R,v} \left( - i v \cdot D -\delta_{\tilde{b}_R} \right) {\tilde b}_{R,v} + {1 \over \sqrt{m_\chi}} ~ \tilde{b}_{R,v} ~ \bar{b} (\alpha_b + \beta_b \gamma^5)~ \x_v + {\rm h.c.} \, ,
\end{align}
where for a right-handed sbottom $\alpha_b = - \beta_b = - g^\prime/3\sqrt{2}$~. The residual mass term is given by the mass splitting $\delta_{\tilde{b}_R} = m_{\tilde{b}_R} - m_\x \ll m_t~$, and the sbottom-bino coupling is the heavy particle version of Eq.~(\ref{eq:fulltheory}). Physically, the heavy particle velocity, $v^\mu$, is conserved in the scattering process. Thus, the sign in the kinetic term denotes a sbottom coming into the vertex, or by using integration by parts, an anti-sbottom coming out of the vertex. In contrast to the relativistic case where $\chi = \chi^c$, the fields $\chi_v$ and $\chi_v^c$ can only be related through the invariance in Eq.~(\ref{eq:invariance}). Hence, the two vertices above are the only ones that contribute to amplitudes involving $\chi_v$ as the initial and final state (e.g., there are no charge-reversed diagrams in Fig.~\ref{fig:MatchLightSbottom2}).  Note from the canonically normalized kinetic term that the heavy sbottom has scaling dimension $3/2$ (hence the factor of $1/\sqrt{m_\chi}$ appearing in the field redefinition in Eq.~(\ref{eq:HPTredefinition3}) and in the sbottom-bino coupling). In the low energy theory the interactions of the heavy bino with the quarks and gluon of 5-flavor QCD are still described by Eq.~(\ref{eq:LWIMPSM}). 

The sbottom-bino interaction introduced in Eq.~(\ref{eq:heavylight}) can be viewed similarly to the so-called ``heavy-light current" in applications for B-meson decays~\cite{Shifman:1986sm,Politzer:1988wp,Politzer:1988bs}. In particular, its running due to QCD corrections from $\mu_t \sim m_t$ down to $\mu_b \sim m_b$ is significant, and we account for this when implementing the RGE down to the bottom quark threshold. Let us discuss in turn the ingredients ${\boldsymbol c}$, ${\boldsymbol R}$, ${\boldsymbol M}$, and ${\boldsymbol f}$ of the factorization presented in Eq.~(\ref{eq:factorize}).

\begin{figure}[t]
\begin{center}
\includegraphics[width=0.9\textwidth]{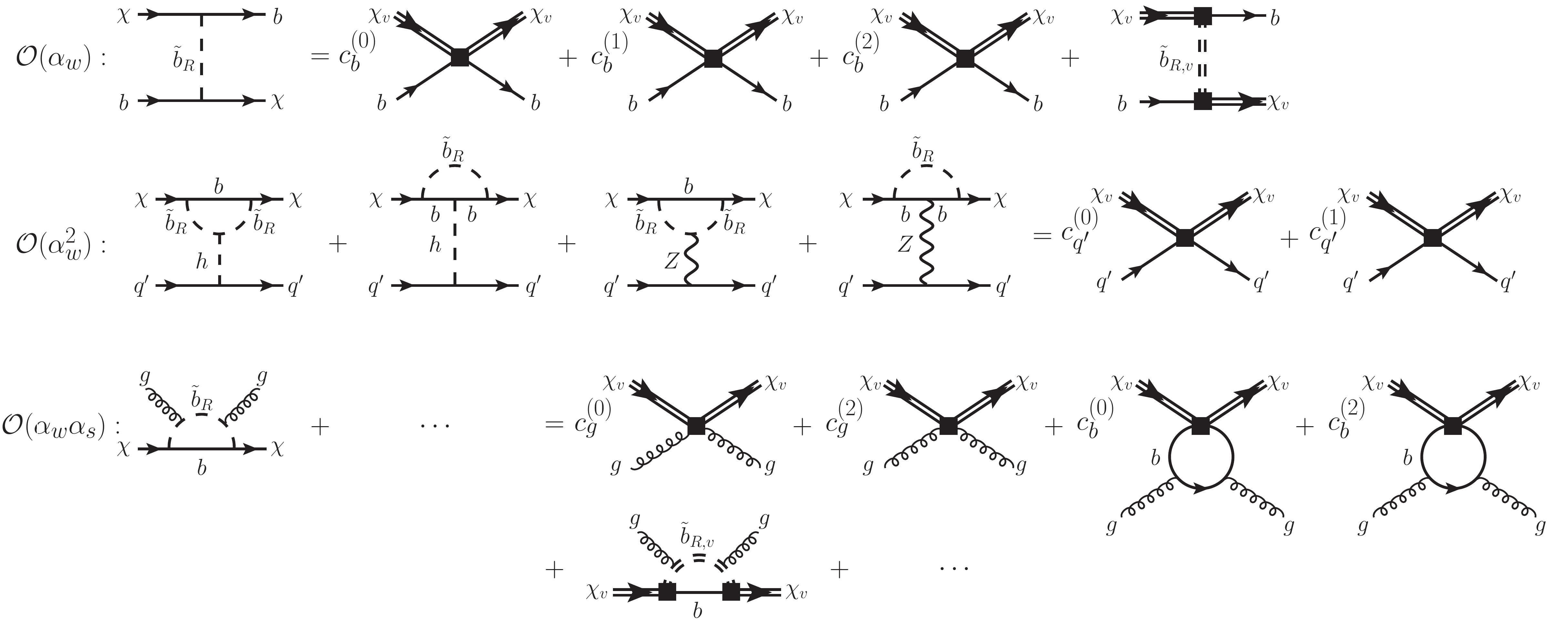} 
\caption{\label{fig:MatchLightSbottom1} Weak scale matching conditions for the case of a right-handed sbottom that is nearly degenerate with the bino. Crossed and charge-reversed diagrams are not shown. In the full theory diagrams, $q^\prime$  refers to $u,d,s,c$~. In the bottom line, the ellipsis denotes similar diagrams where the insertion of the gluon legs vary (see Apps.~\ref{app:glue} and~\ref{app:HPTglue}). Single (double) lines correspond to relativistic (heavy particle) fields. We have omitted the label ``bare" on the coefficients on the right-hand side.}
\end{center}
\end{figure}

\underline{Weak scale coefficients ${\boldsymbol c}(\mu_t)$} : The matching condition at the weak scale $\mu_t \sim m_t$ is shown in Fig.~\ref{fig:MatchLightSbottom1}. The full theory amplitudes are computed using the Lagrangian in Eq.~(\ref{eq:fulltheory}), while the effective theory amplitudes are computed using the Lagrangians in Eqs.~(\ref{eq:LWIMPSM}) and~(\ref{eq:heavylight}). The coefficients $c_{u,d,s,c}^{(0,1,2)}$ are determined by the same $\order(\alpha_w^2)$ loop diagrams of the previous two examples. Since we set all mass scales much lighter than the weak scale to zero, we are implicitly taking the $m_b\, , \delta_{{\tilde{b}}_R} \ll m_t$ limit of both the full theory and effective theory amplitudes. Of course, it is precisely in this limit that the relativistic and heavy particle Feynman rules match. Therefore, the full theory contribution from Eq.~(\ref{eq:fulltheory}) and the effective theory contribution from Eq.~(\ref{eq:heavylight}) cancel in the gluon and bottom quark matching, yielding coefficients $c_g^{(0,2)}$ and $c_b^{(0,1,2)}$ that vanish up to $\order(m_b/m_\x\,, \delta_{\tilde{b}_R}/m_\x)$ corrections. As an explicit example, the relativistic sbottom propagator in the tree-level diagram is expanded as
\begin{align}
\frac{2}{(k-p)^2 - m_{\tilde{b}_R}^2} &= \frac{2}{m_b^2 - 2(m_\x \delta_{\tilde{b}_R} + p \cdot k)}  + \order(\delta_{\tilde{b}_R} / m_\x)
\nl
&=  \frac{1}{m_\chi(- v \cdot k - \delta_{\tilde{b}_R})} + \order(\delta_{\tilde{b}_R} / m_\x\,, m_b / m_\x)
~,
\end{align}
where we have included a factor of $2$ for the crossed diagram, and used $p^\mu = m_\chi v^\mu$. Note that the above result matches the tree-level amplitude obtained from the Feynman rules of Eq.~(\ref{eq:heavylight}). In contrast, the usual expansion of the sbottom propagator in terms of local operators (corresponding to nonzero $c_b^{(0,1,2)}$ coefficients) is valid for $m_b\,, m_\chi \ll m_{{\tilde b}_R}~$. For the gluon matching, we find that the full theory amplitudes vanish at $\order{(1/m_\chi)}$, which must be the case since the gluon coefficients scale as $[ \rm{mass} ]^{-3}$, but the only mass scale is $m_\chi \sim m_{\tilde{b}_R}$ (see Eqs.~(\ref{eq:bareglue})-(\ref{eq:bareglue2}) for the explicit forms of the full theory gluon diagrams in the limit $m_b = \delta_{\tilde{b}_R} = 0$). Similarly, the effective theory loop diagrams are scaleless, and hence vanish, as discussed in Sec.~\ref{sec:HPT} and in the example of Sec.~\ref{sec:case2}. In principle, setting $m_b=0$ introduces IR poles as in Sec.~\ref{sec:case2}, but in this case they appear at $\order (1/m_\x^3)$. Thus, with no spin-2 quark or gluon coefficients generated at $\order(1/m_\x)$, all bare Wilson coefficients are trivially renormalized, i.e., $c(\mu) = c^\text{bare}$. Collecting the Wilson coefficients as in Eq.~(\ref{eq:weakscaleC2}), up to corrections of $\order(m_b/m_\x\,, \delta_{\tilde{b}_R}/m_\x)$, we find
\begin{align}\label{eq:weakscaleC3}
{\boldsymbol c}^T_{\rm SI}(\mu_t) &= \left\{ c_{q}^{(0)}(\mu_t) \,, 0 \,, 0 \,, 0 \right\}  \,, \quad
{\boldsymbol c}^T_{\rm SD}(\mu_t) = \left\{ 0 \right\}  \,.
\end{align}
Note that these two vectors, as in Eq.~(\ref{eq:weakscaleC2}), are 12 and 5 dimensional for SI and SD, respectively.  The coefficient $c_{q}^{(0)}$ is only non-zero for the four quark flavors $q=u,d,s,c$~, and is generated from integrating out the Higgs (corresponding to the full theory diagram where a Higgs is radiated off the sbottom, given in Eq.~(\ref{eq:SbottomHiggs2})). On the other hand, neither $c_b^{(0)}$ nor any of the spin-2 quark and gluon coefficients are generated at $\order(1/m_\chi)$ because the sbottom is kept in the low-energy effective theory below the weak scale. As in the previous case, the contributions from a Higgs radiated off a bottom quark and Z-exchange vanish in the chiral limit $m_b=0$~.

\underline{Running from $\mu_t$ down to $\mu_b$} : 
At leading order in $1/m_\x$~, the only nonvanishing coefficients are those corresponding to the scale invariant current $O_q^{(0)} = m_q {\bar q} q$~, and thus the coefficients in Eq.~(\ref{eq:weakscaleC3}) do not evolve, i.e., ${\boldsymbol c}_{\rm SI}(\mu_b) = {\boldsymbol R}_{\rm SI} (\mu_b, \mu_t) \, {\boldsymbol c}_{\rm SI}(\mu_t)= {\boldsymbol c}_{\rm SI}(\mu_t)$. We must also account for the scale evolution of the sbottom-bino couplings $\alpha_b\,, \beta_b$ in Eq.~(\ref{eq:heavylight}). The anomalous dimension $\gamma$ of the current ${\tilde b}_{R,v} ~ {\bar b} \, \Gamma \chi_v$~, with Dirac structure $\Gamma$, is the same as that of the heavy-light current ${\overline Q}_v \Gamma q$ describing the interaction of a heavy quark $Q_v$ with a light quark $q$~\cite{Shifman:1986sm,Politzer:1988wp,Politzer:1988bs}. It is independent of the Dirac structure $\Gamma$, and is given by $\gamma= -\alpha_s / \pi$. The evolution of the coefficients $c=\alpha_b\,, \beta_b$ is thus 
\be
\label{eq:heavylightRG}
c (\mu_b) = c (\mu_t) \left( \frac{\alpha_s(\mu_b)}{\alpha_s(\mu_t)} \right)^{2/\beta_0} 
~,
\ee
where $\beta_0 = 11-2n_f/3  = 23/3$~. This completely specifies the theory at the scale $\mu_b$\,, given by the Lagrangians in Eqs.~(\ref{eq:LWIMPSM}) and~(\ref{eq:heavylight}). 

\underline{Matching at $\mu_b$} : The matching condition at the bottom quark threshold $\mu_b$ is shown in Fig.~\ref{fig:MatchLightSbottom2}. The diagrams on the left are computed in the theory above the threshold using Eqs.~(\ref{eq:LWIMPSM}) and~(\ref{eq:heavylight}), while the diagrams on the right are computed in the theory below the threshold using Eq.~(\ref{eq:LWIMPSM}) but with four active quark flavors. Since the $q=u,d,s,c$ sectors of the two theories are identical, the only consequence of the matching is to integrate out the bottom and the heavy sbottom, encoding their effects into the scalar and spin-2 gluon coefficients. At this threshold, the mass scales $m_b$ and $\delta_{\tilde b}$ are kept non-zero. It is straightforward to modify the matrix ${\boldsymbol M}(\mu_b)$ in Appendix~\ref{app:R&M} to include the contribution from the heavy sbottom loop. Collecting the Wilson coefficients as in Eq.~(\ref{eq:weakscaleC3}), up to corrections of $\order(m_b/m_\x \,,\delta_{\tilde{b}_R}/m_\x)$, we find 
\begin{align}\label{eq:bottomscaleC3}
{\boldsymbol c}^T_{\rm SI}(\mu_b) &= \left\{ c_{q^\prime}^{(0)}(\mu_t) \,, c_g^{(0)}(\mu_b) \,, 0 \,, c_g^{(2)}(\mu_b) \right\} \,, \quad
{\boldsymbol c}^T_{\rm SD}(\mu_b) = \left\{ 0 \right\} \,,
\end{align}
where $c_{q^\prime}^{(0)}$ is representative of the four quark flavors, i.e., $q^\prime=u,d,s,c$~.  Note that these vectors are 10 and 4 dimensional, respectively, instead of 12 and 5 dimensional as in Eqs.~(\ref{eq:weakscaleC2}) and~(\ref{eq:weakscaleC3}).  Here, integrating out the sbottom and bottom quark at the threshold $\mu_b$ contributes to the scalar and spin-2 gluon coefficients, while the spin-2 quark coefficient is only generated at higher order. The analytic forms of the gluon coefficients are given in Eq.~(\ref{eq:HPTSbottomGlue}). 

\underline{Running and matching matrices ${\boldsymbol R}$ and ${\boldsymbol M}$, and nucleon matrix elements ${\boldsymbol f}(\mu_0)$} : Below the bottom quark threshold, the theory is given by the Lagrangian in Eq.~(\ref{eq:LWIMPSM}) with four quark flavors, and thus, for the remaining analysis down to 3-flavor QCD, we employ the same components ${\boldsymbol R}$, ${\boldsymbol M}$, and ${\boldsymbol f}$ of the previous two examples.


\section{\textbf{Phenomenology}}
\label{sec:pheno}

This section explores the phenomenology of several scenarios for bino DM in the MSSM. Sec.~\ref{sec:RHstop} and \ref{sec:RHsbottom} focus on the specific examples of a right-handed stop ($\tilde{t}_R$) and right-handed sbottom ($\tilde{b}_R$), respectively. In these sections, the matching and running prescription identically follows Sec.~\ref{sec:cases}. In particular, as shown in Eq.~(\ref{eq:LWIMPSM}), our computational scheme follows that of Refs.~\cite{Hill:2011be,Hill:2013hoa,Hill:2014yka,Hill:2014yxa}, employing a matching procedure that includes the leading order contributions for the lowest dimension operators relevant for Majorana DM-nucleon scattering. In Secs.~\ref{sec:mixed} and \ref{sec:slepton}, we present fixed-order calculations involving left-right mixed stops and sbottoms ($\tilde{t}_{1,2}$, $\tilde{b}_{1,2}$), and right-handed charged sleptons ($\tilde{l}_R$), respectively.

\subsection{Right-Handed Stop}
\label{sec:RHstop}

We begin with the simple example of bino-nucleon scattering induced through interactions with a right-handed stop ($\tilde{t}_R$). Note that a fixed-order calculation of this model was presented in Ref.~\cite{Ibarra:2015nca}. We go beyond this calculation by performing the complete leading order matching at the weak scale, and a leading log analysis as described in Sec.~\ref{sec:case1}.
\begin{figure}[t]
\begin{center}
\includegraphics[width=0.495\textwidth]{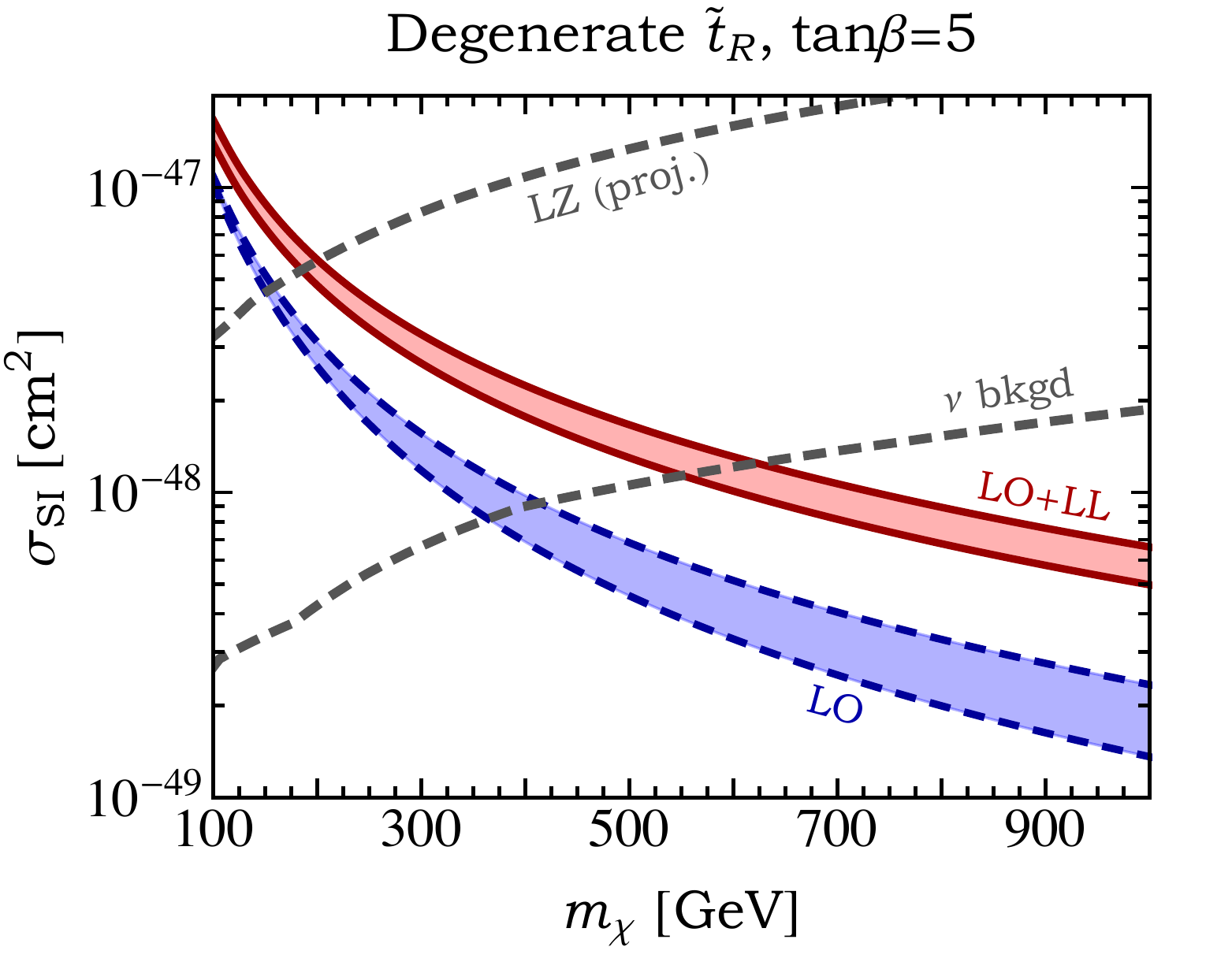} 
\includegraphics[width=0.495\textwidth]{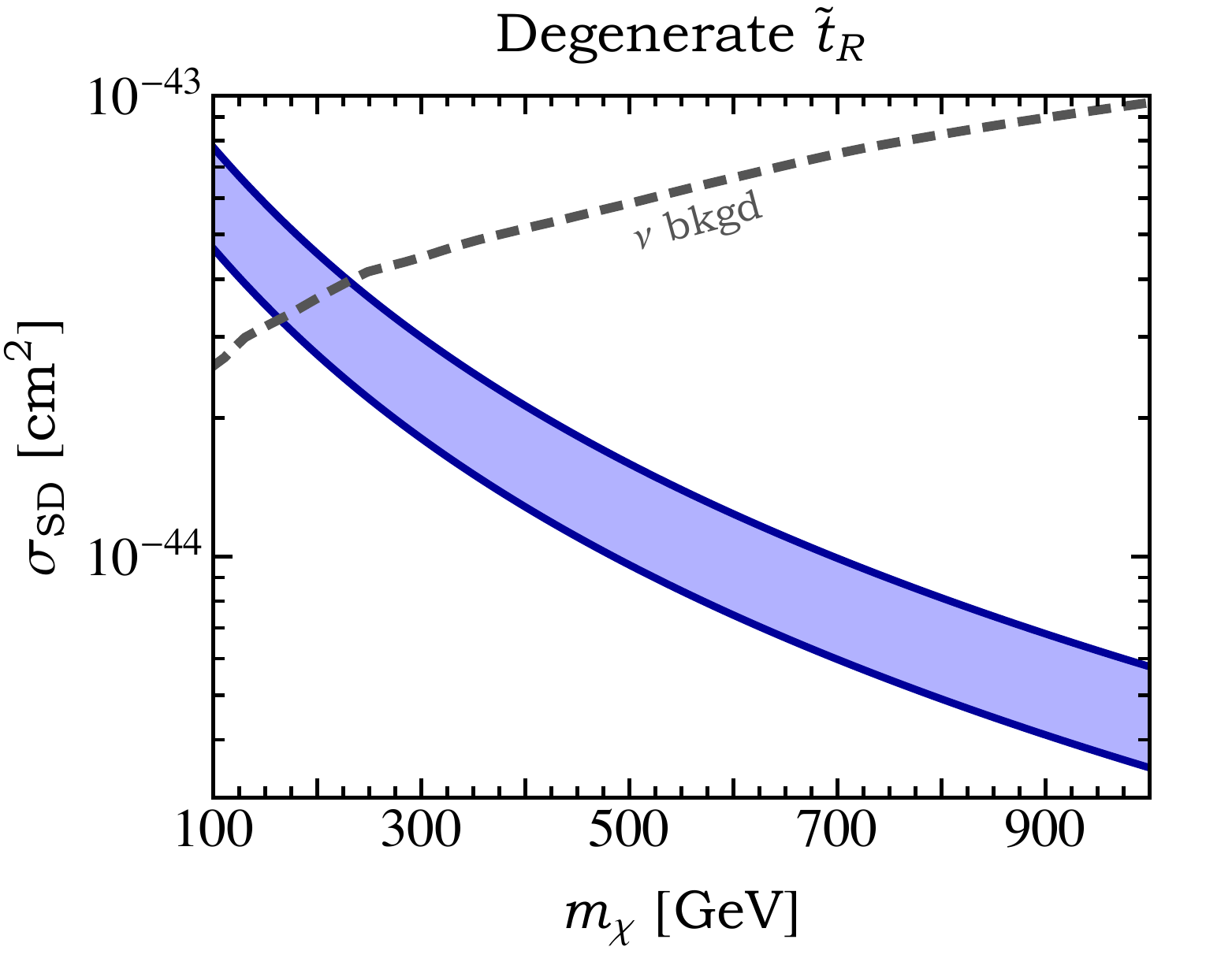} 
\caption{\label{fig:Stop}
\textbf{Left}: The spin-independent cross section (per-nucleon) for the case of a right-handed stop in the optimistic limit that its mass is nearly degenerate with that of the bino, $m_{\x}$. For comparison, we show both the fixed-order result (``LO", blue) and the leading log result from the effective theory analysis (``LO+LL", red). The thickness of the bands corresponds to the combined hadronic input and perturbative uncertainties. The grey dashed lines show the projected sensitivity of the LZ experiment and the neutrino background. \textbf{Right}: The spin-dependent cross section (per-neutron) for the case of a right-handed stop in the optimistic limit that its mass is nearly degenerate with that of the bino, $m_{\x}$. The thickness of the band corresponds to hadronic input uncertainties.}
\end{center}
\end{figure}

Constraints from LHC searches for direct production of stops are ameliorated in the limit of compressed stop spectra (although see Ref.~\cite{An:2015uwa}). For example, monojet searches at a 14 TeV high-luminosity LHC can only exclude binos lighter than 500 GeV~\cite{Low:2014cba}. At the same time, approximate degeneracy avoids power suppression of the amplitudes for bino-nucleon scattering, enhancing the prospects for direct detection. In light of this, we focus on the optimistic scenario that the mass splitting, $\delta_{\tilde{t}_R} = m_{\tilde{t}_R} - m_\x$~, is much less than the weak scale, and hence barring corrections of $\mathcal{O}(\delta_{\tilde{t}_R} / m_t)$, we set $m_{\tilde{t}_R} = m_\x$ when determining weak scale matching coefficients.

The resulting SI and SD cross sections \emph{per nucleon} for scattering on a Xenon target are shown in Fig.~\ref{fig:Stop}. Varying $\tan{\beta}$ would only affect these results at the level of a few percent. For SI scattering, we present a comparison of the ``LO" rate determined from the fixed order analysis described in Sec.~\ref{sec:fixedorder}, and the ``LO+LL" rate determined from the leading log EFT analysis described in Sec.~\ref{sec:cases}. The LO prediction includes the uncertainty from hadronic inputs, while the LO+LL prediction also includes the perturbative uncertainty (added in quadrature), obtained from the variation of renormalization scales $\mu_t$, $\mu_b$, and $\mu_c$~, within the ranges given in Table~\ref{table:scales}. For larger bino masses ($\sim 1$ TeV), the LL corrections enhance the rate by a factor of a few ($\sim 3$), due in part to $\order\left( \alpha_s (\mu_b)  \alpha_w^2 \right)$ corrections that are included in the EFT analysis, but are formally higher order in the fixed order approach. In particular, these are one-loop Higgs exchange diagrams that contribute to $c_g^{(0)}$ at two-loop. While both quark and gluon weak scale coefficients scale as $c_{q,g}^{(0)} \sim 1/v^2 m_\chi$, where $v$ is the SM Higgs vacuum expectation value, the Higgs exchange contributions are enhanced due to a $\log {m_\chi \over m_t}$ factor. The contribution from the spin-2 gluon amplitude is subdominant. For SD scattering, we only consider the ``LO" rate determined from the fixed order analysis described in Sec.~\ref{sec:fixedorder}, since corrections to coefficient running and matching enter at ${\cal O}(\alpha_s^2)$. While the analysis in Ref.~\cite{Ibarra:2015nca} reported destructive interference between the Higgs and gluon diagrams of Fig.~\ref{fig:MatchStop}, we find no such interference, and thus obtain substantially larger rates in Fig.~\ref{fig:Stop}. For more details, see Eqs.~(\ref{eq:piercecomp1}) and (\ref{eq:piercecomp2}).

\begin{table}[t]
\centering
\begin{tabular}{| c || c | c | c |}
\hline
Scale & Central & Range \\ \hline \hline
$\mu_t$ & $(m_W+m_t)/2 = 126$ GeV & $\Big(~m_W / \sqrt{2} ~,~ m_t \sqrt{2} ~\Big)$ \\ \hline
$\mu_b$ & $m_b= 4.75$ GeV & $\Big(~m_b / \sqrt{2} ~,~ m_b \sqrt{2} ~\Big)$ \\ \hline
$\mu_c$ & $m_c= 1.4$ GeV & $\Big(~1 \text{ GeV} ~,~ 2 \text{ GeV} ~\Big)$ \\ \hline
\end{tabular}
\caption{Numerical values used for the variation of renormalization scales of Fig.~\ref{fig:formalism}.}
\label{table:scales}
\end{table}

Although LZ will probe bino masses below $\sim 200$ GeV, Higgs coupling measurements sensitive to deviations in the gluon fusion rate already exclude this region after Run 1 of the LHC~\cite{Ibarra:2015nca}. Future direct detection experiments projected to reach SI cross sections close to the neutrino background will probe bino masses lighter than $\sim 600$ GeV. Furthermore, without an enhancement from coherent scattering, the SD rate from $Z$ exchange is below the neutrino background for masses $\gtrsim 200$ GeV. Note that in order to achieve the observed relic abundance from thermal freeze-out through coannihilation, the bino-stop mass splitting varies between 30 and 40 GeV for sub-TeV bino dark matter and gradually reaches sub-GeV splitting for dark matter mass above 2 TeV \cite{Ibarra:2015nca,deSimone:2014pda}.

\subsection{Right-Handed Sbottom}
\label{sec:RHsbottom}

%
\begin{figure}[t]
\begin{center}
\includegraphics[width=0.495\textwidth]{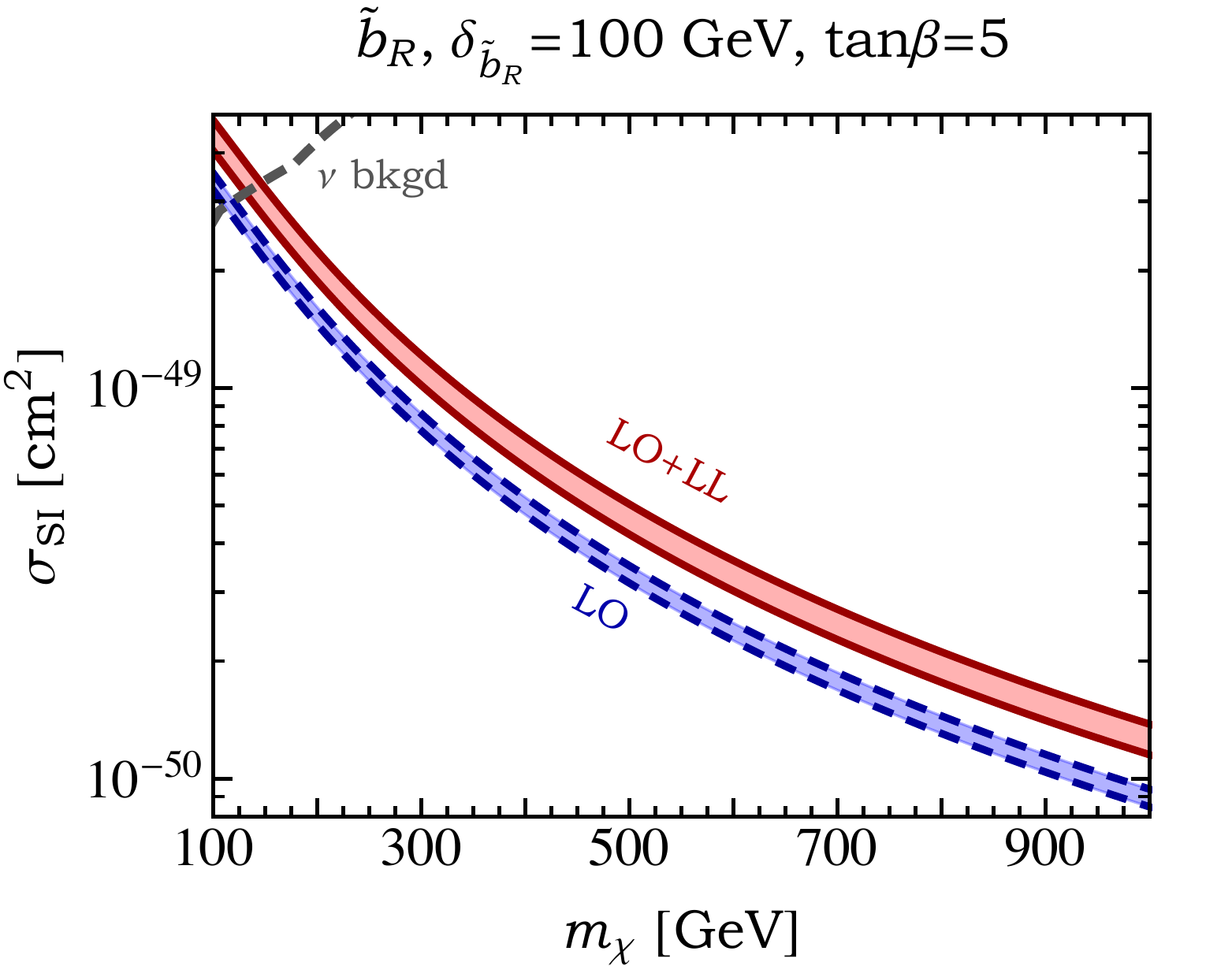} 
\includegraphics[width=0.495\textwidth]{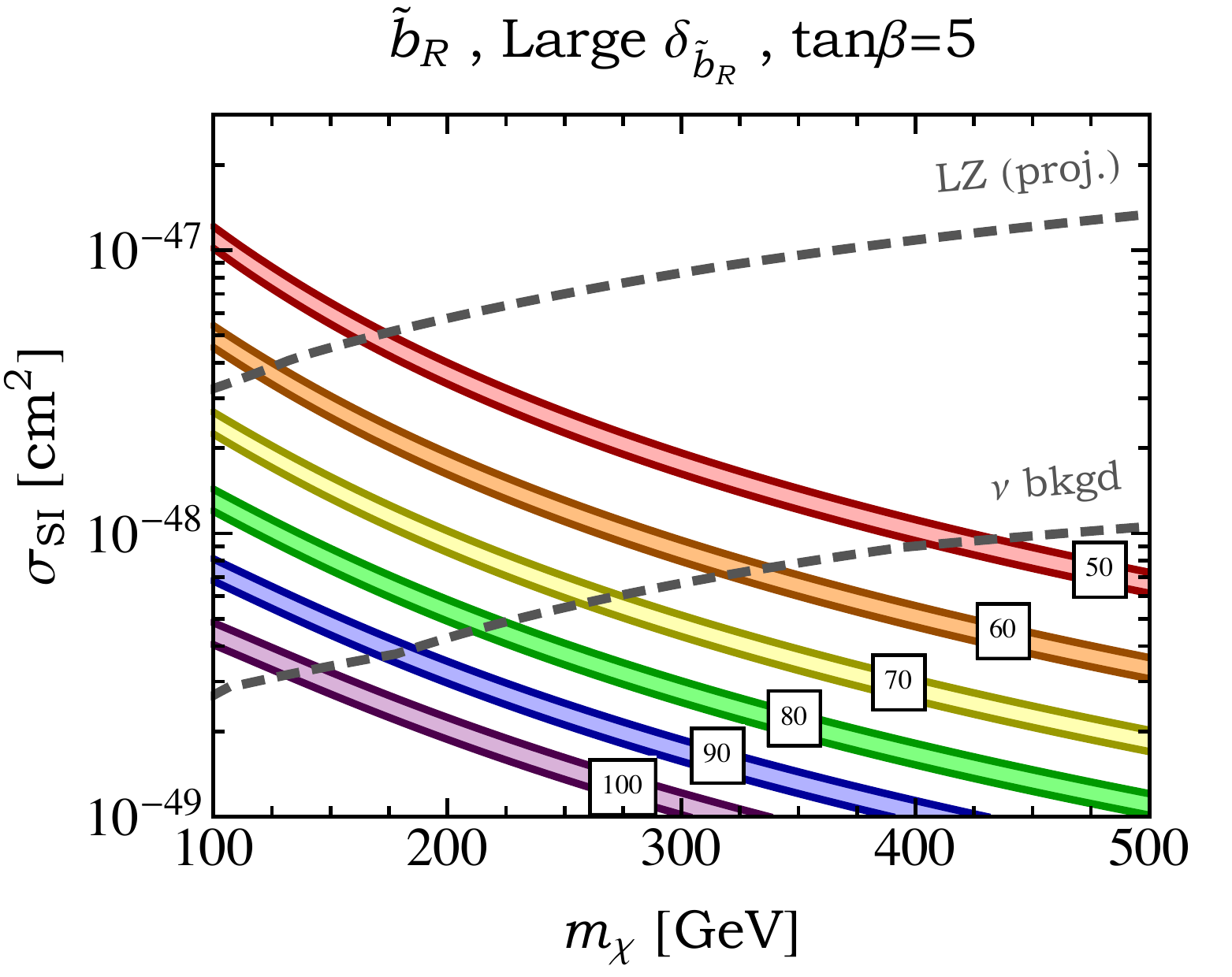} 
\caption{\label{fig:HeavySbottom} \textbf{Left}: The spin-independent cross section (per-nucleon) for the case of a right-handed sbottom and a sbottom-bino mass splitting that is comparable to the weak scale ($\delta_{\tilde{b}_R} = 100$ GeV). 
For comparison, we show both the fixed-order result (``LO", blue) and the leading log result from the effective theory analysis (``LO+LL", red). The thickness of the bands corresponds to combined theoretical and hadronic uncertainties. The gray dashed line shows the point at which the irreducible neutrino background should be relevant. \textbf{Right}: The spin-independent nucleon cross sections as a function of $m_\x$ for various values of the sbottom-bino mass splitting in GeV (white boxes). The calculation is performed using the full ``LO+LL" framework. The width of the bands corresponds to the combined theoretical and hadronic uncertainties.}
\end{center}
\end{figure}

We now examine the direct detection prospects when the bino interacts with a pure right-handed sbottom ($\tilde{b}_R$). We consider the two cases described in Sec.~\ref{sec:cases}, depending on whether the mass splitting, $\delta_{\tilde{b}_R} = m_{\tilde{b}_R} - m_{\x}$~, is of order the weak scale or much smaller. For the complete description of the matching and running procedure, we refer the reader to Secs.~\ref{sec:case2}~and~\ref{sec:lightfsmalldelta}, for the large and small splitting cases, respectively. Assuming that the squark correction to the SM Higgs gluon fusion amplitude is proportional to $\mu_q ~ v / m_{\tilde{q}}^2$ (where $\mu_q$ is the dimensionful trilinear squark-squark-Higgs coupling) and that current LHC Higgs measurements in the gluon fusion channel constrain stops to be heavier than $\sim 300$ GeV, the rescaled limit for sbottoms approaches roughly $\sim 50$ GeV in the large $\tan{\beta}$ limit. Thus, throughout this section, we consider bino and sbottom masses greater than 100 GeV.

\subsubsection{Large Mass Splitting}
\label{sec:HeavyRHSbottom}

We begin with the case where the sbottom is significantly heavier than the bino, $\delta_{\tilde{b}_R} \sim 100 \, {\rm GeV}$. The resulting SI cross sections \emph{per nucleon} for scattering on a Xenon target are shown in Fig.~\ref{fig:HeavySbottom}. On the left panel, we include for comparison predictions for both the LO and LO+LL rates, as determined by the fixed order and EFT analyses, respectively. Perturbative and hadronic uncertainties are calculated as in Sec.~\ref{sec:RHstop}. For this large mass splitting case, the leading log corrections yield a slight enhancement of $\mathcal{O}(50\%)$. On the right panel of Fig.~\ref{fig:HeavySbottom}, we show the SI cross section as a function of $m_{\x}$ for values of the sbottom-bino mass splitting in the range $50-100$ GeV. The rate is dominated by the bino's scalar coupling to gluons. Depending on the particular value of $\delta_{\tilde{b}_R}$, the LZ experiment will probe light binos up to a few hundreds of GeV.

\begin{figure}[t]
\begin{center}
\includegraphics[width=0.495\textwidth]{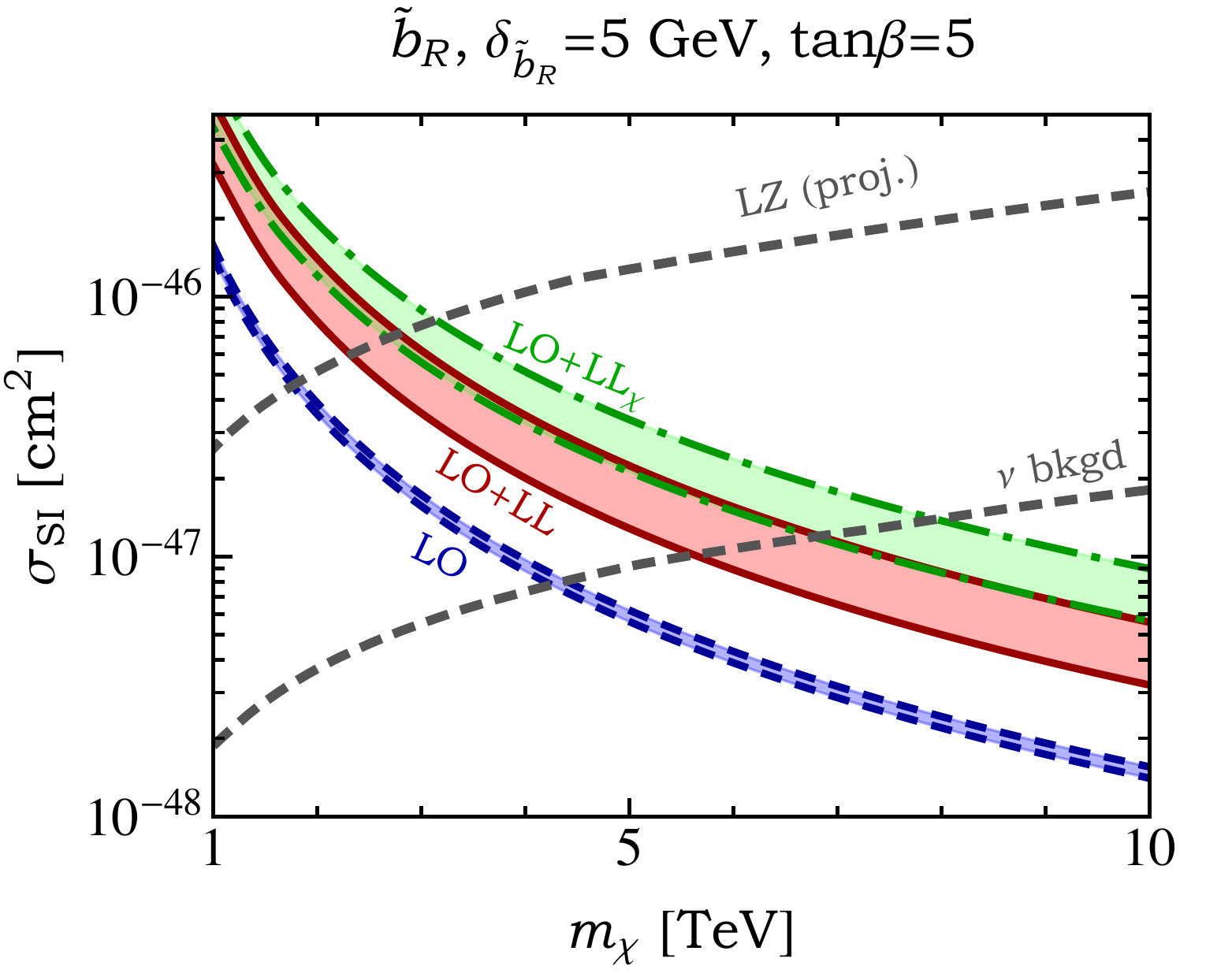} 
\includegraphics[width=0.495\textwidth]{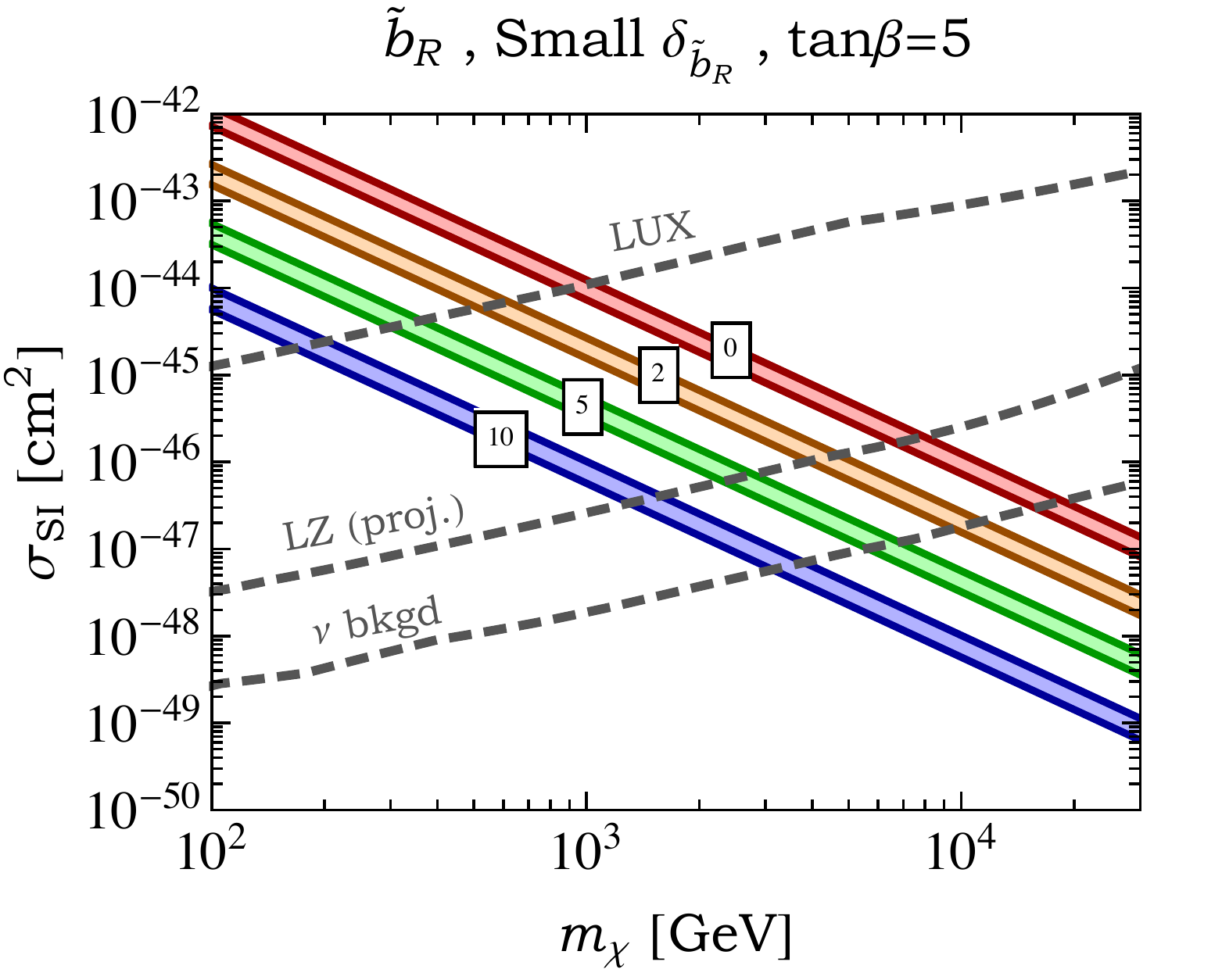} 
\caption{\label{fig:LightSbottom} \textbf{Left}: The spin-independent cross section (per-nucleon) for the case of a right-handed sbottom and a sbottom-bino mass splitting that is much less than the weak scale ($\delta_{\tilde{b}_R} = 5$ GeV). 
For comparison, we show both the fixed-order result (``LO", blue) and the leading log result from the effective theory analysis (``LO+LL", red). We also illustrate the impact of including the running of the $\alpha_f$ and $\beta_f$ coefficients of Eq.~(\ref{eq:fulltheory}) from the scale $\mu_\chi \sim m_\chi$ (``LO+LL${_\chi}$", green). The thickness of the bands corresponds to combined hadronic and theoretical uncertainties. The gray dashed lines show the projected reach of the LZ experiment and the point at which the irreducible neutrino background should be relevant. \textbf{Right}: The spin-independent nucleon cross sections for various values of the sbottom-bino mass splitting in GeV (white boxes). The calculation is performed using the full ``LO+LL" framework. The width of the bands corresponds to the combined theoretical and hadronic uncertainties.}
\end{center}
\end{figure}
%

\subsubsection{Small Mass Splitting}
\label{sec:LightRHSbottom}
Let us now consider the degenerate case, $\delta_{\tilde{b}_R} \ll 100 \, {\rm GeV}$, where the sbottom is kept as an active degree of freedom below the weak scale. The explicit matching and running prescription is detailed in Sec.~\ref{sec:lightfsmalldelta}. The resulting SI cross sections \emph{per nucleon} for scattering on a Xenon target are shown in Fig.~\ref{fig:LightSbottom}. On the left panel, we include predictions for both the LO and LO+LL rates, as determined by the fixed order and EFT analyses, respectively. Perturbative and hadronic uncertainties are calculated as in Sec.~\ref{sec:RHstop}. For this case, the rate receives large contributions from \emph{both} the scalar and spin-2 gluon couplings. 

For small relative mass splittings ($\delta_{\tilde{b}_R}/m_\x \lesssim 10^{-3}$) the enhancement from LL corrections has significant implications for predicting the discovery potential of future experiments. For instance, while the fixed-order approach predicts that bino DM as heavy as $\sim 4$ TeV has a scattering rate above the neutrino background, the complete calculation extends the reach up to $\sim 7$ TeV. In general, incorporating the running of the weak scale Wilson coefficients down to the hadronic scale results in an overall factor of $\sim 3-4$ in the final cross section. As described in Sec.~\ref{sec:lightfsmalldelta}, a significant portion of this enhancement is tied to the RGE of the $\tilde{b}_R \, {\bar b} \, \chi$ heavy-light current of Eq.~(\ref{eq:heavylightRG}), which alone rescales the fixed-order cross section by $[\alpha_s (m_b) / \alpha_s (m_t)]^{24/23} \approx 2$~. 

On the right panel of Fig.~\ref{fig:LightSbottom}, we show the SI cross section for various choices of the small mass splitting $\delta_{\tilde{b}_R}$~. Here, $\delta_{\tilde{b}_R}=0$  corresponds to a sbottom-bino mass splitting that is much smaller than the mass of the bottom quark. Bino DM with mass up to $3-20$ TeV will remain above the neutrino background for $\delta_{\tilde{b}_R} \approx 10-0$ GeV, respectively. Interestingly, such small mass splittings are also needed for standard freeze-out through sbottom co-annihilation, and hence LZ and future experiments will be sensitive to thermal bino DM in the multi-TeV mass range.

In the analysis in Secs.~\ref{sec:fixedorder} and~\ref{sec:cases}, we assumed, for definiteness, that the full theory described in Eq.~(\ref{eq:fulltheory}) was defined at the weak scale, $\mu_t \sim 100 \, {\rm GeV}$. It is interesting to consider the impact of additional RGE for cases where the full theory is defined at a higher scale, e.g., through imposing theoretical constraints of specific ultraviolet completions or observational constraints such as the relic abundance and collider limits. For illustration, let us consider the running of the bino-sfermion-fermion couplings $\alpha_f$ and $\beta_f$ of Eq.~(\ref{eq:fulltheory}) from a scale $\mu_\chi \sim m_\chi$ for the case of a sbottom nearly degenerate with the bino. The effective theory setup is similar to case III described in Sec.~\ref{sec:lightfsmalldelta}: at the scale $\mu_\chi$~, we match the full relativistic theory in Eq.~(\ref{eq:fulltheory}) onto the heavy particle effective theory in Eq.~(\ref{eq:heavylight}), and thus the running of the $\alpha_b$ and $\beta_b$ coefficients are again given by Eq.~(\ref{eq:heavylightRG}). At the weak scale, the contributions from Higgs exchange are $\order(1/m_\chi^2)$, and can be neglected when working to leading order in $1/m_\chi$~. Upon evolving down to the bottom scale $\mu_b$~, the remaining analysis follows that of Sec.~\ref{sec:lightfsmalldelta}. The impact of the additional running is shown in the left panel of Fig.~\ref{fig:LightSbottom} as the green curve labeled ``LO+LL$_\chi$". While the effect on the cross section is only $\sim 60\%$ (the strong coupling asymptotes at high-energies), the implied potential mass reach for an experiment probing cross sections near the neutrino background may be increased by $\sim 1 \, {\rm TeV}$.

\subsection{Mixed Squarks}
\label{sec:mixed}

Left-right mixing in the squark sector can affect the form of the cross sections considerably. In this section, we present a fixed-order estimate for the bino-nucleon scattering rate induced by interactions with mixed third generation squarks. Following the approach of Sec.~\ref{sec:fixedorder}, we match directly to 3-flavor QCD and include contributions from Higgs exchange and gluon diagrams when calculating the SI cross section. In calculating the Wilson coefficients $c_q^{(0)}$, $c_g^{(0)}$, and $c_g^{(2)}$, we substitute the expressions for the interactions in  Appendix~\ref{sec:model} into the general results of Appendix~\ref{sec:app1}. Note that mixing allows for the presence of additional states, resulting in new diagrams where multiple squarks are present in the same loop. Although non-zero, SD nucleon couplings are found to be subdominant throughout the parameter space that we consider, and are therefore omitted from the discussion below.

The lightest neutralino is assumed to be dominantly bino-like. For this to hold true, the Higgsino mass parameter is fixed at $\mu = 10$ TeV, and we refrain from considering bino masses ($m_\x \equiv M_1$) much larger than 1 TeV. In this section, $\mu$ denotes  the Higgsino mass parameter, not to be confused with a renormalization scale. The other gaugino masses are assumed to be completely decoupled from the low energy spectrum. Two parameters independently govern the mixing in the stop and sbottom sectors, 
\begin{align}
&X_t \equiv  A_t - \mu \cot{\beta} ~,~ X_b \equiv A_b - \mu \tan{\beta}
~,
\end{align}
where $X_{t,b} = \pm \sqrt{6} ~ m_{\tilde{t}, \tilde{b}}  \equiv \pm \left( 6 m_{\tilde{Q}_3} m_{\tilde{t}_R,\tilde{b}_R}\right)^{1/2}$ corresponds to maximal left-right mixing in the stop, sbottom sector, respectively. This determines the $A$-terms, $A_{t,b}$, for a given value of $\mu$ and $\tan{\beta}$. At every point in parameter space, we will set the bino mass in terms of the physical squark masses, given in Eq.~(\ref{eq:squarkmass}), such that
\be
\label{eq:masssplitt}
m_\x = \text{Min}(m_{\tilde{t}_{1,2}},m_{\tilde{b}_{1,2}}) - \delta_{\tilde{q}}
~,
\ee
which effectively defines the minimal mass splitting $\delta_{\tilde{q}}$~. 

Since left-right mixing introduces several new degrees of freedom compared to the models of the previous sections, we assume simplifying relations to reduce the size of the parameter space. In particular, we focus on two different schemes in parametrizing left-right mixing. In the first scheme, we set the third-generation left and right soft squark masses and mixing parameters equal:
\be
\label{eq:massparam}
m_{\tilde{q}} \equiv m_{\tilde{Q}_3} = m_{\tilde{t}_R} = m_{\tilde{b}_R} ~,~ X_q \equiv X_t = X_b 
~.
\ee
In the second scheme, we decouple the right-handed sbottom to 10 TeV and focus on left-right mixing in the stop sector alone:
\be
\label{eq:massparam2}
m_{\tilde{b}_R} \gg m_{\tilde{Q}_3} , m_{\tilde{t}_R} ~,~ m_{\tilde{Q}_3} \neq m_{\tilde{t}_R}
~.
\ee
\begin{figure}[t]
\begin{center}
\includegraphics[width=0.495\textwidth]{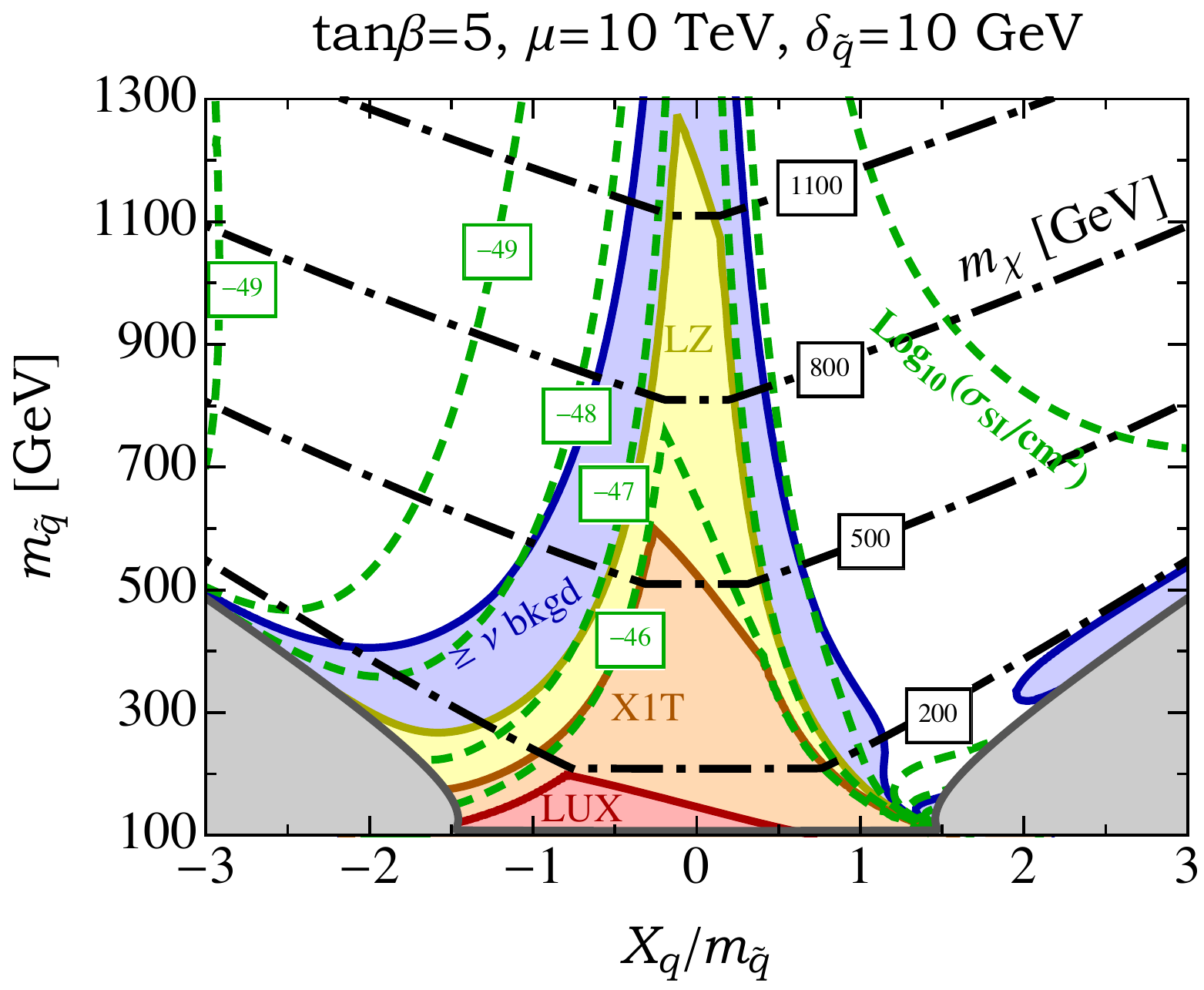} 
\includegraphics[width=0.495\textwidth]{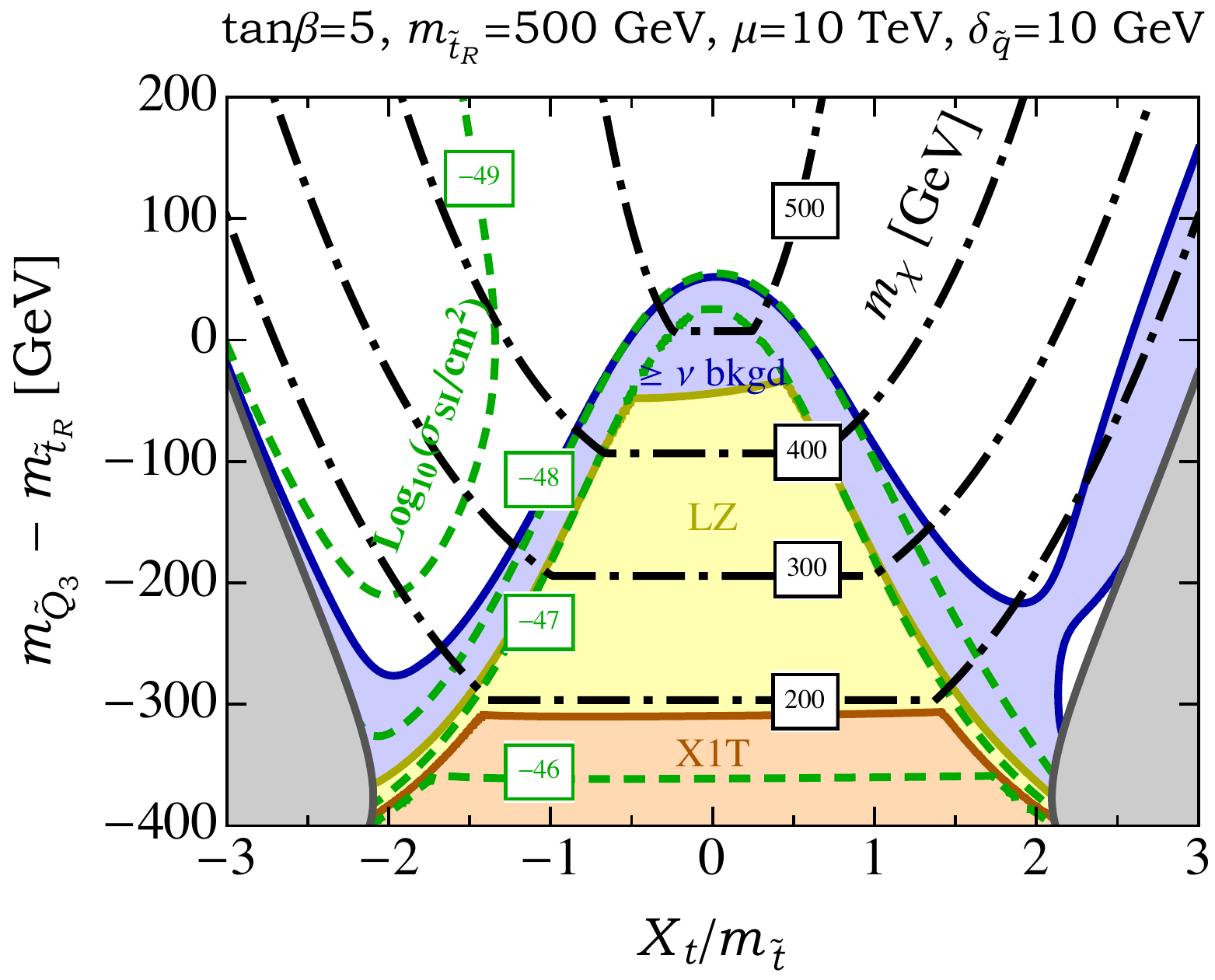} 
\caption{\label{fig:MixedStopSbottom} Results from a fixed-order calculation when the bino interacts with mixed stops and sbottoms using the parametrization of Eq.~(\ref{eq:massparam}) (left) and Eq.~(\ref{eq:massparam2}) (right). The Higgsino mass is fixed to $\mu =10$ TeV (not to be confused with a renormalization scale). The filled contours correspond to rates that are currently excluded by LUX (red) or will be probed by future experiments like XENON1T (orange) and LZ (yellow). Also shown are regions with cross sections greater than the neutrino background (blue). We do not consider bino masses lighter than 100 GeV or tachyonic squarks (both grey). For reference, we also show contours of fixed bino mass in GeV (black dot-dashed) and the spin-independent nucleon cross section in units of $\log_{10}(\sigma_\text{SI}/\text{cm}^2)$ (green dashed).}
\end{center}
\end{figure}

The prospects for detecting bino-nucleon scattering induced by its interactions with mixed stops and sbottoms are shown in Fig.~\ref{fig:MixedStopSbottom}. The left and right panels employ the parametrization of Eqs.~(\ref{eq:massparam}) and~(\ref{eq:massparam2}), respectively. Here, we fix the squark-bino mass splitting of Eq.~(\ref{eq:masssplitt}) to be $\delta_{\tilde{q}} = 10$ GeV, and $\tan{\beta}=5$~. We show the region currently excluded by LUX (red), the projected reach of XENON1T (orange) and LZ (yellow), and the parameter space with cross sections above the neutrino background (blue). We do not consider values of parameters where the bino is very light ($m_\x < 100$ GeV) or the mass of one or more squarks is tachyonic (both in grey). 

For both cases, left-right mixing tends to diminish the overall scattering rate, but for different reasons. For the left panel, corresponding to Eq.~(\ref{eq:massparam}), due to the small squark-bino mass splitting, the dominant scattering diagrams correspond to one-loop couplings to gluons through the exchange of the light sbottoms $\tilde{b}_1$ and $\tilde{b}_2$ (see Fig.~\ref{fig:glue}). These diagrams add coherently when the degree of mixing is small, i.e., $X_q \ll m_{\sq}$~. On the other hand, as soon as $| X_q / m_{\sq} | \gtrsim 0.5$, the diagrams involving $\tilde{b}_1$ or $\tilde{b}_2$ tend to interfere deconstructively, vastly lowering the scattering rate. This explains the sharp peak in the cross section near $X_q = 0$. 

For the right panel, corresponding to Eq.~(\ref{eq:massparam2}), larger mixing lowers the mass of the lightest stop relative to the left-handed sbottom, which decouples the lightest sbottom from the bino for a fixed mass splitting, $\delta_{\tilde{q}}$~, and suppresses potential contributions from sbottom induced gluon couplings. We find that, for these scenarios, squark mixing generally tends to reduce the reach of future direct detection experiments.

Mixing also strongly affects the stop sector. For brevity, we focus the discussion on the left panel of Fig.~\ref{fig:MixedStopSbottom}, corresponding to Eq.~(\ref{eq:massparam}); the behavior is similar for the right panel, corresponding to the parametrization of Eq.~(\ref{eq:massparam2}). Due to the large mass of the top quark, coupling to gluons through the exchange of $\tilde{t}_{1,2}$ does not see the enhancement at $X_q =0$, and instead Higgs exchange is the dominant process that involves stops. When $| X_q / m_{\sq} | $ is somewhat large, the Higgs-stop interaction grows and the mass splitting between the two stops is several hundreds of GeV, effectively decoupling $\tilde{t}_2$~. In this limit, we find that different behavior emerges depending on the sign of $X_q$~. In particular, for large and positive $X_q$~, the two Higgs exchange diagrams where $h$ is emitted off either an intermediate $\tilde{t}_1$ or top quark (Fig.~\ref{fig:higgsexchange}) interfere slightly, while for large and negative $X_q$ this pair of diagrams tends to add coherently. Hence, even though large mixing stifles the contribution from sbottom-gluon diagrams, Higgs exchange via virtual stops is able to somewhat lift this suppression for large and negative $X_q$~. This feature is clearly seen on the left-hand side of Fig.~\ref{fig:MixedStopSbottom}, which shows less diminished scattering rates near $ (- X_q/m_{\sq}) \sim 2-3$.

\subsection{Charged Sleptons}
\label{sec:slepton}

%
\begin{figure}[t]
\begin{center}
\includegraphics[width=0.7\textwidth]{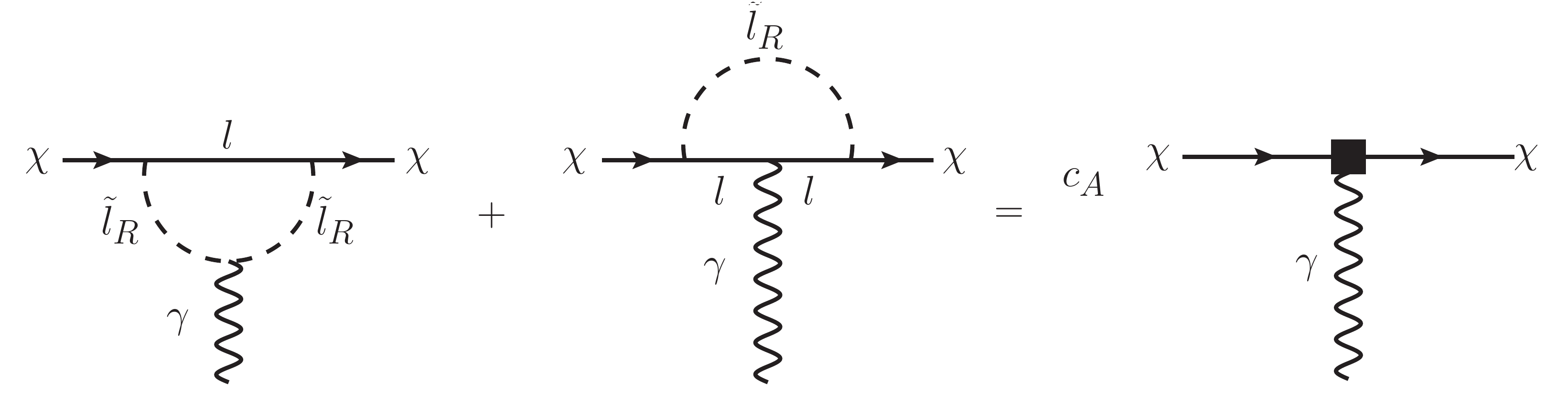} 
\caption{\label{fig:MatchSlepton} Matching procedure for the case of a single right-handed slepton. Charge-reversed diagrams are not shown.}
\end{center}
\end{figure}

In Secs.~\ref{sec:RHstop}-\ref{sec:mixed}, we presented examples where the fermion in Eq.~(\ref{eq:fulltheory}) is either sufficiently heavy such that Higgs exchange is the primary scattering process, or sufficiently light and colored such that coupling to gluons dominates the cross section. If the fermion is both light and uncolored, e.g., a charged lepton ($l$), one must reconsider the processes that contribute to elastic nucleon scattering. In this section, we will focus on the case where the bino ($\x$) interacts with a single right-handed selectron ($\tilde{e}_R$) or stau ($\tilde{\tau}_R$). Simplified models related to this scenario have been studied in~\cite{Bai:2014osa,Chang:2014tea}.

At leading order, SI scattering is given solely, to a good approximation, by loop diagrams coupling $\x$ to the electromagnetic current, $J_\mu^\text{EM}\equiv \partial^\nu F_{\nu \mu}$~, where $F_{\mu \nu}$ is the photon field strength. In particular, at dimension six, gauge invariance dictates that a Majorana fermion may only couple to the photon via the anapole operator, defined to be $\xbar \gamma^\mu \gamma^5 \x ~ J_\mu^\text{EM}$. Therefore, at low energies, in place of Eq.~(\ref{eq:lowbasis}), we consider the effective Lagrangian
\be
\label{eq:lowbasisslepton}
\mathcal{L} = c_A ~ \xbar \gamma^\mu \gamma^5 \x ~ J_\mu^\text{EM}
~.
\ee
From the definition of the current, it is apparent that this interaction must vanish in the limit of zero momentum transfer. This is also seen explicitly in the amplitude, where the contact interaction above leads to the effective form for the photon-amputated amplitude

\be
\label{eq:photonamp}
{\cal M}^\mu = 2 c_A ~ \bar{u}(p_f) \left( q^\mu \slashed{q} - q^2 \gamma^\mu \right) \gamma^5 u(p_i)
~ ,
\ee
where $p_i$, $p_f$ are the incoming and outgoing bino momenta, $u(p_{i,f})$ are the associated 4-component spinors, and $q \equiv p_f - p_i$ is the momentum transfer. The factor of two in the above expression accounts for the Majorana nature of $\chi$~.  

The prescription for matching Eq.~(\ref{eq:fulltheory}) onto the anapole operator Eq.~(\ref{eq:lowbasisslepton}) is shown in Fig.~\ref{fig:MatchSlepton}. We regulate IR poles with finite lepton masses. In this scheme, no divergences emerge, and hence $c_A$ is trivially renormalized. Explicit forms for $c_A$ are given in Sec.~\ref{app:anapole}. The proton matrix element of the current corresponds to the counting operator and is given by
\be
\langle p(k) | J_\mu^\text{EM} (\mu) | p(k) \rangle \equiv e(\mu) ~ \bar{u}(k) \gamma_\mu u(k)
~,
\ee
where the running of the electric coupling, $e (\mu)$~, is the only source of scale dependence. Unlike the scalar form factors in Sec.~\ref{eq:matrixelements}, the nucleon matrix element above is easily evaluated at the weak scale, and hence a fixed-order calculation suffices. After taking matrix elements, the SI bino-proton cross section is then given by
\be
\sigma_\text{SI} = \frac{e^2}{2 \pi} ~ m_p ~ E_R ~ c_A^2
~,
\ee
where $m_p$ is the proton mass and $E_R$ is the recoil energy~\cite{Bai:2014osa}. As a representative value we set  $E_R=10$ keV. 

\begin{figure}[t]
\begin{center}
\includegraphics[width=0.495\textwidth]{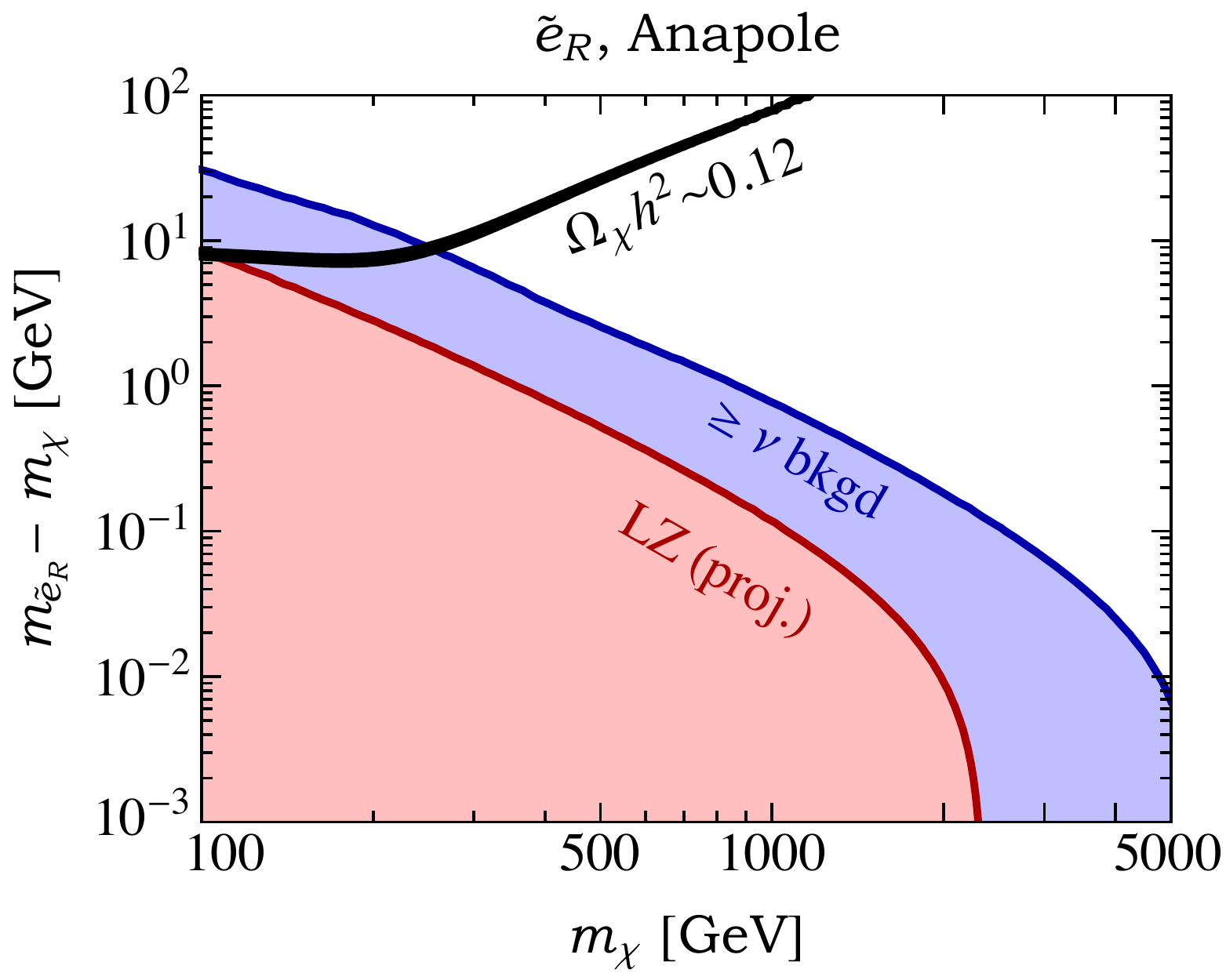} 
\includegraphics[width=0.495\textwidth]{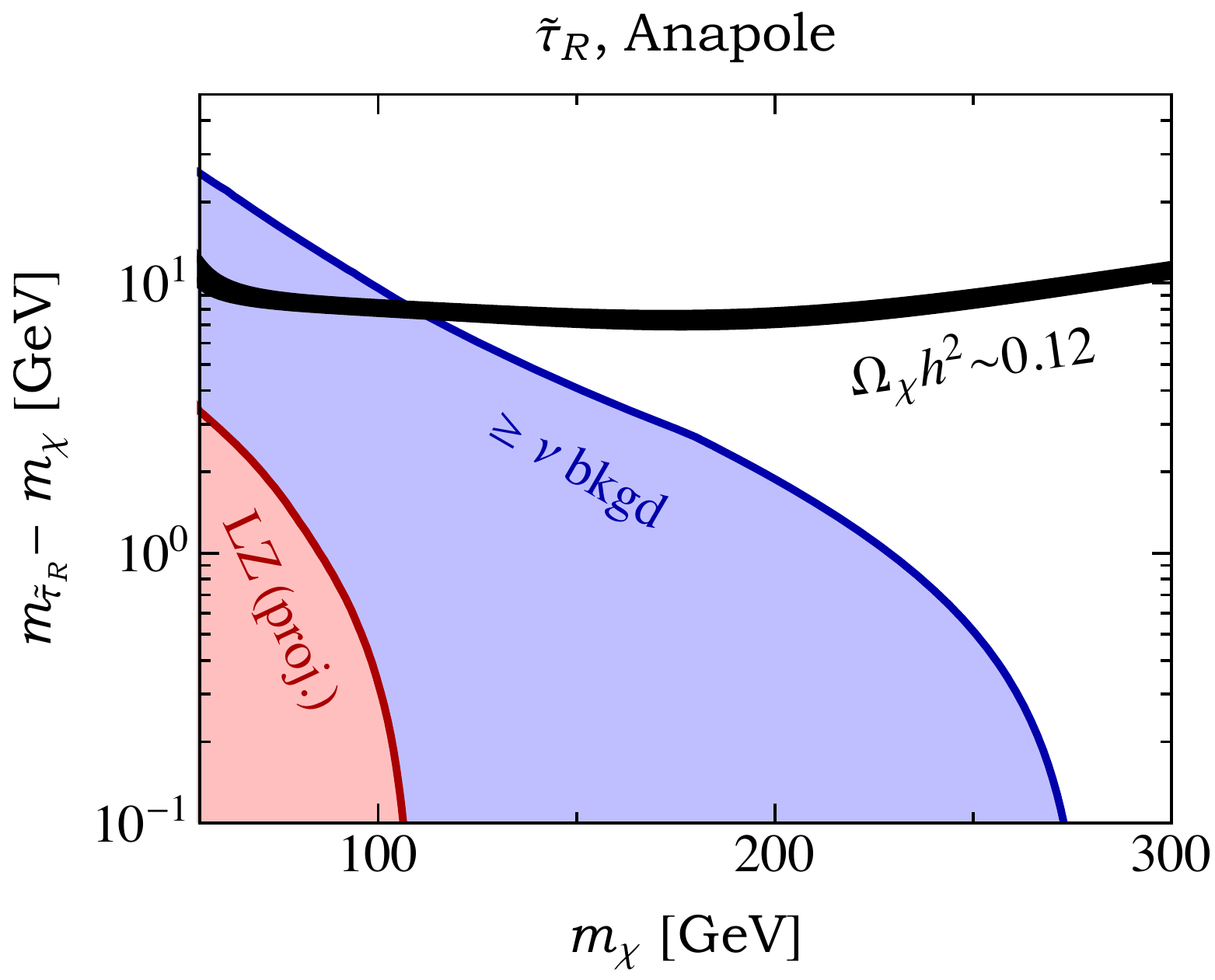} 
\caption{\label{fig:anapole} Results from a fixed-order calculation for the case of SI bino-proton scattering mediated by either a right-handed selectron (left) or stau (right). The filled contours correspond to regions that will be probed by LZ (red) or future direct detection experiments sensitive to rates above the irreducible neutrino background (blue). For reference, we also show regions where the calculated relic abundance matches the observed dark matter density (black).}
\end{center}
\end{figure}

The reach in the SI cross section is shown in Fig.~\ref{fig:anapole} for a single right-handed selectron or stau. As the anapole Wilson coefficient is strongly enhanced when the lepton mass $m_l$ is much smaller than $m_\x$~, selectron mediated scattering benefits from large cross sections compared to those mediated by a right-handed stau. Also in Fig.~\ref{fig:anapole}, we overlay the region of parameter space where the relic abundance of $\x$ matches the observed dark matter density. Interactions relevant for annihilations and co-annihilations to SM particles are built in FeynRules~\cite{Alloul:2013bka} and implemented in micrOMEGAs~\cite{Belanger:2001fz}. Sommerfeld effects are not included as photon exchange in the initial state is expected to only slightly alter the final calculated abundance~\cite{Hisano:2006nn,Hryczuk:2011tq}. While LZ will only be able to probe thermal coannihilating selectrons and binos for $m_\x \lesssim 100$ GeV, future direct detection experiments will be able to probe selectron (stau) mediated scenarios for thermal bino masses $m_\x \lesssim 300~(100)$ GeV. Left-right mixing introduces an additional slepton of opposite hypercharge and therefore tends to diminish the overall scattering rate.

%

\section{\textbf{Conclusions}}
\label{sec:conclusions}

We have presented EFT methods for computing direct detection rates, focusing on bino DM scattering through loops mediated by heavy-flavor squarks or sleptons. In the presence of large hierarchies between mass scales, such as the weak and hadronic scales, large logarithms can substantially contribute to the total scattering cross section. A sequence of effective theories, linked together by matching computations and renormalization group analysis, provides a systematically improvable framework for incorporating such contributions and assessing the impact of perturbative uncertainty. 

Including these effects from running enhances the scattering cross section by a factor of $\sim 3 - 4$ in some cases, and significantly improves the DM mass reach of direct detection experiments. The specific sources of these effects vary for different models. For example, as explained in Sec.~\ref{sec:RHstop}, in our calculation for bino DM interacting with a right-handed stop, leading log corrections increase the rate through the inclusion of $\order (\alpha_s)$ threshold terms for the scalar quark coefficient when evaluated near the hadronic scale. Alternatively, in the fixed-order approach, these contributions are formally higher order and are not included, highlighting one of the advantages of our scheme. On the other hand, in the case of bino DM coupled to a nearly degenerate right-handed sbottom, the mass reach increases from $\sim 4$ to $\sim 7$ TeV for an experiment such as LZ. This is largely due to the fact that RG evolution significantly enhances the bino-sbottom-bottom interaction at low energies. Interestingly, if relic density constraints also require such small mass splittings, this implies that much of the co-annihilation region may be constrained through direct detection experiments. This motivates a careful investigation of the correlation between relic density and direct detection observables, including, e.g., higher order QCD corrections, and a complete treatment of thermally induced masses and Sommerfeld enhancement (see, e.g., Refs.~\cite{Ibarra:2015nca,Berger:2008ti,deSimone:2014pda,Harz:2014tma}). Assuming an experiment sensitive to cross sections close to the neutrino background, for the case of a stop mediator, the mass reach is around 500 GeV, while for the case of a selectron mediator, the thermal mass reach is round 300 GeV.

In the dark matter context, heavy particle effective theories are efficient for parametrizing unknown interactions of heavy (or nonrelativistic) DM particles with the SM degrees of freedom at a given energy scale, and for factorizing amplitudes into contributions from the hard and soft modes of the process, necessary for resumming large logarithms. These methods have been applied for investigating universal behavior in the scattering of heavy WIMPs~\cite{Hill:2011be,Hill:2013hoa}, the impact of large Sudakov logarithms and Sommerfeld enhancement on the annihilation rate of heavy WIMPs~\cite{Baumgart:2014vma,Bauer:2014ula,Ovanesyan:2014fwa,Beneke:2012tg}, and the low-energy interactions of DM with QCD and nucleons~\cite{Fan:2010gt,Fitzpatrick:2012ix,Hill:2014yxa}. 

In this work, we applied heavy particle techniques to three generic scenarios for bino scattering, depending on the mass hierarchy between the bino, the sfermion, and its partner fermion. In the first and second scenarios, where the sfermion is integrated out of the theory below the weak scale, heavy particle theory was employed for writing the basis of low-energy operators in Eq.~(\ref{eq:LWIMPSM}). We have employed a matching procedure that includes the leading order contributions for the lowest dimension operators relevant for Majorana DM-nucleon scattering, including spin-2 couplings to gluons. Compared to the relativistic basis in Eq.~(\ref{eq:lowbasis}), this basis has manifest power counting, and thus redundant or suppressed operators in the $m_\chi \gg m_b$ limit are easily avoided. Nonetheless, the results for the running and matching matrices, ${\boldsymbol R}$ and ${\boldsymbol M}$, are properties of the QCD currents in Eq.~(\ref{eq:ops}), and can be applied regardless of whether the DM is taken to be a relativistic or heavy particle field. In the third scenario, where the sfermion is kept as a degree of freedom below the weak scale, we used heavy particle theory to systematically separate the full theory amplitudes into contributions that are either encoded in the coefficients of contact operators defined at the weak scale, or are matched by the heavy-light current ${\tilde b}_{R,v} ~ {\bar b} \, \Gamma \chi_v$. The running of this current down to low energies is the dominant source of enhancement to the rate for bino scattering mediated by a nearly degenerate squark. 

In Sec.~\ref{sec:cases}, we focused on simple models with only a few parameters such that definite predictions can be made, and radiative corrections become important not only for determining robust scattering rates, but also for correlating different constraints. Our analysis for nearly degenerate sleptons and mixed squarks is new, and, within our simplifying assumptions, we find that mixing generally tends to reduce the scattering rates. Furthermore, since these rates already depend on several free parameters, we have not studied the impact of radiative corrections for this example. We still note that, similar to the models presented in Sec.~\ref{sec:cases}, it would be interesting to consider leading log corrections for the mixed case, since they may have substantial impact on the estimated reach of future experiments like XENON-1T and LZ. 

Aside from providing robust estimates of benchmark cross sections, employing EFT methods also allows for making the connection between parameters of a high scale theory to low energy observables. We focused here on the starting point where the high-energy theory is defined at the weak scale. It is interesting to further consider the impact of RGE from an even higher scale, where the parameters may be constrained by theoretical UV considerations or from other phenomenological inputs such as collider limits and the DM relic density. Moreover, while we have focused here on the effects from QCD corrections, previous studies have shown that electroweak corrections may also have impact~\cite{Crivellin:2014qxa,D'Eramo:2014aba}. A complete picture of the complementarity between DM observables, e.g., the correlation between parameters determined from relic density, collider limits, and direct detection, should incorporate the connection between different scales in the physical processes.

\vspace{0.2in}
\noindent {\em Acknowledgments:} We thank Richard Hill and Jason Kumar for valuable discussions. AB is supported by the Kavli Institute for Cosmological Physics at the University of Chicago through grant NSF PHY-1125897. DR, MS and KZ are supported by the DoE under contract DE-AC02-05CH11231. DR is supported by the S\~ao Paulo Research Foundation (FAPESP).

\newpage
\section{\textbf{Appendix}}
\appendix


\section{\textbf{Model}}
\label{sec:model}

Following the conventions in Ref.~\cite{Baer:2006rs}, we specify the masses of the bino/sfermion sector of the MSSM. The sfermion is assumed to be either a squark or a slepton. The soft hypercharge gaugino mass parameter ($M_1$) is taken to be positive so that no chiral field redefinitions are necessary to ensure the positivity of the physical bino mass. In particular, the bino and its physical mass will be denoted by $\x$ and $m_\x$~. The mass matrices that relate the sfermion mass eigenstates ($\sfe_{1,2}$) to the gauge eigenstates ($\sfe_{L,R}$) are given by
\be
\label{eq:SfermionMass}
\begin{pmatrix} \sfe_1 \\ \sfe_2 \end{pmatrix} 
= \begin{pmatrix} \cos{\theta_f} & -\sin{\theta_f} \\ \sin{\theta_f} & \cos{\theta_f} \end{pmatrix} 
\begin{pmatrix} \sfe_L \\ \sfe_R \end{pmatrix}
~.
\ee

In the case that the sfermions are stops ($\tilde{t}$) or sbottoms ($\tilde{b}$), the mixing angles are given explicitly by the tree-level expressions
\begin{align}
\tan{\theta_t} &= \frac{m_{\tilde{Q}_3}^2 + m_t^2 + m_Z^2 \cos{2 \beta} \left( \frac{1}{2} - \frac{2}{3} \sin^2{\theta_w} \right) - m_{\tilde{t}_1}^2}{m_t \left( -A_t + \mu \cot{\beta} \right)} \,,
\nl
\tan{\theta_b} &= \frac{m_{\tilde{Q}_3}^2 + m_b^2 + m_Z^2 \cos{2 \beta} \left( -\frac{1}{2} + \frac{1}{3} \sin^2{\theta_w} \right) - m_{\tilde{b}_1}^2}{m_b \left( -A_b + \mu \tan{\beta} \right)}
~.
\end{align}
Above, $m_{\tilde{Q}_3}$ is the left-handed squark soft mass parameter, $A_{t,b}$ are the soft trilinear couplings to the Higgs, $\tan{\beta}$ is the ratio of the up and down type Higgs vacuum expectation values, $\mu$ is the supersymmetric Higgs mass parameter, $\theta_w$ is the Weinberg angle, and $m_{\sfe_{1,2}}$ are the physical masses of the lightest, heaviest sfermion, respectively. These physical tree-level masses, obtained by diagonalizing the sfermion mass matrix, are 
\begin{align}
\label{eq:squarkmass}
m_{\tilde{t}_{1,2}}^2 &= \frac{1}{2} \left( m_{\tilde{Q}_3}^2 + m_{\tilde{t}_R}^2 \right) + \frac{1}{4} m_Z^2 \cos{2 \beta} + m_t^2 
\nl
&\mp \Bigg\{ \bigg[ \frac{1}{2} \left( m_{\tilde{Q}_3}^2 - m_{\tilde{t}_R}^2 \right) + m_Z^2 \cos{2 \beta} \left( \frac{1}{4} - \frac{2}{3} \sin^2{\theta_w} \right) \bigg]^2 + m_t^2 \left( \mu \cot{\beta} - A_t \right)^2 \Bigg\}^{\frac{1}{2}}
\nl
m_{\tilde{b}_{1,2}}^2 &= \frac{1}{2} \left( m_{\tilde{Q}_3}^2 + m_{\tilde{b}_R}^2 \right) - \frac{1}{4} m_Z^2 \cos{2 \beta} + m_b^2 
\nl
&\mp \Bigg\{ \bigg[ \frac{1}{2} \left( m_{\tilde{Q}_3}^2 - m_{\tilde{b}_R}^2 \right) - m_Z^2 \cos{2 \beta} \left( \frac{1}{4} - \frac{1}{3} \sin^2{\theta_w} \right) \bigg]^2 + m_b^2 \left( \mu \tan{\beta} - A_b \right)^2 \Bigg\}^{\frac{1}{2}}
~,
\end{align}
where $m_{\sfe_R}$ are the right-handed sfermion soft masses. Note that we have chosen the sign convention for $\mu$ where the Higgsino contributions to the neutralino and chargino mass matrices are given by $+ \mu$ and $- \mu$, respectively. Radiative corrections at one-loop can significantly alter the forms of the tree-level expressions above~\cite{Pierce:1996zz,Carena:1998gk,Carena:2002es}. For example, the correction to the bottom Yukawa can be parametrized in terms of a quantity $\Delta_b$ as
\be
y_b \to \frac{m_b}{\sqrt{2}~v \cos{\beta}\left( 1+ \Delta_b \right)} \,, 
\ee
with the effect that in the sbottom mass matrix, $A_b$ and $\tan{\beta}$ are replaced by the effective parameters 
\be
A_{b, \text{eff}} = \frac{A_b}{1+\Delta_b} \quad , \quad \tan{\beta}_\text{eff} = \frac{\tan{\beta}}{1+\Delta_b}
~,
\ee
as in Ref.~\cite{Batell:2013psa}. Here we take the SM Higgs vacuum expectation value to be $v=174$ GeV. The trilinear coupling $A_t$ may be defined similarly in the stop sector, for which the masses and Higgs interactions are independent of $\tan{\beta}$ in the large $\tan{\beta}$ limit. From here on out, we will drop the ``eff" subscript with the understanding that the squark masses and interactions are defined in terms of these ``effective" inputs at the weak scale.

When dealing with sleptons, we will choose to ignore intra-generational mixing since first and second generation lepton masses are very small compared to the soft masses. For the example of a single right-handed selectron ($\tilde{e}_R$), its tree-level mass, $m_{\tilde{e}_R}^2 \approx m_{\tilde{e}_1}^2 - m_Z^2 \cos{2 \beta} \sin^2{\theta_w}$~, receives negligible corrections at one-loop, and is essentially a free parameter controlled by the first generation right-handed slepton soft mass $m_{\tilde{e}_1}$~.

The interactions of a pair of sfermions $\sfe_{1,2}$ with a bino LSP ($\x$) and SM fermion ($f$) are parametrized in terms of the SM hypercharge coupling $g^\prime$ and the sfermion mixing angles of Eq.~(\ref{eq:SfermionMass}). We adopt the following notation for these interactions,
\be
\mathcal{L} \supset \sum\limits_{i=1,2}  \tilde{f}_i ~ \bar{f} \left( \alpha_f^{(i)} + \beta_f^{(i)} \gamma^5 \right) \x + \text{ h.c.}
~,
\ee
where the effective couplings for the stop/sbottom sector are given by
\begin{align}
\label{eq:couplings}
\alpha_t^{(1)} &\equiv \frac{-g^\prime}{3 \sqrt{2}} \left( \frac{1}{2} \cos{\theta_t} + 2 \sin{\theta_t} \right) \quad,\quad \beta_t^{(1)} \equiv \frac{-g^\prime}{3 \sqrt{2}} \left( \frac{1}{2} \cos{\theta_t} - 2 \sin{\theta_t}\right)
\nl
\alpha_t^{(2)} &\equiv \frac{-g^\prime}{3 \sqrt{2}} \left( \frac{1}{2} \sin{\theta_t} - 2 \cos{\theta_t}\right) \quad,\quad \beta_t^{(2)} \equiv \frac{-g^\prime}{3 \sqrt{2}} \left( \frac{1}{2} \sin{\theta_t} + 2 \cos{\theta_t} \right)
\nl
\nl
\alpha_b^{(1)} &\equiv \frac{-g^\prime}{3 \sqrt{2}} \left( \frac{1}{2} \cos{\theta_b} - \sin{\theta_b} \right) \quad,\quad \beta_b^{(1)} \equiv \frac{-g^\prime}{3 \sqrt{2}} \left( \frac{1}{2} \cos{\theta_b} + \sin{\theta_b} \right)
\nl
\alpha_b^{(2)} &\equiv \frac{-g^\prime}{3 \sqrt{2}} \left(  \frac{1}{2} \sin{\theta_b} + \cos{\theta_b} \right) \quad,\quad \beta_b^{(2)} \equiv \frac{-g^\prime}{3 \sqrt{2}} \left( \frac{1}{2} \sin{\theta_b} - \cos{\theta_b} \right)
~.
\end{align}
The effective couplings for a single right-handed slepton are similarly defined, with $\alpha_\ell = -\beta_\ell = -g^\prime/\sqrt{2}$ . 

Although a pure bino possesses no tree-level interactions with the electroweak bosons of the SM, the sfermions and their associated SM fermion partners interact with the $Z$, photon, and SM Higgs ($h$) through terms that we parametrize as
\begin{align}
\mathcal{L}_Z &\supset Z^\mu ~ i \Big[ \left( g_f^v - g_f^a \cos{2\theta_f} \right) \left( \tilde{f}_1^\dagger \partial_\mu \tilde{f}_1 - \tilde{f}_1 \partial_\mu \tilde{f}_1^\dagger \right) +  \left( g_f^v + g_f^a \cos{2\theta_f} \right) \left( \tilde{f}_2^\dagger \partial_\mu \tilde{f}_2 - \tilde{f}_2 \partial_\mu \tilde{f}_2^\dagger \right)
\nl
&- g_f^a \sin{2 \theta_f} \left( \tilde{f}_1^\dagger \partial_\mu \tilde{f}_2 - \tilde{f}_1 \partial_\mu \tilde{f}_2^\dagger + \tilde{f}_2^\dagger \partial_\mu \tilde{f}_1 -  \tilde{f}_2 \partial_\mu \tilde{f}_1^\dagger \right) \Big] + Z_\mu ~ \bar{f} \gamma^\mu \left( g_f^v + g_f^a \gamma^5 \right) f
\nl
\nl
\mathcal{L}_\gamma &\supset - \sum_{\substack{i=1,2}} i e Q_f ~ A^\mu ~ \left( \tilde{f}_i^\dagger \partial_\mu \tilde{f}_i - \tilde{f}_i \partial_\mu \tilde{f}_i^\dagger \right) - e Q_f ~ A_\mu ~ \bar{f} \gamma^\mu f 
\nl
\nl
\mathcal{L}_h &\supset \sum_{\substack{i=1,2}} \Big( \mu_f^{(i)} h ~ \tilde{f}_i^\dagger \tilde{f}_i \Big) + \mu_f^{(12)} h ~ \left( \tilde{f}_1^\dagger \tilde{f}_2 + \tilde{f}_1 \tilde{f}_2^\dagger \right) - \frac{m_f}{\sqrt{2}~v} ~ h \bar{f} f
~,
\end{align}
where the effective parameters above for (s)tops, (s)bottoms, and (s)leptons are
\begin{align}
\label{eq:explcoup}
g_t^v &= \frac{-5e}{12} \tan{\theta_w} + \frac{e}{4} \cot{\theta_w} \quad,\quad g_t^a = \frac{-e}{4} \left( \tan{\theta_w} + \cot{\theta_w} \right)
\nl
g_b^v &= \frac{e}{12} \tan{\theta_w} - \frac{e}{4} \cot{\theta_w} \quad,\quad g_b^a = \frac{e}{4} \left( \tan{\theta_w} + \cot{\theta_w} \right)
\nl
g_\ell^v &= \frac{e}{4} \left( 3 \tan{\theta_w} - \cot{\theta_w} \right) \quad,\quad g_\ell^a = \frac{e}{4} \left( \tan{\theta_w} + \cot{\theta_w} \right)
\nl
Q_t &= 2/3 \quad,\quad Q_b = \frac{-1}{3} \quad,\quad Q_\ell = -1
\end{align}

\begin{align}
\mu_t^{(1)} &= \frac{-\sqrt{2} m_t}{v} \left[m_t + \frac{1}{2} \sin{2 \theta_t} ( A_t - \mu  \cot{\beta} ) \right] - \frac{g^2 v \cos{2 \beta}}{6 \sqrt{2}}\Big[ 4 \sin^2{\theta_t} \tan^2{\theta_w} \nl &\qquad + \cos^2{\theta_t} \left(3-\tan^2{\theta_w}\right)\Big] \,, 
\nl
\mu_t^{(2)} &= \frac{-\sqrt{2} m_t}{v} \left[m_t - \frac{1}{2} \sin{2 \theta_t} ( A_t - \mu  \cot{\beta} ) \right] - \frac{g^2 v \cos{2 \beta}}{6 \sqrt{2}}\Big[ 4 \cos^2{\theta_t} \tan^2{\theta_w} \nl &\qquad + \sin^2{\theta_t} \left(3-\tan^2{\theta_w}\right)\Big]\, ,
\nl
\mu_t^{(12)} &= \frac{m_t}{\sqrt{2} v} \cos{2 \theta_t} (A_t - \mu \cot{\beta}) + \frac{g^2 (1-4 \cos{2 \theta_w}) \sec^2{\theta_w} ~v \cos{2 \beta}}{12 \sqrt{2}} \sin{2 \theta_t} \,, 
\nl
\mu_b^{(1)} &= \frac{-\sqrt{2} m_b}{v} \Big[ m_b + \frac{1}{2} \sin {2 \theta_b} (A_b-\mu  \tan{\beta})\Big] +  \frac{g^2 v \cos{2 \beta} \sec^2{\theta_w}}{12 \sqrt{2}} \Big[3 + \cos{2 \theta_b} (1+2 \cos{2 \theta_w})\Big] \,, 
\nl
\mu_b^{(2)} &= \frac{-\sqrt{2} m_b}{v} \Big[m_b - \frac{1}{2} \sin {2 \theta_b} (A_b - \mu  \tan{\beta})\Big] + \frac{g^2 v \cos{2 \beta} \sec^2{\theta_w}}{12 \sqrt{2}} \Big[3-\cos {2 \theta_b} (1 + 2 \cos{2 \theta_w})\Big] \,,
\nl
\mu_b^{(12)} &= \frac{m_b}{\sqrt{2} v}\cos {2 \theta_b} (A_b-\mu  \tan{\beta})+\frac{g^2 v \cos{2 \beta} (1+2 \cos{2 \theta_w}) \sec ^2{\theta_w}}{12 \sqrt{2}} \sin {2 \theta_b}
~,
\end{align}
such that $g$ is the $SU(2)_w$ coupling, $e$ is the electromagnetic coupling, $v = 174$ GeV, and we have worked in the alignment limit where the Higgs is SM-like.

\section{\textbf{Hadronic Inputs}}
\label{app:hadinputs}

In this section, we present the numerical values for the hadronic form factors defined in Eq.~(\ref{eq:matrixelements}). More detailed discussion on the determination of these quantities can be found in Sec.~4 of Ref.~\cite{Hill:2014yxa}.  

The up and down quark scalar form factors are determined from the nucleon sigma terms,
\begin{align}\label{eq:sigmas}
\Sigma_{\pi N} &= {m_u+m_d\over 2} \langle N | (\bar{u} u + \bar{d} d) | N \rangle = 
44 (13) \,{\rm MeV}  \,,
\nl
\Sigma_{-} &= (m_d - m_u) \langle N | (\bar{u} u - \bar{d} d) | N \rangle =  \pm 2(2) \,{\rm MeV} \,,
\end{align}
where the upper (lower) sign in $\Sigma_-$ is for the proton (neutron) (see also Ref.~\cite{Alarcon:2011zs}). For the strange quark, we use $m_N f^{(0)}_{s,N} = 40 \pm 20 \, {\rm MeV}$. The up and down quark scalar form factors are then
\begin{align}\label{eq:fufd}
f^{(0)}_{u,N} = {R_{ud} \over 1 + R_{ud}} \, {\Sigma_{\pi N} \over m_N} (1 + \xi) ~,~  f^{(0)}_{d,N} = {1 \over 1 + R_{ud}} \, {\Sigma_{\pi N} \over m_N} (1 - \xi) ~,~  \xi = {1 + R_{ud} \over 1 - R_{ud}}  ~ { \Sigma_-  \over 2 \Sigma_{\pi N} } ~,
\end{align}
where the ratios of quark masses are
\begin{align}\label{eq:massratios}
R_{ud} \equiv {m_u\over m_d} = 0.49 \pm 0.13 ~,~ R_{sd} \equiv {m_s \over m_d} = 19.5 \pm 2.5 ~ .
\end{align}
The gluon scalar form factor is determined from the next-to-leading order terms of Eq.~(\ref{eq:gluemassfrac2}). For our leading log analysis, we take as default
\be
\label{eq:glueNLO}
f_{g,N}^{(0)} (\mu_0)= 1 - \left( 1 + \frac{2 \alpha_s(\mu_0)}{\pi} \right) \sum\limits_{q=u,d,s} f_{q,N}^{(0)}  \, .
\ee
Note that, as long as $\alpha_s$ terms are consistently kept in the functions ${\tilde \beta}(\mu_0)$ and ${\gamma_m}(\mu_0)$ appearing in ${\boldsymbol f}(\mu_0)$ and ${\boldsymbol R}(\mu_0,\mu_c)$, the dependence on the low scale $\mu_0$ cancels in the product ${\boldsymbol f}^T(\mu_0) {\boldsymbol R}(\mu_0,\mu_c)$. We may thus simplify the analysis by taking $\mu_0 = \mu_c \sim m_c$~. 
\begin{table}[t]
\begin{center}
\begin{tabular}{|c||c|c|c|c|}
\hline
 $\mu \ (\rm GeV) $ & $f_{u,p}^{(2)}(\mu)$ & $f_{d,p}^{(2)}(\mu)$ & $f_{s,p}^{(2)}(\mu)$ & $f_{g,p}^{(2)}(\mu)$ \\
\hline 
\hline
1 & 0.404(9) & 0.217(8) & 0.024(4)  & 0.356(29) \\
\hline
1.4 & 0.370(8) & 0.202(7) & 0.030(4)  & 0.398(23) \\
\hline
2 & 0.346(7) & 0.192(6) & 0.034(3)  & 0.419(19) \\
\hline
\end{tabular} 
\end{center}
\caption{\label{tab:momfraction}
Proton form factors for spin-2 operators at different values of $\mu$. The neutron form factors follow from approximate isospin symmetry ($u \leftrightarrow d$).
} 
\end{table}

Spin-2 form factors are derived from the second moments of parton distribution functions,  
\be
f_{q,N}^{(2)}(\mu) = \int_0^1 dx \, x ~[ q(x,\mu) + {\bar{q}}(x,\mu) ] ~,~  f_{g,N}^{(2)}(\mu) = \int_0^1 dx \, x  ~g(x,\mu) 
~,
\ee
where $q(x,\mu)$, $\bar{q}(x,\mu)$, $g(x,\mu)$ are the quark, anti-quark, and gluon parton distribution functions evaluated at the scale $\mu$, respectively. Table~\ref{tab:momfraction} lists values for renormalization scales $\mu=1\,, 1.4\,, 2 \ {\rm GeV}$. Finally, for the spin-1 axial-vector form factors of the proton, we take
\be
f_{u,p}^{(1)} = 0.75(8) ~,~ f_{d,p}^{(1)} = -0.51(8) ~,~ f_{s,p}^{(1)} = -0.15(8)
~,
\ee
where neutron matrix elements follow from isospin symmetry ($u \leftrightarrow d$).

\section{\textbf{Running and Matching Matrices}}
\label{app:R&M}

In this section, we present the analytic forms for the leading order RGE and threshold matching matrices (${\boldsymbol R}$ and ${\boldsymbol M}$) appearing in Eq.~(\ref{eq:factorize}). Since the scalar and spin-2 operators do not mix with each other under RGE, in the basis of Eq.~(\ref{eq:weakscaleC}), these matrices have the block diagonal forms
\begin{align}
\boldsymbol R = \left\{ {\boldsymbol R}^{(0)}  \,, {\boldsymbol R}^{(2)}  \right\} \,, \quad \boldsymbol M = \left\{ {\boldsymbol M}^{(0)}  \,, {\boldsymbol M}^{(2)}  \right\} \,, 
\end{align}
where ${\boldsymbol R}^{(S)}$ and ${\boldsymbol M}^{(S)}$, for $S=0\,, 2$, are the running and matching matrices for the scalar ($S=0$) and spin-2 ($S=2$) operators. Detailed discussion on the derivation of these quantities can be found in Sec.~3 of Ref.~\cite{Hill:2014yxa}.  
\begin{table}[t]
\begin{center}
\small
\begin{tabular}[t]{|c||c|}
\hline
Operator & Running Matrix \\
\hline
\hline
$O_q^{(0)}\, , O_g^{(0)}$  & $R^{(0)}_{qq} =1 \, , \quad R^{(0)}_{qq^\prime} =0 \, , \quad  R^{(0)}_{qg} = 16 ~ \frac{\alpha_s(\mu_l)/\alpha_s(\mu_h)-1}{\frac{2}{3} n_f - 11} \, ,$\\
&$ R^{(0)}_{gq} = 0  
\, , \quad R^{(0)}_{gg} =  \frac{\alpha_s(\mu_l)}{\alpha_s(\mu_h)} $\\
\hline
$O_{q}^{(2)}\, , O_{g}^{(2)}$  & $ R^{(2)}_{qq} - R^{(2)}_{qq^\prime} = r(0) \,, \quad R^{(2)}_{qq^\prime} =  {1 \over n_f} \Big[ {16 r(n_f) +3 n_f \over 16 + 3n_f} - r(0) \Big] \, ,$ \\
&$ R^{(2)}_{qg} = {16[1-r(n_f)] \over 16 + 3n_f}  \, ,$ \\
&$ R^{(2)}_{gq}  = {3[1-r(n_f)] \over 16 + 3n_f} \, , \quad R^{(2)}_{gg} =  {16+3n_f r(n_f) \over 16 + 3n_f}$\\
\hline
\end{tabular}
\end{center}
\caption{\label{tab:QCDrunning}
Running matrices at leading order in $\alpha_s$ for scalar and spin-2 quark and gluon operators in $n_f$-flavor QCD. Spin-2 operators are given in terms of the function $r(t)$ (see Eq.~(\ref{eq:roft})). 
}
\end{table}

The running and heavy quark threshold matching for the spin-1 axial-vector operators are trivial at leading order in $\alpha_s$~, and hence we take ${\boldsymbol R}= {\boldsymbol M} =\mathbb{1}$ in evolving the coefficients $c_q^{(1)}$. For scalar and spin-2 operators, the running matrix ${\boldsymbol R}^{(S)}(\mu_l, \mu_h)$ from a high scale ($\mu_h$) to a low scale ($\mu_l$) in the basis $(u,d,s,\dots | g)$ with $n_f$ flavors of quarks has the form
\renewcommand{\arraystretch}{1.4}
\begin{align}\label{eq:Rsol1}
{\boldsymbol R}^{(S)}(\mu_l, \mu_h)
&=  \left( \begin{array}{ccc|c} 
 & & & R_{qg}^{(S)} \\
&  \mathbb{1} (R_{qq}^{(S)} - R_{qq^\prime}^{(S)} ) + \mathbb{J}  R_{qq^\prime}^{(S)} & & \vdots \\
&&  & R_{qg}^{(S)} \\
\hline
R_{gq}^{(S)} & \cdots & R_{gq}^{(S)} & R_{gg}^{(S)}
\end{array} 
\right) \,,
\end{align}
where $\mathbb{1}$ and $\mathbb{J}$ are $n_f \times n_f$ matrices corresponding to the identity matrix and the matrix with all elements equal to unity, respectively. The elements $R_{ij}^{(S)}$  are specified in Table~\ref{tab:QCDrunning}. The elements for the spin-2 operator involve the function
\be
\label{eq:roft}
r(t) \equiv \left(\alpha_s(\mu_l) \over \alpha_s(\mu_h) \right)^{ \frac{32/9  + 2t/3}{2 n_f / 3 - 11}}\,.
\ee
\begin{table}[t]
\begin{center}
\small
\begin{tabular}[t]{|c||c|}
\hline
Operator & Matching Matrix \\
\hline
\hline
$O_q^{(0)}\, , O_g^{(0)}$  & $M^{(0)}_{gQ} = {-\alpha_s^{\prime}(\mu_Q)\over 12\pi}  \, , \quad M^{(0)}_{gg} =1  $ \\
\hline
$O_{q}^{(2)}\, , O_{g}^{(2)}$  & $M^{(2)}_{gQ} =  {\alpha_s^{\prime} \over 3\pi}  \log {\mu_Q \over m_Q}   \, , \quad   M^{(2)}_{gg} = 1  $ \\
\hline
\end{tabular}
\end{center}
\caption{\label{tab:HQmatching}
Heavy quark threshold matching matrices at leading order in $\alpha_s$ for scalar and spin-2 operators.  The strong coupling in the $(n_f+1)$-flavor theory is denoted $\alpha_s^{\prime}$~. $m_Q$ and $\mu_Q$ correspond to the mass of the heavy quark and the scale at which it is integrated out, respectively.}
\end{table}

For the scalar and spin-2 operators, the heavy quark ($Q$) threshold matching between $n_f+1$ and $n_f$-flavor QCD involves the $(n_f+1) \times (n_f+2)$ matrix ${\boldsymbol M}^{(S)}$, which is given in the basis $(u,d,s, \dots |Q|g)$ by
\renewcommand{\arraystretch}{1.2}
\begin{align}\label{eq:Mlead}
{\boldsymbol M}^{(S)} = \left( \begin{array}{ccc|c|c} 
 1 &  & & 0 & 0 \\
&  \ddots  & &\vdots & \vdots \\
&  &  1 & 0 & 0 \\
\hline
0 & \cdots & 0 & M_{gQ}^{(S)} & M_{gg}^{(S)} 
\end{array} 
\right) \, ,
\end{align}
with the elements $M_{ij}^{(S)}$ given in Table~\ref{tab:HQmatching}.

\section{\textbf{Collection of Wilson Coefficients}}
\label{sec:app1}

In this appendix, we present the Wilson coefficients (in the notation of Eqs.~(\ref{eq:lowbasis}) and (\ref{eq:LWIMPSM})) that are obtained from integrating out charged scalars as well as electroweak vector and scalar bosons. We will begin with diagrams that allow for $\chi$ to scatter with quarks at tree-level, and will then proceed to various loop-level processes in order of increasing complexity. Throughout, we will parametrize the Lagrangian governing the ultraviolet couplings (denoted as $\mathcal{L}_\text{UV}$) in a generic manner, although a simple mapping to the MSSM can be performed by comparing to the particular couplings of Appendix~\ref{sec:model}. In this sense, the results presented in this appendix can easily be applied to other models involving Majorana DM and charged scalars. In the case that two different sfermions can be present in the same loop, Latin subscripts are used to denote the fields. For example, we will often denote a single sfermion of a generation as $\sfe_i$ where $i=1,2$. Alternatively, if the sfermion must be colored (as in the one-loop couplings to gluons), we will write $\sq_i$ to denote squarks. 

For most of this appendix, the Wilson coefficients are presented as integrals over Feynman and Schwinger parameters, since they can be written in compact forms that are easy to evaluate numerically. Of course, given certain assumptions for the couplings and mass spectrum, these integrals can be evaluated to obtain analytic forms. For each diagram, in addition to the general model-independent result, we will provide limiting forms for three distinct cases, namely a right-handed stop degenerate with a bino (Secs.~\ref{sec:case1} and \ref{sec:RHstop}), a right-handed sbottom much heavier than the bino (Secs.~\ref{sec:case2} and \ref{sec:HeavyRHSbottom}), and a right-handed sbottom nearly degenerate with the bino (Secs.~\ref{sec:lightfsmalldelta} and \ref{sec:LightRHSbottom}).

\begin{figure}[t]
\begin{center}
\includegraphics[width=0.25\textwidth]{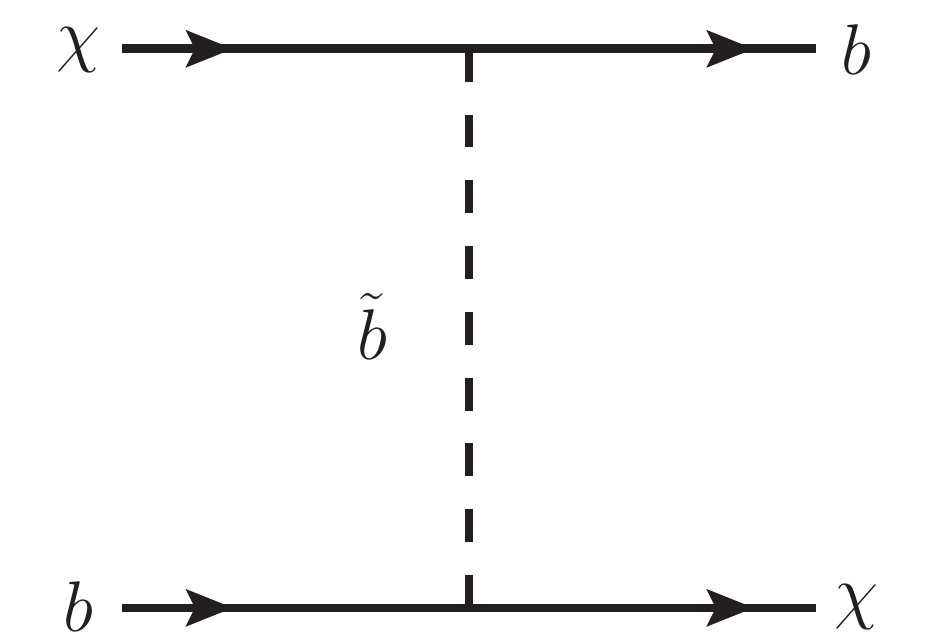} 
\caption{\label{fig:treelevel}
Feynman diagram responsible for tree-level scattering. Crossed diagram not shown.}
\end{center}
\end{figure}
%
\subsection{Tree-Level}
\label{app:tree}

The simplest process that allows $\chi$ to scatter with nuclei is tree-level exchange of a squark. Here, we have in mind the exchange of a sbottom (as shown in Fig.~\ref{fig:treelevel}) for matching onto 5-flavor QCD, although these results may be applied to first-generation squarks as well. Parametrizing the UV couplings as,
\be
\mathcal{L}_\text{UV} \supset \tilde{b} ~ \bar{b} \left( \alpha + \beta \gamma^5 \right) \x + \text{h.c.}
\quad ,
\ee
and applying the appropriate Fierz transformations, we obtain the Wilson coefficients in the limit that $m_{\tilde{b}} \gg m_\x$~,
\begin{align}
c_b^{(0)\text{bare}} &=  \frac{- \left(\alpha^2-\beta^2 \right)}{4 m_b ( m_{\tilde{b}}^2 - m_\x^2 )} + \frac{m_\x \left( \alpha^2 + \beta^2 \right)}{8 ( m_{\tilde{b}}^2 - m_\x^2 )^2}
\quad , \quad
c_b^{(2)\text{bare}} = \frac{m_\x \left( \alpha^2 + \beta^2 \right)}{2( m_{\tilde{b}}^2 - m_\x^2 )^2}
\quad .
\end{align}
The above expressions agree with those presented in Ref.~\cite{Hisano:2010ct}. For a right-handed sbottom, the coefficients reduce to 
\begin{align}
\label{eq:HeavySbottomTree}
\nl
&c_b^{(0) \text{bare}} = \frac{(g^\prime)^2 m_\x}{72 (m_{\tilde{b}_R}^2 - m_\x^2)^2} \quad , \quad c_b^{(2) \text{bare}} = \frac{(g^\prime)^2 m_\x}{18 (m_{\tilde{b}_R}^2 - m_\x^2)^2} 
~.
\end{align}
\begin{figure}[t]
\begin{center}
\includegraphics[width=0.45\textwidth]{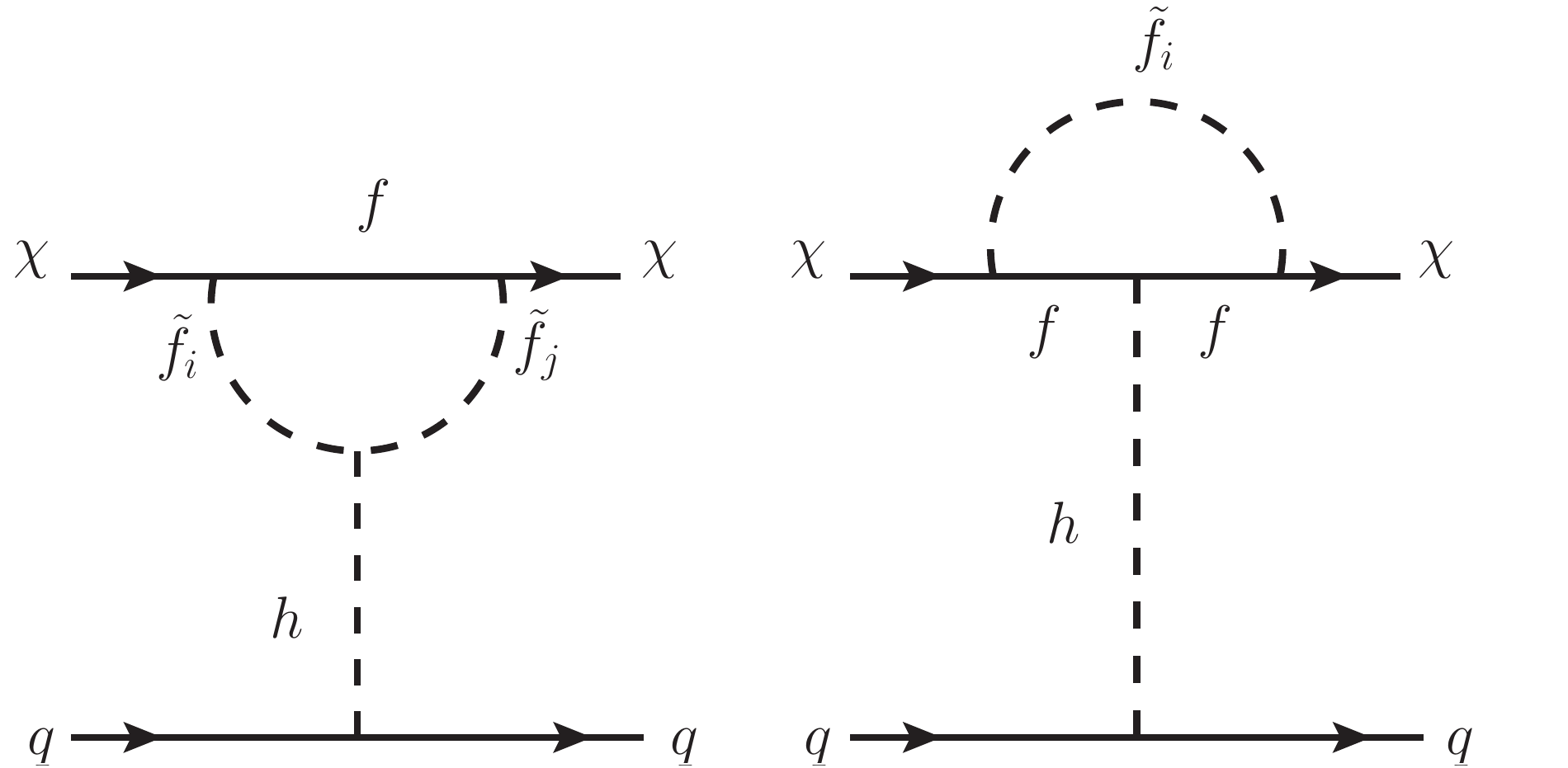} 
\caption{\label{fig:higgsexchange}
Feynman diagrams responsible for couplings to quarks through Higgs exchange. Charge-reversed diagrams not shown.}
\end{center}
\end{figure}
%

\subsection{Higgs Exchange}
\label{app:Higgs}

We now proceed to one-loop couplings to quarks. The simplest example of this process is through the t-channel exchange of a Higgs that is radiated off of an intermediate fermion or sfermion as shown in Fig.~\ref{fig:higgsexchange}. For general sfermion mixing, the Higgs possesses both diagonal and off-diagonal couplings to the sfermions $\sfe_{1,2}$ as discussed in Appendix~\ref{sec:model}. We will parametrize the interactions responsible for this process as
\be
\mathcal{L}_\text{UV} \supset \sum\limits_{i} \Big[ \tilde{f}_i ~ \bar{f} \left( \alpha_f^{(i)} + \beta_f^{(i)} \gamma^5 \right) \x + \text{h.c.} \Big] +  \sum\limits_{i \leq j}\frac{\mu_h^{(ij)}}{1+\delta_{i j}} h \left( \tilde{f}_i \tilde{f}_j^\dagger + \text{h.c.} \right) + \lambda_f^h ~h~ \bar{f} f
\quad ,
\ee
where the sum runs over a complete gauge multiplet of sfermions (e.g., for a single right-handed stop, i=1, while for mixed left and right-handed stops and sbottoms $i=1, 2, 3, 4$). The bare scalar Wilson coefficient is given by
\begin{align}
c_q^{(0)\text{bare}} &= \frac{(\lambda_q^h/m_q) n_c}{8 \pi^2 m_h^2} ~\Big(~ \sum\limits_{i \leq j} M_{ij} + \sum\limits_i M_i ~\Big)
\quad ,
\end{align}
where $n_c$ is the number of colors of $\tilde{f}_i$~. The contributions $M_{ij}$ and $M_i$ correspond to the left and right diagrams of Fig.~\ref{fig:higgsexchange}, respectively, and are given by 
\begin{align}
M_{ij} &\equiv \frac{1}{1+ \delta_{ij}} ~ \frac{\mu_h^{(ij)}}{m_{\tilde{f}_i}^2 - m_{\tilde{f}_j}^2} \int_0^1 dx ~ \bigg[ \alpha_f^{(i)} \alpha_f^{(j)} \big( (1-x )~ m_\x + m_f \big) \nl &\qquad + \beta_f^{(i)} \beta_f^{(j)} \big( (1-x )~ m_\x - m_f \big) \bigg] \log{\frac{\Delta_i}{\Delta_j}} \,, 
\nl
M_i &\equiv - \lambda_f^h \Big( (\alpha_f^{(i)})^2 -  (\beta_f^{(i)})^2 \Big) \int_0^1 dx ~ x ~ \log{\frac{\mu^2}{\Delta_i}} ~+~ \frac{\lambda_f^h}{2} \int_0^1 dx ~ \frac{x}{\Delta_i} \bigg[  (\alpha_f^{(i)})^2 \big( (1-x) m_\x + m_f \big)^2 
\nl & \qquad - (\beta_f^{(i)})^2 \big( (1-x) m_\x - m_f \big)^2 \bigg] \,,
\nl
\Delta_i &\equiv \left(x-1\right)\big(x ~ m_\x^2 - m_{\tilde{f}_i}^2\big) + x ~ m_f^2
\quad .
\end{align}
Above, we have dropped terms, such as UV poles, that vanish when summed over a complete gauge multiplet of sfermions. Accordingly, the dependence on the renormalization scale, $\mu$, should also vanish, but we have kept it to allow for a more compact form.

In the case of a degenerate right-handed stop ($m_{\tilde{t}_R} = m_\x$), the coefficient reduces to 
\begin{align}
\label{eq:DegStopHiggs}
&c_q^{(0)\text{bare}} =\frac{(g^\prime)^2}{36 \pi ^2 m_h^2 m_\x}\Bigg\{ g^2 \cos{2 \beta} \tan^2{\theta_w}+\frac{1}{2} m_t^2 \log{\frac{m_\x^2}{m_t^2}} \left(\frac{3}{v^2} + \frac{1}{m_\x^2}g^2 \cos{2 \beta} \tan^2{\theta_w}\right)
\nl
&+(4 m_\x^2/m_t^2-1)^{-1/2}\tan^{-1}(4 m_\x^2/m_t^2-1)^{1/2} \left[ \frac{3 m_t^2}{v^2} - \left(2-\frac{m_t^2}{m_\x^2}\right) g^2  \cos{2 \beta} \tan^2{\theta_w}\right]\Bigg\}
~.
\end{align}
Similarly, for a non-degenerate right-handed sbottom, the general results simplify to
\begin{align}
\label{eq:SbottomHiggs}
&c_q^{(0) \text{bare}} = \frac{-(g^\prime)^2 g^2  \cos{2 \beta} \tan^2{\theta_w}}{288 \pi ^2 m_h^2 m_\x^3}  \left[m_\x^2 - \left(m_{\tilde{b}_R}^2 - m_\x^2\right) \log{\frac{m_{\tilde{b}_R}^2}{m_{\tilde{b}_R}^2-m_\x^2}}\right] 
~,
\end{align}
while for a degenerate right-handed sbottom we find
\begin{align}
\label{eq:SbottomHiggs2}
&c_q^{(0) \text{bare}} = \frac{-(g^\prime)^2 g^2 \cos{2 \beta} \tan^2{\theta_w}}{288 \pi^2 m_h^2 m_\x}
~.
\end{align}

Let us compare these results to limiting forms presented in Ref.~\cite{Ibarra:2015nca}. For the case of a right-handed stop that is much heavier than the bino, the Wilson coefficient reduces to the approximate form
\be
\label{eq:piercecomp1}
c_q^{(0)\text{bare}} \approx \frac{(g^\prime)^2 m_t^2 m_\x}{12 \pi^2 v^2 m_h^2 m_{\tilde{t}_R}^2}
~,
\ee
while for a right-handed stop that is degenerate with the bino but much heavier than the top quark,
\be
\label{eq:piercecomp2}
c_q^{(0)\text{bare}} \approx \frac{(g^\prime)^2 m_t^2 \log{(m_\x^2 / m_t^2)}}{24 \pi^2 v^2 m_h^2 m_\x}
~.
\ee
Both Eqs.~(\ref{eq:piercecomp1}) and (\ref{eq:piercecomp2}) agree with the limiting forms presented in Ref.~\cite{Ibarra:2015nca}, up to an overall sign. On the other hand, we agree with the full result in Ref.~\cite{Ibarra:2015fqa}, and also check that our expressions are consistent with low energy Higgs theorems~\cite{Kniehl:1995tn}.  As a result, we find that Higgs exchange adds constructively with the gluon diagrams of Sec.~\ref{app:glue} when evaluating the scattering amplitude. The phenomenological impact of this is discussed in Sec.~\ref{sec:RHstop}.

\begin{figure}[t]
\begin{center}
\includegraphics[width=0.45\textwidth]{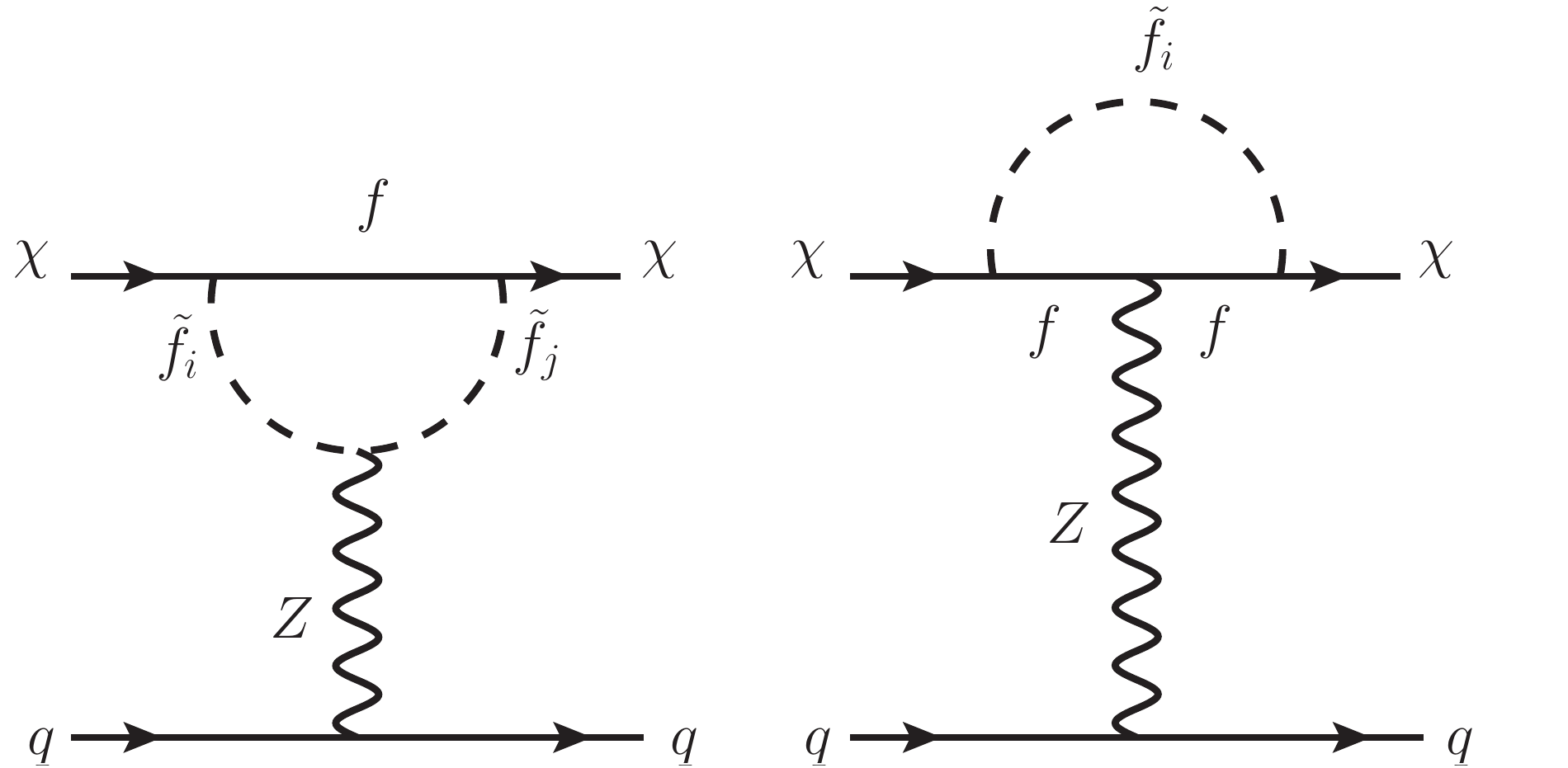} 
\caption{\label{fig:zexchange}
Feynman diagrams responsible for couplings to quarks through $Z$ exchange. Charge-reversed diagrams not shown.}
\end{center}
\end{figure}
%

\subsection{$Z$ Exchange}
\label{app:Z}

Analogous to Higgs exchange, quark scattering can also occur at one-loop order through the t-channel exchange of a $Z$ boson. As shown in Fig.~\ref{fig:zexchange}, both diagonal and off-diagonal Z-sfermion couplings contribute. The UV couplings are parametrized as
\begin{align}
\mathcal{L}_\text{UV} &\supset \sum\limits_{i} \Big[ \tilde{f}_i ~ \bar{f} \left( \alpha_f^{(i)} + \beta_f^{(i)} \gamma^5 \right) \x + \text{h.c.} \Big] + \sum\limits_{i \leq j} \Big[ \frac{i g_{\tilde{f}}^{(ij)}}{1+\delta_{ij}} Z^\mu \left( \tilde{f}_i^\dagger \partial_\mu \tilde{f}_j + \tilde{f}_j^\dagger \partial_\mu \tilde{f}_i \right) + \text{h.c.} \Big] 
\nl
&+ Z_\mu ~ \bar{f} \gamma^\mu (g_f^v + g_f^a \gamma^5) f
\quad ,
\end{align}
where the sum over $i$ runs over a complete gauge multiplet of sfermions. Integrating out the $Z$ in Fig.~\ref{fig:zexchange} generates the axial-vector Wilson coefficient
\be
c_q^{(1)\text{bare}} = \frac{g_q^a n_c}{16 \pi^2 m_Z^2} ~\Big(~ \sum\limits_{i \leq j} M_{ij} - \sum\limits_i M_i ~\Big)
\quad,
\ee
where $n_c$ is the number of colors of $\tilde{f}_i$~. The contributions $M_{ij}$ and $M_i$ correspond to the left and right diagrams of Fig.~\ref{fig:zexchange}, respectively, and are given by
\begin{align}
M_{ij} &\equiv \frac{g_{\tilde{f}}^{(ij)} \left(  \alpha_f^{(i)} \beta_f^{(j)} + \beta_f^{(i)} \alpha_f^{(j)} \right)}{1+\delta_{ij}} \left[ 1+ \frac{2}{m_{\tilde{f}_i}^2-m_{\tilde{f}_j}^2} \int_0^1 dx ~ \left( \Delta_i \log{\frac{\mu^2}{\Delta_i}} - \Delta_j \log{\frac{\mu^2}{\Delta_j}} \right) \right] \,, 
\nl
M_i &\equiv  \left[ g_f^a \left( (\alpha_f^{(i)})^2 + (\beta_f^{(i)})^2 \right) - 2 g_f^v \alpha_f^{(i)} \beta_f^{(i)} \right] \left[~ \frac{1}{2} - \int_0^1 dx ~ x  \log{\frac{\mu^2}{\Delta_i}} ~\right]
\nl
&+ \int_0^1 dx ~ \frac{x}{\Delta_i} \Bigg\{ g_f^a \left[ (\alpha_f^{(i)})^2 \Big(  (1-x) ~m_\x + m_f \Big)^2 + (\beta_f^{(i)})^2 \Big(  (1-x) ~m_\x - m_f \Big)^2 \right] 
\nl
&+ 2 g_f^v \alpha_f^{(i)} \beta_f^{(i)} \left( m_f^2 - (1-x)^2 ~ m_\x^2 \right) \Bigg\} \,, 
\nl
\Delta_i &\equiv \left(x-1\right)\big(x ~ m_\x^2 - m_{\tilde{f}_i}^2\big) + x ~ m_f^2
\quad .
\end{align}
Here, we have dropped terms, such as divergent pieces, that vanish due to gauge invariance when summed over a complete multiplet of sfermions. Although the dependence on the renormalization scale, $\mu$, also drops out, we have kept it explicit above to allow for a more compact form.

In the case of a degenerate right-handed stop ($m_{\tilde{t}_R} = m_\x$), the above result simplifies to
\begin{align}
\label{eq:DegStopZ}
&c_q^{(1)\text{bare}} =\frac{(g^\prime)^2 g_u^a g_q^a m_t^2}{12 \pi ^2 m_\x^2 m_Z^2} \left[ \log{\frac{m_\x^2}{m_t^2}}+2 \left(1-2 m_\x^2/m_t^2\right) (4 m_\x^2/m_t^2-1)^{-1/2} \tan^{-1}{(4 m_\x^2/m_t^2-1)^{1/2}}\right]
~.
\end{align}
\begin{figure}[t]
\begin{center}
\includegraphics[width=0.45\textwidth]{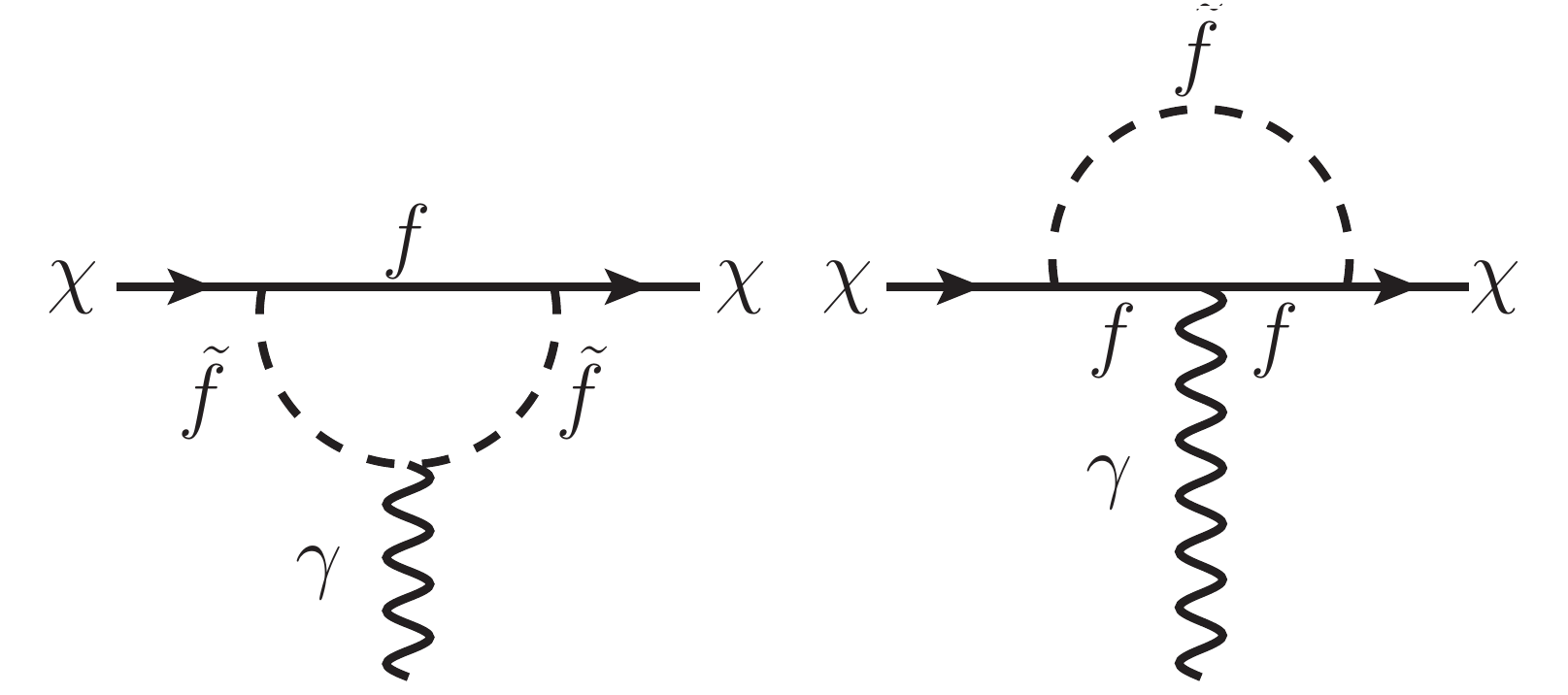} 
\caption{\label{fig:photonexchange}
Feynman diagrams responsible for couplings to the electromagnetic current. Charge-reversed diagrams not shown.}
\end{center}
\end{figure}

For a right-handed sbottom, we find
\begin{align}
\label{eq:SbottomZ}
&c_q^{(1)\text{bare}} = 0
~.
\end{align}
This can be understood in the following manner. We have set the bottom quark mass to zero when matching at the weak scale, and when the sbottom is purely right-handed, this corresponds to an enhanced chiral symmetry. In this case, the bino's coupling to the $Z$ is proportional to photon exchange, which vanishes at zero-momentum transfer due to gauge invariance (as shown in the next section). 

\subsection{Photon Exchange}
\label{app:anapole}

The bino's interactions with charged scalars also generate an effective coupling to the electromagnetic current $J_\mu^\text{EM}\equiv \partial^\nu F_{\nu \mu}$, where $F_{\mu \nu}$ is the photon field strength. The relevant Feynman diagrams are shown in Fig.~\ref{fig:photonexchange}. A Majorana fermion may couple to the photon via the anapole operator, defined in Eq.~(\ref{eq:lowbasisslepton}) of Sec.~\ref{sec:slepton}.

The relevant interactions are parametrized as
\be
\mathcal{L}_{\text{UV}} \supset \Big[ \tilde{f}~ \bar{f} \left( \alpha + \beta \gamma^5 \right) \x -  i e Q_f A^\mu \tilde{f}^\dagger \partial_\mu \tilde{f} + \text{h.c.} \Big] - e Q_f A_\mu \bar{f} \gamma^\mu f  \,,
\ee
where $e$ is the electromagnetic coupling constant and $Q_f$ is the electric charge of $f$ (in units of $e$). Note that conservation of the electromagnetic current ($\partial^\mu J_\mu^\text{EM}=0$) implies that the photon cannot couple sfermions of different mass. Therefore, left-right sfermion mixing does not affect the form of these interactions. In the limit that the fermion mass, $m_f$, is much larger than the typical momentum transfer of scattering events ($\sqrt{-q^2} \sim 50$ MeV), we find
\be
c_A^\text{bare} = \sum\limits_i \frac{-n_c e Q_f \alpha \beta}{48 \pi^2} \int_0^1 dx ~ \frac{3x-2}{\Delta_i} \quad \quad \Big( m_f^2 \gg - q^2 \Big)
\quad ,
\ee
where $n_c$ is the number of colors of $\tilde{f}$, and

\be
\Delta \equiv x(x-1)~m_\x^2 + x ~ m_{\tilde{f}}^2 + (1-x) ~m_f^2
\quad .
\ee
This contribution has already been presented in particular limits. For example, in the limit that $m_f, m_\x \ll m_{\tilde{f}}$ and $\alpha = - \beta = \lambda/2$, the above form reduces to
\be
c_A^\text{bare} = \frac{n_c e Q_f \lambda^2}{96 \pi ^2 m_{\tilde{f}}^2} \log{\frac{m_f^2}{m_{\tilde{f}}^2}}
~,
\ee
which agrees with Ref.~\cite{Bai:2014osa}.

However, in the limit that $m_f^2 \lesssim -q^2$, the light fermion cannot be integrated out, and instead of doing a simple matching to the local anapole operator, we keep the full $q^2$ dependence in $c_A^\text{bare}$. In this case,  $c_A^\text{bare}$ takes a more general form, 

\be
c_A^\text{bare} = \sum\limits_i \frac{-e Q_f \alpha \beta}{8 \pi ^2} \int_0^1 d x_1 ~ \int_0^{1-x_1} d x_2 ~ \Bigg[ \frac{x_2 (2 x_2+x_1-2)}{\Delta}+\frac{(2 x_2-1) (2 x_2+x_1-1)}{2 \tilde{\Delta}}\Bigg] 
~,
\ee
with
\begin{align}
\Delta &\equiv x_1 (x_1-1)m_\x^2 + x_1 m_{\tilde{f}}^2 + (1-x_1) m_f^2 + x_2(x_1+x_2-1)~ q^2 \,,
\nl
\tilde{\Delta} &\equiv x_1 (x_1-1)m_\x^2 + (1-x_1) m_{\tilde{f}}^2 + x_1 m_f^2 + x_2(x_1+x_2-1)~ q^2 \, .
\end{align}

\subsection{Box Diagrams}
\label{app:Box}

\begin{figure}[t]
\begin{center}
\includegraphics[width=0.6\textwidth]{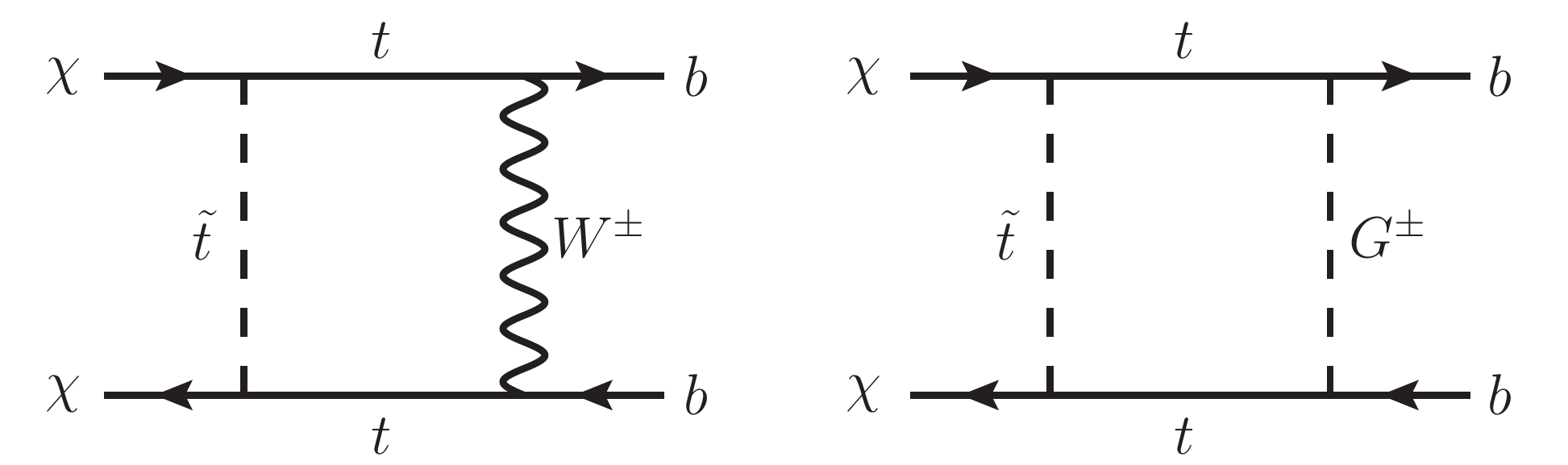} 
\caption{\label{fig:box}
Box-type Feynman diagrams responsible for one-loop couplings to quarks in 5-flavor QCD. Crossed diagrams not shown. Here, $G^\pm$ is the charged Goldstone.}
\end{center}
\end{figure}

In Sec.~\ref{sec:case1}, box diagrams involving gauge and Goldstone bosons contribute to one-loop couplings to bottom quarks. The relevant diagrams are shown in Fig.~\ref{fig:box}. We parametrize the bino-stop interactions as
\be
\mathcal{L}_\text{UV} \supset \tilde{t} ~ \bar{t} \left( \alpha + \beta \gamma^5 \right) \x + \text{h.c.}
~ .
\ee
Similar to the tree-level calculations of Appendix~\ref{app:tree}, obtaining the Wilson coefficients from the diagrams in Fig.~\ref{fig:box} requires the application of Fierz transformations. Working in Feynman gauge, we find
\begin{align}
c_b^{(0)\text{bare}} &=\frac{g_w^2}{16 \pi ^2} \int_0^1 dx_1 ~ \int_0^{1-x_1} dx_2 ~ x_2 (1-x_1-x_2) \Bigg\{ ~ \frac{-(\alpha +\beta)}{2 \Delta^2}\Big[ 3 x_1 m_\x (\alpha +\beta ) + 4 m_t (\alpha -\beta )\Big]
\nl
&+\frac{x_1 m_\x }{\Delta ^3}\Big[ x_1 m_\x (\alpha +\beta ) + m_t (\alpha -\beta )\Big]^2 ~ \Bigg\}
\nl
&~+ \frac{\lambda_G^2}{32 \pi ^2}  \int_0^1 dx_1 ~ \int_0^{1-x_1} dx_2 ~  x_2 (1-x_1-x_2) \Bigg\{\frac{-(\alpha -\beta )}{2 \Delta^2} \Big[ 3 x_1 m_\x (\alpha -\beta ) + 4 m_t (\alpha +\beta )\Big] 
\nl
&+ \frac{x_1 m_\x}{\Delta^3}  \Big[ x_1 m_\x (\alpha -\beta ) + m_t (\alpha +\beta) \Big]^2 ~ \Bigg\}
\end{align}
\begin{align}
c_b^{(2)\text{bare}} &= \frac{g_w^2 m_\x}{4 \pi ^2} \int_0^1 dx_1 ~ \int_0^{1-x_1} dx_2 ~ x_1 x_2 (1-x_1-x_2)  \Bigg\{\frac{1}{2 \Delta ^2}(\alpha +\beta )^2
\nl
&+\frac{1}{\Delta ^3}\Big[ x_1 m_\x (\alpha +\beta ) + m_t (\alpha -\beta )\Big]^2 \Bigg\}
\nl
&~+ \frac{\lambda_G^2 m_\x}{8 \pi ^2} \int_0^1 dx_1 ~ \int_0^{1-x_1} dx_2 ~ x_1 x_2 (1-x_1-x_2) \Bigg\{ \frac{1}{2 \Delta ^2} (\alpha -\beta )^2 
\nl
&+\frac{1}{\Delta ^3} \Big[ x_1 m_\x (\alpha -\beta ) + m_t (\alpha +\beta ) \Big]^2 \Bigg\}
~,
\end{align}
where $g_w = -g/2\sqrt{2}$, $\lambda_G = m_t/2v$, and we have defined
\be
\Delta \equiv x_1 (x_1-1) m_\x^2 + x_1 m_{\tilde{t}}^2 + x_2 m_W^2 + (1-x_1-x_2)m_t^2 \, .
\ee
In the case of a degenerate right-handed stop ($m_\x = m_{\tilde{t}_R}$), the above form simplifies to
\begin{align}
\label{eq:DegStopBox}
&c_b^{(0)\text{bare}} = \frac{g^2 (g^\prime)^2 m_t^2 m_\x  }{48 \pi ^2} \int_0^1 dx_1 ~ \int_0^{1-x_1} dx_2 ~ x_1 x_2 (x_1+x_2-1)\left[\frac{1}{4 \Delta ^2 m_W^2}-\frac{1}{3 \Delta ^3} \bigg( 1 + \frac{x_1^2 m_\x^2}{2 m_W^2}\bigg)\right] 
\nl
&c_b^{(2) \text{bare}} = \frac{-g^2 (g^\prime)^2 m_t^2 m_\x }{36 \pi ^2} \int_0^1 dx_1~ \int_0^{1-x_1} dx_2 ~ x_1 x_2 (x_1+x_2-1) \left[\frac{1}{4 \Delta ^2 m_W^2} + \frac{1}{\Delta^3} \bigg( 1 + \frac{x_1^2 m_\x^2}{2 m_W^2}\bigg)\right] 
\nl
&\Delta \equiv x_1^2 m_\x^2 + x_2 m_W^2 + (1-x_1-x_2)m_t^2
~.
\end{align}
\begin{figure}[t]
\begin{center}
\includegraphics[width=0.5\textwidth]{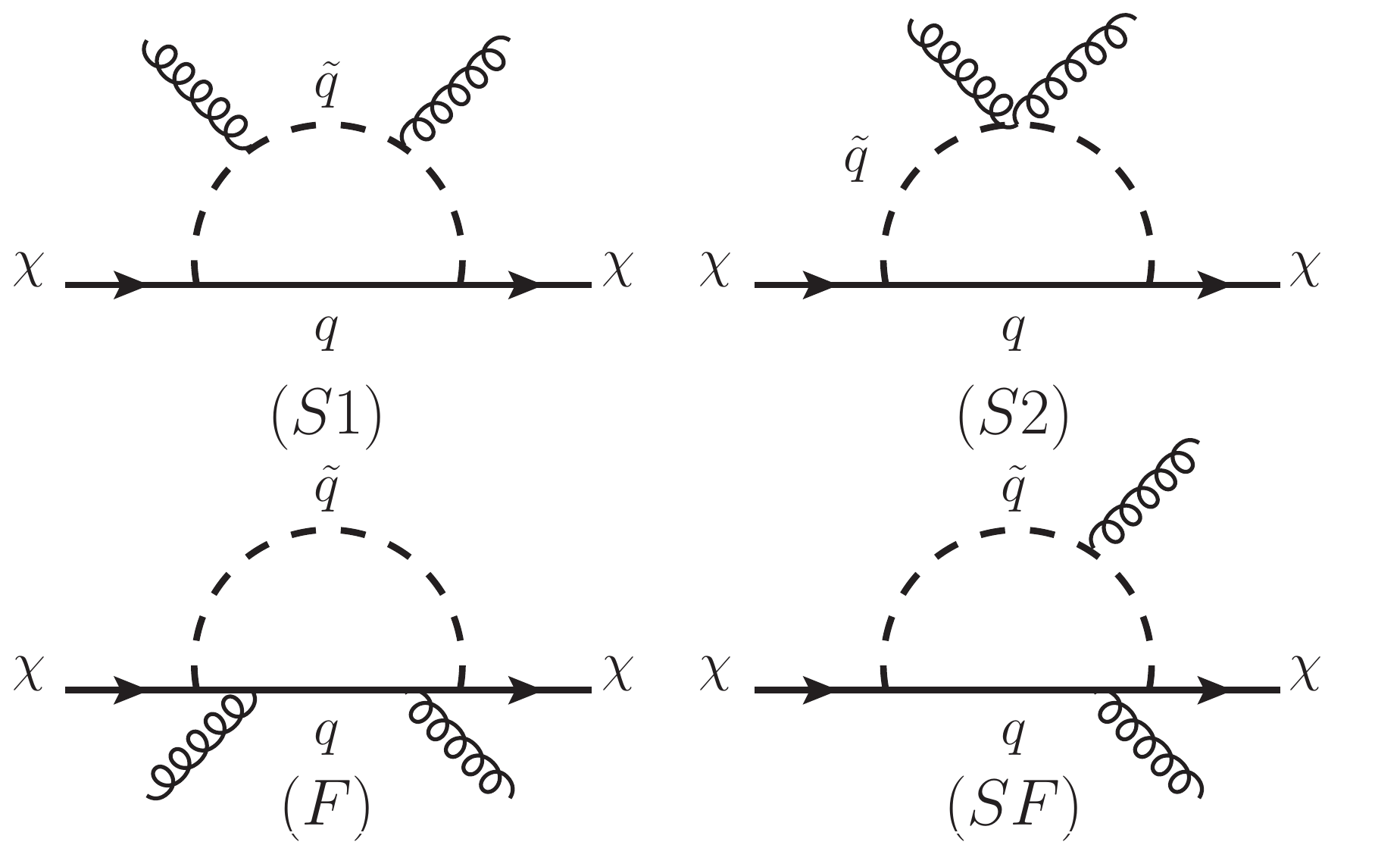} 
\caption{\label{fig:glue}
Full theory Feynman diagrams responsible for one-loop couplings to gluons. Charge-reversed diagrams not shown.}
\end{center}
\end{figure}
%

\subsection{Gluon Couplings in the Full Theory}
\label{app:glue}

In addition to coupling to quarks, $\x$ may scatter off gluons at the one-loop level. The set of Feynman diagrams for this process is shown in Fig.~\ref{fig:glue}. This calculation is simplified by working with Fock-Schwinger gauge in a background gluon field, where gauge invariance is made manifest by expressing the gluon field directly in terms of the field strength. However, this simplification comes at the cost of breaking translational invariance, and, as a result, different forms of colored propagators are needed for the non-reversed and charge-reversed diagrams. We refer the reader to Refs.~\cite{Novikov:1983gd, Hisano:2010ct,Hill:2014yxa} for a detailed discussion. We have cross-checked our results by computing also in Feynman gauge. In this section, IR poles are regulated with dimensional regularization in $d=4-2\epsilon$ dimensions.  Parametrizing the squark-bino interaction Lagrangian as 
\be
\mathcal{L}_\text{UV} \supset \tilde{q} ~ \bar{q} \left( \alpha_q + \beta_q \gamma^5 \right) \x + \text{h.c.}
~,
\ee
we find that the gluon Wilson coefficients corresponding to each diagram (in Fock-Schwinger gauge) are given by

\be
c_g^{(0)\text{bare}}\text{(S1)} = c_g^{(2) \text{bare}}\text{(S1)} = c_g^{(0) \text{bare}}\text{(SF)} = c_g^{(2) \text{bare}}\text{(SF)} = 0
\ee
\begin{align}
\label{eq:bareglue}
c_g^{(0)\text{bare}}\text{(S2)} &= 
\begin{dcases} \frac{-\alpha_s m_{\tilde{q}}^2}{48 \pi } \int_0^1 dx ~ \frac{x^3}{\Delta^3} \left[ x \lambda_q^{(+)} m_\x + \lambda_q^{(-)} m_q \right] \quad \Big( m_{\tilde{q}} > m_\x \text{ or } m_q > 0 \Big)
\\  
\frac{5 \alpha_s}{384 \pi} \frac{\lambda_q^{(+)}}{m_\x^3} \quad \Big( m_{\tilde{q}} = m_\x \text{ and } m_q = 0 \Big) 
\end{dcases}
\nl
\nl
c_g^{(2)\text{bare}}\text{(S2)} &= 
\begin{dcases} \frac{\alpha_s}{12 \pi}  \int_0^1 dx ~ \frac{(1-x) x^3}{\Delta^2} \Big[ \frac{1}{2} \lambda_q^{(+)} m_\x 
\\
+\frac{ 1-x}{\Delta}\left(x \lambda_q^{(+)} m_\x + \lambda_q^{(-)} m_q \right) m_\x^2 \Big] \quad  \Big( m_{\tilde{q}} > m_\x \text{ or } m_q > 0 \Big) 
\\ 
\frac{\alpha_s}{16 \pi} \frac{\lambda_q^{(+)}}{m_\x^3}  \left(\frac{1}{\epsilon_\text{IR}} + \log{\frac{\mu^2}{m_\x^2}} + 1 \right) \quad \Big( m_{\tilde{q}} = m_\x \text{ and } m_q = 0 \Big) 
\end{dcases}
\nl
\nl
c_g^{(0) \text{bare}}\text{(F)} &= \begin{dcases} \frac{\alpha_s}{128 \pi} \int_0^1 dx~\frac{(x-1)^2}{\Delta^2} \Bigg\{\lambda_q^{(-)} m_q \Big(x (3 x-2)+3\Big)+2 \lambda_q^{(+)} m_\x  (3 x-1) x^2 
\\
+ \frac{1}{\Delta} \Bigg[\frac{2}{3} \lambda_q^{(-)} m_q (x-1) \Big(m_q^2 (3 x+1)+3 m_\x ^2 (1-3 x) x^2\Big)
\\
-\frac{1}{3}\lambda_q^{(+)} m_\x  (x-1) x \Big(4 m_\x ^2 x^2 (6 x-1)-m_q^2 (9 x+7)\Big)\Bigg]
\\
+\frac{3}{\Delta^2}(x-1)^2 x^2 (x^2 m_\x^2  -m_q^2) m_\x ^2 \Big[x \lambda_q^{(+)} m_\x  + \lambda_q^{(-)} m_q \Big]\Bigg\} \quad \Big( m_q > 0 \Big)  
\\
0 \quad \Big( m_q = 0 \Big) \end{dcases}
\end{align}
\begin{align}
\label{eq:bareglue2}
c_g^{(2) \text{bare}} \text{(F)} &= \begin{dcases} \frac{\alpha_s m_\x}{16 \pi} \int_0^1 dx~\frac{(x-1)^3 x}{\Delta^2} \Bigg\{-\lambda_q^{(+)} \left(x+\frac{4}{3}\right)+\frac{1}{\Delta} \bigg[\lambda_q^{(+)} \Big(m_\x ^2 x^2 (3 x-1)
\\
-\frac{1}{6} m_q^2 (3 x+5)\Big)+\frac{2}{3}\lambda_q^{(-)} m_q m_\x  x (3 x-2)\bigg]
\\ +\frac{3}{2\Delta^2} (x-1)x (m_q^2-x^2 m_\x^2) m_\x \bigg[x \lambda_q^{(+)} m_\x +\lambda_q^{(-)} m_q  \bigg] \Bigg\} \quad \Big( m_{\tilde{q}} > m_\x \text{ and } m_q > 0 \Big) 
\\
\frac{-\alpha_s \lambda_q^{(+)}}{12 \pi }\Bigg\{ \frac{m_\x}{\left(m_{\tilde{q}}^2-m_\x^2\right)^2}\bigg[ \frac{1}{\epsilon_\text{IR}} +  \log{\frac{\mu^2}{m_{\tilde{q}}^2-m_\x^2}}+\frac{3}{2} 
\\
+ \left( \partial_1 - \partial_2 - \partial_3 \right) \prescript{}{2}{F}_1 \left(4,0;4;\frac{m_\x^2}{m_\x^2-m_{\tilde{q}}^2}\right) \bigg] 
\\
-\frac{\left(2 m_{\tilde{q}}^2+m_\x^2\right)}{m_\x^5}  \log{\frac{m_{\tilde{q}}^2}{m_{\tilde{q}}^2-m_\x^2}} + \frac{2 m_{\tilde{q}}^2}{m_\x^3 \left(m_{\tilde{q}}^2-m_\x^2\right)}\Bigg\} \quad \Big( m_{\tilde{q}} > m_\x \text{ and } m_q = 0 \Big) 
\\ 
\frac{-\alpha_s \lambda_q^{(+)}}{8 \pi m_\x^3} \left( \frac{1}{\epsilon_\text{IR}} + \log{\frac{\mu^2}{m_\x^2}} - \frac{1}{2}\right) \quad \Big( m_{\tilde{q}} = m_\x \text{ and } m_q = 0 \Big) 
~,
\end{dcases}
\end{align}
where $\lambda_q^{(\pm)}$ and $\Delta$ are defined to be
\be
\lambda_q^{(\pm)} \equiv \alpha_q^2 \pm \beta_q^2 ~~,~~ \Delta \equiv x(x-1) m_\x^2 + x m_{\tilde{q}}^2 + (1-x) m_q^2 ~,
\ee
and $\partial_i ~ (\prescript{}{2}{F}_1)$ corresponds to differentiation of the hypergeometric function $\prescript{}{2}{F}_1(a,b;c;d)$ in its $i$-th argument. The above expressions for $c_g^{(0)}$ agree with the results presented in Ref.~\cite{Hisano:2010ct}. 

In the case of a degenerate right-handed stop ($m_\x = m_{\tilde{t}_R}$), the total contribution from the above expressions is
\begin{align}
\label{eq:DegStopGlue}
c_g^{(0)\text{bare}} &= \frac{\alpha_s (g^\prime)^2 m_\x}{18 \pi  \left(4 m_\x^2-m_t^2\right)^2} \bigg[\frac{3}{4}-\frac{m_t^2}{12 m_\x^2}-\frac{2 m_\x^2}{3 m_t^2}
\nl &\quad -\frac{m_\x^2 }{m_t^2} (4 m_\x^2/m_t^2-1)^{-1/2} \tan^{-1}{(4 m_\x^2/m_t^2-1)^{1/2}}\bigg]  \,,
\nl
c_g^{(2)\text{bare}} &= \frac{\alpha_s (g^\prime)^2}{18 \pi  m_\x^3} \Bigg[\frac{1}{2} \log{\frac{m_\x^2}{m_t^2}}+\bigg(\frac{m_\x^2}{m_t^2}-\frac{7m_\x^4}{m_t^4} + \frac{8 m_\x^6}{m_t^6}\bigg) \left(4 m_\x^2/m_t^2-1\right)^{-2}
\nl &\quad +\bigg(1-\frac{10 m_\x^2}{m_t^2}+\frac{92 m_\x^4}{3 m_t^4}-\frac{68 m_\x^6}{3m_t^6}\bigg) \left(4 m_\x^2/m_t^2-1\right)^{-5/2} \tan^{-1}{(4m_\x^2/m_t^2-1)^{1/2}}\Bigg]
~.
\end{align}
Similarly, for a non-degenerate right-handed sbottom ($m_\x \neq m_{\tilde{b}_R}$), we find
\begin{align}
\label{eq:SbottomGlue}
c_g^{(0)\text{bare}} &=  \frac{-\alpha_s (g^\prime)^2}{864 \pi  } \frac{ m_\x}{m_{\tilde{b}_R}^2 \left(m_{\tilde{b}_R}^2-m_\x^2\right)} \, ,
\nl
c_g^{(2)\text{bare}} &= \frac{-\alpha_s (g^\prime)^2 }{216 \pi m_\x^3} \left( \frac{4 m_{\tilde{b}_R}^2-3 m_\x^2}{m_\x^2-m_{\tilde{b}_R}^2} + \frac{4 m_{\tilde{b}_R}^2-m_\x^2}{m_\x^2}\log{\frac{m_{\tilde{b}_R}^2}{m_{\tilde{b}_R}^2-m_\x^2}}\right)
 -\frac{\alpha_s (g^\prime)^2}{108 \pi }\Bigg\{ \frac{m_\x}{\left(m_{\tilde{b}_R}^2-m_\x^2\right)^2}\bigg[ \frac{1}{\epsilon_\text{IR}}
\nl & \quad  +  \log{\frac{\mu^2}{m_{\tilde{b}_R}^2-m_\x^2}}+\frac{3}{2} + \left( \partial_1 - \partial_2 - \partial_3 \right) \prescript{}{2}{F}_1 \left(4,0;4;\frac{m_\x^2}{m_\x^2-m_{\tilde{b}_R}^2}\right) \bigg] 
\nl
&\quad -\frac{\left(2 m_{\tilde{b}_R}^2+m_\x^2\right)}{m_\x^5}  \log{\frac{m_{\tilde{b}_R}^2}{m_{\tilde{b}_R}^2-m_\x^2}} + \frac{2 m_{\tilde{b}_R}^2}{m_\x^3 \left(m_{\tilde{b}_R}^2-m_\x^2\right)}\Bigg\} 
~.
\end{align}
Note that $c_g^{(2) \text{bare}}$ has an IR divergence in the full theory, which arises as a singularity in the integration over Feynman parameters. Upon performing weak scale matching, this is identified as an UV pole of the low-energy theory that is renormalized according to Eq.~(\ref{eq:renormalize}). The contribution from $c_q^{(2)\text{bare}}$, using Eq.~(\ref{eq:HeavySbottomTree}), cancels the divergence precisely, yielding the finite renormalized coefficient
\begin{align}
\label{eq:RenSbottomGlue}
&c_g^{(2)} = \frac{-\alpha_s (g^\prime)^2 }{216 \pi m_\x^3} \left( \frac{4 m_{\tilde{b}_R}^2-3 m_\x^2}{m_\x^2-m_{\tilde{b}_R}^2} + \frac{4 m_{\tilde{b}_R}^2-m_\x^2}{m_\x^2}\log{\frac{m_{\tilde{b}_R}^2}{m_{\tilde{b}_R}^2-m_\x^2}}\right)
\nl
&-\frac{\alpha_s (g^\prime)^2}{108 \pi }\Bigg\{ \frac{m_\x}{\left(m_{\tilde{b}_R}^2-m_\x^2\right)^2}\bigg[  \log{\frac{\mu^2}{m_{\tilde{b}_R}^2-m_\x^2}}+\frac{3}{2} + \left( \partial_1 - \partial_2 - \partial_3 \right) \prescript{}{2}{F}_1 \left(4,0;4;\frac{m_\x^2}{m_\x^2-m_{\tilde{b}_R}^2}\right) \bigg]
\nl
&-\frac{\left(2 m_{\tilde{b}_R}^2+m_\x^2\right)}{m_\x^5}  \log{\frac{m_{\tilde{b}_R}^2}{m_{\tilde{b}_R}^2-m_\x^2}} + \frac{2 m_{\tilde{b}_R}^2}{m_\x^3 \left(m_{\tilde{b}_R}^2-m_\x^2\right)}\Bigg\}
~,
\end{align} 
as explained in Eq.~(\ref{eq:poles}).

\subsection{Gluon Couplings in Heavy Particle Theory}
\label{app:HPTglue}

%
\begin{figure}[t]
\begin{center}
\includegraphics[width=0.25\textwidth]{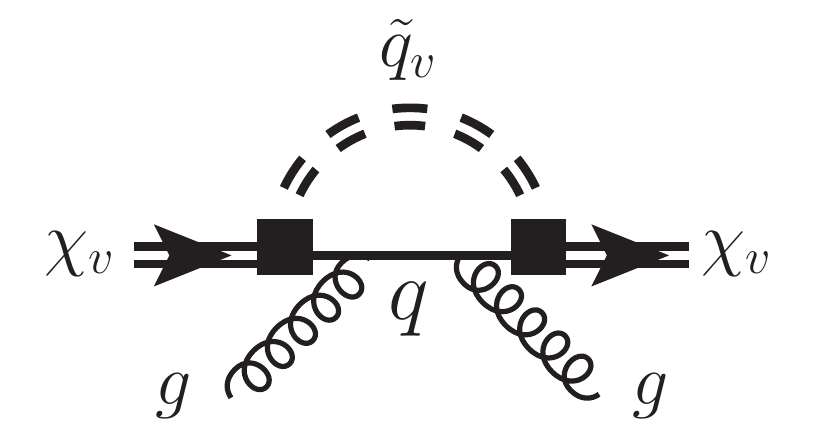} 
\caption{\label{fig:HPTglue}
Leading order Feynman diagram responsible for one-loop couplings to gluons in heavy particle theory. Single (double) lines correspond to relativistic (heavy particle theory) fields.}
\end{center}
\end{figure}

In Sec.~\ref{sec:lightfsmalldelta}, we discussed the modified matching prescription when dealing with a light quark and nearly degenerate squark ($m_q, \delta_{\tilde{q}} \ll m_t$). In this section, we will present the gluon calculation using Fock-Schwinger gauge in the framework of heavy particle theory. We have checked our results in Feynman gauge. The relevant leading order Feynman diagram is shown in Fig.~\ref{fig:HPTglue}. Two of the four possible gluon diagrams vanish exactly in Fock-Schwinger gauge as in Appendix~\ref{app:glue}. These are the heavy particle theory equivalents of diagrams ``S1" and ``SF" in Fig.~\ref{fig:glue}. Furthermore, the diagram involving the 4-point $\tilde{q}-\tilde{q}-g-g$ vertex is found to be subleading in $1/m_\x$~, and therefore, only the diagram of Fig.~\ref{fig:HPTglue} contributes at leading order. Parametrizing the heavy particle Lagrangian as 
\be
\mathcal{L}_\text{HPT} \supset \frac{1}{\sqrt{m_\x}} ~ \tilde{q}_v ~ \bar{q} \left( \alpha_q + \beta_q \gamma^5 \right) \x_v + \text{h.c.}
~,
\ee
the diagram in Fig.~\ref{fig:HPTglue} is written in Fock-Schwinger gauge as
\begin{align}
\label{eq:hptamp1}
i {\cal M} &= \frac{- \pi \alpha_s}{2 m_\x} ~ G_{\alpha \mu}^A G_{\beta \nu}^A \int \frac{d^d l}{(2\pi)^d} \frac{i}{-v \cdot l - \delta_{\tilde{q}}} ~ \frac{\partial}{\partial k_{1 \mu}}  ~ \frac{\partial}{\partial k_{2 \nu}} ~ \bigg[ \bar{u}(k) (\alpha_q - \beta_q \gamma^5) i S^{(0)} (l) \gamma^\alpha i S^{(0)} (l-k_1) \gamma^\beta 
\nl
& \quad \quad \quad \quad \quad \quad \quad \quad \quad \quad \quad \quad \quad \quad \quad \quad \quad 
\times i S^{(0)} (l-k_1-k_2) (\alpha_q + \beta_q \gamma^5) u(k) \bigg] \Bigg|_{k_{1,2}=0}  
~,
\end{align}
where $k$ is the residual bino momentum ($p= m_\x v + k$), and we have defined the free quark propagator
\be
iS^{(0)} (p) \equiv \frac{i(\slashed{p}+m_q)}{p^2 - m_q^2}
~.
\ee
The tensor $G_{\alpha \mu}^A G_{\beta \nu}^A$ can be projected onto the scalar and spin-2 gluon currents defined in Eq.~(\ref{eq:ops}) as\footnote{Note that the last two terms of the second line of Eq.~(\ref{eq:glueproj}) differ by a sign from those of Eq.~(50) in \cite{Hisano:2010ct}.}
\begin{align}
\label{eq:glueproj}
G_{\alpha \mu}^A G_{\beta \nu}^A &= \frac{1}{d (d-1)} (g_{\alpha \beta} g_{\mu \nu} - g_{\alpha \nu} g_{\beta \mu}) ~ O_g^{(0)}
\nl
&+ \frac{1}{d-2} \left( - g_{\alpha \beta} ~ O_{g \mu \nu}^{(2)} - g_{\mu \nu} ~ O_{g \alpha \beta}^{(2)} + g_{\alpha \nu} ~ O_{g \beta \mu}^{(2)} + g_{\beta \mu} ~ O_{g \alpha \nu}^{(2)}\right) + \cdots
~,
\end{align}
where the ellipsis denotes higher spin tensor contributions. 

Using the top line of Eq.~(\ref{eq:glueproj}) in Eq.~(\ref{eq:hptamp1}) leads to
\be
c_g^{(0)\text{bare}} = \frac{2 i \pi \alpha_s m_q}{d ~ m_\x} ~\int \frac{d^d l}{(2\pi)^d} ~ \frac{\lambda_q^{(+)} ~ m_q ~ \slashed{l} + \lambda_q^{(-)} ~ l^2}{(l^2-m_q^2)^4 ~ (v \cdot l + \delta_{\tilde{q}})}
~,
\ee 
where $\lambda_q^{(\pm)} \equiv \left( \alpha_q^2 \pm \beta_q^2 \right)$~. The coefficient $c_g^{(2)}$ is similarly evaluated by using the bottom line of Eq.~(\ref{eq:glueproj}) in Eq.~(\ref{eq:hptamp1}). We find

\begin{align}
\label{eq:HPTSbottomGlue}
c_g^{(0)\text{bare}} &= \frac{\alpha_s m_q}{96 \pi m_\x}  ~\int_0^\infty dx ~ \Bigg[ \frac{\lambda_q^{(-)}}{\Delta^2} - \frac{x}{4 \Delta^3} \big( 2 \lambda_q^{(+)} ~ m_q + \lambda_q^{(-)} ~ x \big) \Bigg]
\nl
\nl
c_g^{(2)\text{bare}} &= \frac{\alpha_s}{16 \pi m_\x} ~\int_0^\infty dx ~ \Bigg[ \frac{\lambda_q^{(+)} x}{3 \Delta^2} +\frac{x}{48 \Delta ^3} \left(10 \lambda_q^{(+)} m_q^2+ 3 \lambda_q^{(+)} x^2 + 8 \lambda_q^{(-)} m_q~x \right)
\nl 
& + \frac{3 x^2}{128 \Delta ^4} \left(4 m_q^2 - x^2\right) \left(2 \lambda_q^{(-)} m_q+\lambda_q^{(+)} x \right)\Bigg]
~,
\end{align}
where $x$ is a dimensionful Schwinger parameter and $\Delta \equiv \frac{1}{4} x^2 + \delta_{\tilde{q}}~x + m_q^2$~.

\vspace{-0.03\textwidth}

\newpage
\bibliography{1Loopbino}

\end{document}